\documentclass[showpacs,prc,nofootinbib,floatfix]{revtex4}

\usepackage{graphicx}
\usepackage{latexsym}
\usepackage{hhline}
\usepackage{amssymb}
\usepackage{wasysym}

\sloppypar

\newcommand{\be}{\begin{equation}}
\newcommand{\ee}{\end{equation}}
\newcommand{\bea}{\begin{eqnarray}}
\newcommand{\eea}{\end{eqnarray}}
\newcommand{\ba}{\begin{array}}
\newcommand{\ea}{\end{array}}
\newcommand{\bi}{\begin{itemize}}
\newcommand{\ei}{\end{itemize}}
\newcommand{\mi}{\mbox i}

\newcommand{\refe}[1]{(\ref{#1})}
\newcommand{\mca}{{\mathcal A}}

\newcommand{\mcf}{{\mathcal F}}
\newcommand{\mci}{{\mathcal I}}
\newcommand{\mck}{{\mathcal K}}
\newcommand{\mcl}{{\mathcal L}}
\newcommand{\mcm}{{\mathcal M}}
\newcommand{\mcp}{{\mathcal P}}

\newcommand{\mct}{{\mathcal T}}
\newcommand{\mcv}{{\mathcal V}}
\newcommand{\difd}{\mathrm d}
\newcommand{\vt}{\vartheta}
\newcommand{\vth}{\frac{\vartheta}{2}}
\newcommand{\vp}{\varphi}
\newcommand{\ve}{\varepsilon}
\newcommand{\ra}{\rightarrow}

\newcommand{\bigbracket}{\begin{array}{c} \\ \\ \end{array} \hspace{-4mm}}
\renewcommand{\vec}[1]{\mbox{\boldmath $#1 \!\!$ \unboldmath}}
\renewcommand{\slash}{/ \!\!\!\!\,}
\newcommand{\smallfrac}[2]{\mbox {$\frac{#1}{#2}$}}
\newcommand{\avec}[1]{\: \: \! \! {\mathrm #1}}
\newcommand{\foh}{\frac{1}{2}}
\newcommand{\fot}{\frac{1}{3}}
\newcommand{\fof}{\frac{1}{4}}

\newcommand{\fth}{\frac{3}{2}}
\newcommand{\ffh}{\frac{5}{2}}
\newcommand{\fvh}{\frac{\vartheta}{2}}
\newcommand{\sqt}{\sqrt 2}
\newcommand{\sqth}{\sqrt 3}
\newcommand{\sfoh}{\smallfrac{1}{2}}
\newcommand{\sfot}{\smallfrac{1}{3}}

\newcommand{\sfth}{\smallfrac{3}{2}}
\newcommand{\sftt}{\smallfrac{2}{3}}
\newcommand{\umat}{1 \! \! 1}
\newcommand{\ncdot}{\! \cdot \!}
\newcommand{\etp}{\, \varepsilon \! \cdot \! p \,}
\newcommand{\etpp}{\, \varepsilon \! \cdot \! p' \,}
\newcommand{\etep}{\, \varepsilon \! \cdot \! \varepsilon' \,}
\newcommand{\eptp}{\, \varepsilon' \! \cdot \! p \,}
\newcommand{\eptpp}{\, \varepsilon' \! \cdot \! p' \,}
\newcommand{\eptk}{\, \varepsilon' \! \cdot \! k \,}
\newcommand{\JPR}[3]{Phys. Rev. {\bf #1}, #2 (#3)}
\newcommand{\JPS}[3]{Phys. Scr. {\bf #1}, #2 (#3)}
\newcommand{\JPRL}[3]{Phys. Rev. Lett. {\bf #1}, #2 (#3)}
\newcommand{\JPL}[3]{Phys. Lett. {\bf #1}, #2 (#3)}
\newcommand{\JPRC}[3]{Phys. Rev. C {\bf #1}, #2 (#3)}
\newcommand{\JPRD}[3]{Phys. Rev. D {\bf #1}, #2 (#3)}
\newcommand{\JNP}[3]{Nucl. Phys. {\bf #1}, #2 (#3)}
\newcommand{\JNPP}[3]{Nucl. Phys. B, Proc. Suppl. {\bf #1}, #2 (#3)}

\newcommand{\JAP}[3]{Ann. Phys. (N.Y.) {\bf #1}, #2 (#3)}
\newcommand{\JAPP}[3]{Acta Phys. Pol. {\bf #1}, #2 (#3)}
\newcommand{\JEPJA}[3]{Eur. Phys. J. A {\bf #1}, #2 (#3)}
\newcommand{\JEPJC}[3]{Eur. Phys. J. C {\bf #1}, #2 (#3)}
\newcommand{\JNCA}[3]{Nuovo Cimento A {\bf #1}, #2 (#3)}
\newcommand{\JNC}[3]{Nuovo Cimento {\bf #1}, #2 (#3)}
\newcommand{\JZP}[3]{Z. Phys. {\bf #1}, #2 (#3)}

\newcommand{\Jpin}[3]{$\pi N$-Newsletter {\bf #1}, #2 (#3)}
\newcommand{\JPPNP}[3]{Prog. Part. Nucl. Phys. {\bf #1}, #2 (#3)}
\newcommand{\ibid}[3]{{\it ibid.} {\bf #1}, #2 (#3)}
\newcommand{\JEPJ}[3]{Eur. Phys. J. {\bf #1}, #2 (#3)}
\newcommand{\JPRep}[3]{Phys. Rep. {\bf #1}, #2 (#3)}

\begin{document}

\title{Vector meson production and nucleon resonance analysis in a
  coupled-channel approach for energies $m_N < \sqrt s < 2$ GeV \\
  I: pion-induced results and hadronic parameters}

\author{G. Penner}
\email{gregor.penner@theo.physik.uni-giessen.de}
\author{U. Mosel}
\affiliation{Institut f\"ur Theoretische Physik, Universit\"at Giessen, D-35392
Giessen, Germany}


\begin{abstract}
We present a nucleon resonance analysis by simultaneously considering
all pion- and photon-induced experimental data on the final states
$\gamma N$, $\pi N$, $2\pi N$, $\eta N$, $K\Lambda$, $K\Sigma$, and
$\omega N$ for energies from the nucleon mass up to $\sqrt s = 2$
GeV. In this analysis we find strong evidence for the resonances
$P_{31}(1750)$, $P_{13}(1900)$, $P_{33}(1920)$, and
$D_{13}(1950)$. The $\omega N$ production mechanism is dominated by
large $P_{11}(1710)$ and $P_{13}(1900)$ contributions. In this first
part, we present the results of the pion-induced reactions and the
extracted resonance and background properties with emphasis on the
difference between global and purely hadronic fits.
\end{abstract}


\maketitle

\section{Introduction}

The reliable extraction of nucleon resonance properties from experiments 
where the nucleon is excited via either hadronic or electromagnetic 
probes is one of the major issues of hadron physics. The goal is 
to be finally able to compare the extracted masses and partial-decay widths 
to predictions from lattice QCD (e.g., Ref. \cite{flee}) and/or quark models 
(e.g., Refs. \cite{capstick,riska}).

Basically all information about nucleon resonances identified so far
from experiment \cite{pdg} stems from analyses of pion-induced $\pi
N$ and $2\pi 
N$ production \cite{manley92,arndt95,vrana}, and also from pion
photoproduction \cite{arndt96,arndt02}. However, it is well known
that, for example, in the case of the $S_{11}(1535)$ the consideration
of the $\eta N$ final state is inevitable to extract its properties
reliably, and similar effects can be expected for higher lying
resonances and different thresholds. Only in the analysis of Vrana {\it et
al.} \cite{vrana} the model space has been extended to also include
information on $\pi N \ra \eta N$ in the comparison with experimental
data. On the other side, quark models
predict a much richer resonance spectrum than has been found in $\pi
N$ and $2\pi N$ production so far, giving rise to speculations that
many of these resonance states only become visible in other reaction
channels. This is the basis for a wealth of analyses concentrating on
identifying these ``missing'' or ``hidden'' resonances in the
production of other final states as $\eta N$, $K\Lambda$, $K\Sigma$,
or $\omega N$. For a consistent identification of those resonances and 
their properties, the consideration of unitarity effects are
inevitable and as many final states as possible have to be taken into
account simultaneously. With this aim in mind we developed in Refs. 
\cite{feusti98,feusti99} a unitary coupled-channel effective
Lagrangian model (\textit{Giessen  model}) that 
incorporated the final states $\gamma N$, $\pi N$, $2\pi N$, $\eta N$,
and $K \Lambda$ and was used for a simultaneous analysis of all
available experimental data on photon- and pion-induced reactions on the 
nucleon. In later studies the model was used to also analyze 
kaon-induced reactions \cite{matze} and for a first investigation on
$\pi N \ra K\Sigma$ \cite{agung}. The premise is to use the \textit{same
  Lagrangians} for the pion- and photon-induced reactions allowing for
a consistent analysis, thereby generating the background dynamically
from $u$- and $t$-channel contributions without new parameters.

In an extension of the model to higher center-of-mass energies, i.e., 
up to c.m. energies of $\sqrt s = 2$ GeV for the
investigation of higher and hidden or missing nucleon
resonances, the consideration of the $\omega N$ state in a unitary
model is mandatory. Furthermore, $\omega$ production on the nucleon
represents a possibility to project out $I = \foh$ resonances in the
reaction mechanism. The inclusion of $K\Sigma$ gives additional
information on resonance properties, since especially in the pure
$I=\fth$ reaction $\pi^+ p \ra K^+ \Sigma^+$ many data have been 
taken in the $1960$s and $1970$s. It is also known \cite{nobby} that the
inclusion of the $K\Sigma$ final state can have an important influence
on the description of $K\Lambda$ observables. Hence we have extended
the model of Refs. \cite{feusti98,feusti99} to also include $\omega N$
and $K \Sigma$.

For the newly incorporated channels $\omega N$ and $K\Sigma$, almost
all models in the literature are based on single-channel effective
Lagrangian calculations, ignoring rescattering effects (often called
``$T$-matrix models'') and thereby the influence of the extracted
resonance properties on other reaction channels. 
This problem can only be circumvented if all channels are compared
simultaneously to experimental data thereby restricting the freedom
severely; this is done in the model underlying the present
calculation. To our knowledge, the only other calculation considering
the $\omega N$ channel in a coupled-channel approach is the model by
Lutz {\it et al.} \cite{lutz}, where pointlike interactions are
used. There, the 
complexity of the vector-meson nucleon states is further simplified by
the use of only one specific combination of the $VN$ helicity states
(cf. Appendix \ref{appnot}). Due to the lack of $J^P=\foh^+$ and
$J^P=\fth^+$ $(P)$ waves in their model, these authors are 
only able to compare to production cross sections at energies very
close to the corresponding threshold by assuming $S$-wave
dominance. The photon coupling is implemented via strict vector meson
dominance (VMD), i.e., the photon can only couple to any other particle
via its ``hadronic'' components, the $\rho$ and $\omega$
mesons. 

In Ref. \cite{gregi} we have presented our first results on the
analysis of the pion-induced reactions. In this work, we give a
comprehensive discussion of the results for the pion-induced
reactions, both with and without additionally taking into account the 
photoproduction data, which allow us to pin down the resonant
contributions even more reliably. The results of the photoproduction
reactions themselves are presented in the succeeding paper \cite{pm2}, 
called PMII in the following. Hence this
analysis differs from all other resonance analyses by its larger
channel space. For the investigation of the $\pi N \ra 
\omega N$ channel, this calculation is different from other models in
the following respects: First, a larger energy region is considered,
which also means there are more restrictions from experiment, second,
the reaction process is influenced by all other channels and vice
versa, and third, also a large $\omega$ photoproduction data base is
taken into account. This leads to strong constraints in the choice of
$\omega N$ contributions, and it is therefore possible to extract
these more reliably.

We start in Sec. \ref{model} with a review of the model of
Ref. \cite{feusti98,feusti99,gregi} with special 
emphasis on the extensions. In Sec. \ref{secresultsintro} the
performed calculations are described and in Sec.
\ref{secpiindresults} these calculations are compared to the available
experimental data. We conclude with a summary. In the appendices, we
give a self-contained summary of the full formalism underlying the
present model; more details can be found in Ref. \cite{gregiphd}. The
formalism and the results for the photon-induced reactions are given
in PMII \cite{pm2}.

\section{\label{model}Giessen Model}

The scattering equation that needs to be solved is the Bethe-Salpeter
(BS) equation for the scattering amplitude:
\bea
M(p',p;\sqrt s) = V(p',p;\sqrt s) 
+ \int \frac{\mathrm d^4 q}{(2\pi)^4} V(p',q;\sqrt s)
G_{BS}(q;\sqrt s) M(q,p;\sqrt s) 
\label{bse}
\eea
in the notation given in Appendix \ref{appnot}. 
Here, $p$ ($k$) and $p'$ ($k'$) are the incoming and outgoing baryon
(meson) four-momenta. After splitting up the two-particle BS propagator 
$G_{BS}$ into its real and imaginary parts, one can introduce the $K$ matrix 
via (in a schematical notation) $K = V + \int V \mathrm{Re} G_{BS}
M$. Then $M$ is given by $M = K + i \int M \mathrm{Im} G_{BS}
K$. Since the imaginary part of $G_{BS}$ just contains its on-shell
part
\bea
\mi \mathrm{Im} (G_{BS}(q)) =
- \mi \pi^2 
\frac{m_{B_q} \sum_{\lambda_{B_q}} 
u(p_q, \lambda_{B_q}) \bar u(p_q,\lambda_{B_q})}
{E_{B_q} E_{M_q}}
\delta (k_q^0 - E_{M_q}) \delta (p_q^0 - E_{B_q}) 
\label{kmatprop} \; ,
\eea
the BS equation simplifies to
\bea
\mct^{fi}_{\lambda' \lambda} =
\mck^{fi}_{\lambda' \lambda} + \mi 
\int \difd \Omega_a \sum_a
\sum_{\lambda_a} \mct^{fa}_{\lambda' \lambda_a} 
\mck^{ai}_{\lambda_a \lambda} \; ,
\label{tkrelwint}
\eea
where we have introduced the $\mct$ and $\mck$ amplitudes defined in
Appendix \ref{appnot}. $a$ represents the intermediate two-particle
state. As shown in Appendix \ref{apppwd} this can be
further simplified for parity conserving and rotationally invariant
interactions by a partial-wave decomposition in $J$, $P$, and $I$ and
one arrives at an algebraic equation relating the decomposed
$\mct^{fi}$and $\mck^{fi}$:
\bea
\mct^{IJ\pm}_{fi} =
\left[ \frac{\mck^{IJ\pm}}{1 - \mi \mck^{IJ\pm}} \right]_{fi} \; .
\label{bsematinv}
\eea
Hence unitarity is fulfilled as long as $\mck$ is Hermitian. 

To date, a full solution of the BS Equation \refe{bse} in the meson-baryon
sector has only been possible for low-energy $\pi N$ scattering
\cite{lahiff}, i.e., where no other channels are
important. Consequently, various approximations to the BS Equation  
\refe{bse} preserving unitarity can be found in the literature. Many of 
these approximations reduce the four-dimensional BS Equation \refe{bse} to a
three-dimensional Lippmann-Schwinger equation. However, due to
technical feasibility, most of them are restricted to elastic
pion-nucleon scattering, while only a few ones also include inelastic
channels \cite{gross,krehl}. A general problem of the
three-dimensional (3D) reduction
is the way the reduction is performed. There is no unique method
\cite{gross}; it can even be shown that the 3D reduction can be
achieved in an infinite number of ways, all of which satisfy Lorentz
invariance and elastic two-body unitarity \cite{yaes}. In view of the
number of parameters that have to be determined by comparison of our
effective Lagrangian calculation with experimental data, we apply the
so-called $K$-matrix Born approximation, which is the only feasible
method that still satisfies the important condition of unitarity. In the
$K$-matrix Born approximation, the real part of $G_{BS}$ is 
neglected and thus $K$ reduces to $K = V$. 

The validity of the effective Lagrangian $K$-matrix method as compared
to calculations accounting also for analyticity has first been tested
by Pearce and Jennings \cite{pearce}. By fitting the $\pi N$ elastic
phase shifts up to $\approx 1.38$ GeV with various approximations to
the intermediate two-particle propagator $G_{BS}$, these authors have
found no significant differences in the parameters extracted in the
various schemes. It has also been deduced that the contributions of
$G_{BS}$ to the principal value part of the scattering equation are of
minor importance, since they have been reduced by a very soft cutoff
dictated by experimental data. It has been concluded that --- in order
to fulfill the low-energy theorems --- an important feature of the
reduced intermediate two-particle propagator is a delta function on
the energy transfer. It has also been argued in \cite{goudsmit}, that
for $\pi N$ scattering the main effect of the real part of the
intermediate loop integrals is a renormalization of the coupling
constants and masses of the involved particles. Therefore in the
present $K$-matrix calculation these are taken to be physical values
and are either taken from other reliable sources (if available) or to
be determined by comparison with experimental data.

It should be mentioned that within the $K$-matrix method
the nature of a resonance as a genuine three-quark excitation or dynamic
scattering resonance cannot be determined. There are, e.g., hints,
that the Roper $P_{11}(1440)$ resonance is a quasibound $\sigma N$
state \cite{krehl}. In addition, in the chiral models of Refs. 
\cite{nobby} and \cite{inoue} the $S_{11}(1535)$ can be explained as a 
quasibound meson-baryon ($K\Sigma$ and $\eta N$) state. Moreover, it
has been shown in Ref. \cite{denschlag} by using a 
generalized separable Lee model, that explicit $S_{11}(1535)$
resonance contributions might not play a large role if the
coupled-state system $\pi N \oplus \eta N$ is treated analytically,
i.e., the real part of the Bethe-Salpeter propagator $G_{BS}$
is taken into account. Because of the neglect of the real part of
$G_{BS}$ in the $K$-matrix approximation, these resonances cannot be
generated dynamically as quasibound meson-baryon states, but have to
be put into the potential explicitly. We note, however, that a clean
distinction between dynamic and quark-state resonances is very
difficult, if not impossible. If at all possible, it may require more
and other data than analyzed here, in particular also from
electroproduction (see, e.g., Ref. \cite{burkert}), where 
information on the spatial extent of the states can be obtained.

\subsection{Potential}
\label{secpot}

The interaction potential in the Giessen model is determined by the
inclusion of $s$-, $u$-, and $t$-channel contributions generated by
means of an effective generic Lagrangian,
\bea
\mcl = \mcl_{Born} + \mcl_t + \mcl_{Res} \; ,
\label{genlagr}
\eea
where $\mcl_{Born} + \mcl_t$ is given fully in Eq. \refe{lagback} and
the resonance Lagrangians are summarized in Appendices
\ref{appres12lagr} and \ref{appres32lagr}. 
Consequently, the background is dynamically generated by the Born
terms ($\mcl_{Born}$), the $t$-channel exchanges ($\mcl_t$), and the
$u$-channel contributions of the resonance couplings
($\mcl_{Res}$). Since these background terms give contributions to all
partial waves simultaneously, the number of free parameters is largely
reduced.

\subsubsection{Background contributions}
\label{secnuclchi}

In this section, we discuss the ingredients of the Born and
$t$-channel Lagrangian $\mcl_{Born} + \mcl_t$ of Eq. \refe{genlagr},
where the $\pi N$ part underlies special constraints due to chiral
symmetry.

Since an effective hadronic interaction Lagrangian should resemble the
underlying fundamental theory QCD as closely as possible, the
interaction  also should be in conformity with chiral symmetry, which
is known to be important for low-energy $\pi N$ physics. We choose 
Weinberg's nonlinear realization \cite{weinberg68} and thus
pseudovector pion-nucleon coupling: $\gamma_5 \gamma_\mu
\partial^\mu \vec \pi \ncdot \vec \tau$ and identify the
Weinberg-Tomazawa contact term \cite{weinbergtomazawa,weinberg68},
which automatically accounts for the values of the $\pi N$ scattering
lengths, with a $\rho$ meson exchange. Thus the $\rho$ couplings
should be fixed by the Kawarabayashi-Suzuki-Riazuddin-Fayyazuddin 
(KSRF) relation \cite{ksrf}: $\sqrt{g_\rho
  g_{\rho \pi \pi}} = m_\rho / (2 f_\pi )$ with the pion-decay
constant $f_\pi = 93$ MeV, which gives $g_\rho \approx 
2.84$ using the value $g_{\rho \pi \pi} = 6.02$. It should be remarked 
that this equivalence only holds at threshold, while the energy
dependence of the $\rho$ exchange is different from the
Weinberg-Tomazawa contact term. Since the aim of the present
calculation is the analysis of a wide energy region, we
allow for deviations from the KSRF relation by varying the $\rho$
nucleon coupling $g_\rho$.

In the nonlinear chiral symmetry realization the $\sigma$ meson is not 
needed. Nevertheless, a $t$-channel $\sigma$ exchange can be used to  
model an effective interaction, representing higher-order processes
such as the correlated $2\pi$ exchange in the scalar-isoscalar wave,
which is not explicitly included in our potential. In order to keep
the agreement with chiral symmetry and the soft-pion theorem, the
derivative coupling of the sigma to the pion $(\sigma \partial_\mu
\vec \pi \partial^\mu \vec \pi )$ should be used. Indeed, in the $\pi
N$ sector the background part of $\mcl$ of Eq. \refe{genlagr} respects 
chiral symmetry and is identical to that used in Refs.
\cite{pearce,lahiff,pascatjon}:
\bea
\mcl_{\chi} = - \bar u \left[ 
\frac{g_\pi}{2 m_N} \gamma_5 \gamma_\mu (\partial^\mu \vec \pi ) 
\vec \tau  + g_\sigma \sigma +
g_\rho \left( \gamma_\mu - \frac{\kappa_\rho}{2 m_N} \sigma_{\mu \nu}
  \partial^\nu_\rho \right) \vec \rho^\mu \vec \tau 
\right] u
- \frac{g_{\sigma \pi \pi}}{2 m_\pi} (\partial_\mu \vec \pi )
(\partial^\mu \vec \pi') 
\sigma 
- g_{\rho \pi \pi} \left( \vec \pi \times (\partial_\mu \vec \pi') \right)
\vec \rho^\mu \; .
\label{lagrpin}
\eea
Note that in Refs. \cite{feusti98,feusti99} the sigma meson had not been
included. To investigate the effects of chiral symmetry breaking, we
have also performed a calculation using a direct $\sigma \vec \pi \vec
\pi$ coupling as in Refs. \cite{pascatjon,goudsmit}.

Since the $\sigma$ meson is supposed to model the scalar-isoscalar
two-pion correlated exchange, its mass $m_\sigma$ is {\it a priori} not
fixed. In Ref. \cite{pearce,lahiff} $m_\sigma$ was thus used as a free
parameter and fitted to $\pi N \ra \pi N$ data. In our calculation, it
turns out that the final quality of the fit is almost independent of
the actual value. As long as it is in a reasonable range of $m_\sigma
\approx 450-750$ MeV a change in $m_\sigma$ can be compensated by a
change in $g_{\sigma NN} g_{\sigma \pi \pi}$. For example, a mass
change from $m_\sigma = 650$ to $560$ MeV leads to a coupling
reduction of about 30\% while all other $\pi N$ parameters change at
most by a few percent. The mass of the sigma meson has thus been
chosen as $650$ MeV, which was also used in Ref. \cite{krehl}. There, the
correlated two-pion exchange in the scalar-isoscalar channel was also
parametrized by a $\sigma$ meson exchange and $m_\sigma$ was
determined by comparison to the $\pi \pi$ dynamical model of Ref. 
\cite{durso}. The value for $m_\sigma$ is in line with the values
found by Refs. \cite{pearce} and \cite{lahiff}, and also in
the range of $\pi \pi$ calculations and predictions
\cite{surovtsev,tornqvist}.

Several investigations on
$\eta$ production \cite{tiator,ben95,sauermann,feusti98,feusti99} have
found $\eta  NN$ couplings five to ten times smaller compared to $\pi NN$,
leading to a minor significance of the choice for the $\eta NN$
coupling. In particular, this has been demonstrated in \cite{ben95},
where several fits on $\eta$ photoproduction data using pseudoscalar
(PS) and pseudovector (PV)
eta-nucleon coupling have been performed, showing that the resulting
magnitude of the $\eta NN$ coupling and the quality of the fit hardly
differ. In the case of $K\Lambda N$, however, from SU(3)
considerations, the coupling is expected to be larger. Thus one would
expect observable differences between PS and PV coupling. This point
has been examined in the Giessen model \cite{feusti99} and in a
single-channel effective Lagrangian model \cite{cheoun}. Performing 
calculations with both coupling schemes, however, has revealed that
neither the magnitude of $g_{K\Lambda N}$ nor the quality of the fit
differ significantly in both cases as long as form factors are 
used. Therefore here the same PS-PV choice is made as in Ref.
\cite{feusti99}, i.e., using PV coupling for all Born couplings besides
$\eta NN$. Note that as in Refs. \cite{feusti98,feusti99} no $u$-channel
Born diagrams are taken into account in $K\Lambda$ and $K\Sigma$
production.

To circumvent the problem of the inclusion of the full $2 \pi N$
complexity ($\pi \Delta$, $\rho N$, $\sigma N$, ...), we continue to
parametrize the $2\pi N$ channel effectively by the $\zeta N$ channel
\cite{sauermann,feusti98,feusti99}. Here, $\zeta$ is treated as a
scalar-isovector meson of mass $m_\zeta = 2 m_\pi$. A consistent
description of background contributions for the $2\pi N$ channels is
hence difficult, since each background diagram would introduce
meaningless coupling parameters. In the case of the baryon resonances,
however, the situation is different because the decay into $\zeta N$
can be interpreted as the total ($\sigma N + \pi \Delta + \rho N +
\dots$) $2\pi N$ width. As it turns out, a qualitative description of
the $2\pi N$ partial-wave flux data from Manley {\it et al.} \cite{manley84}
(see Sec. \ref{secresp2}) is indeed possible by allowing for the
$2\pi N$ production only via baryon resonances. Therefore no 
$t$-channel and Born contributions to $2\pi N$ are
included in the model. 

The nucleon couplings to the $\omega$ meson are chosen in analogy to
the $\gamma NN$ and $\rho NN$ couplings and are the same as used in
Refs. \cite{feusti99,gregi}.

The properties of all considered $t$-channel mesons (and asymptotic
particles) are given in Table \ref{tabparticleprops}.
\begin{table}
  \begin{center}
    \begin{tabular}
      {l|c|r|r|r|c}
      \hhline{======}
      & mass [GeV] & $S$ & $P$ & $I$ & $t$-channel contributions \\ 
      \hhline{======}
      $N$ & $0.939$ & $\foh$ & $+$ & $\foh$ & \\
      $\Lambda$ & $1.116$ & $\foh$ & $+$ & $0$ & \\
      $\Sigma$ & $1.193$ & $\foh$ & $+$ & $1$ & \\
      \hline 
      $\pi$ & 0.138 & $0$ & $-$ & $1$ & $(\gamma,\gamma),(\gamma,\pi),(\gamma,\omega)$ \\
      $\zeta$ & 0.276 & $0$ & $+$ & $1$ & \\
      $K$ & 0.496 & $0$ & $-$ & $\foh$ & $(\gamma,\Lambda),(\gamma,\Sigma)$ \\
      $\eta$ & 0.547 & $0$ & $-$ & $0$ & $(\gamma,\gamma),(\gamma,\omega)$ \\
      $\omega$ & 0.783 & $1$ & $-$ & $0$ & $(\gamma,\pi),(\gamma,\eta)$ \\
      \hline
      $\sigma$ & 0.650 & $0$ & $+$ & $0$ & $(\pi,\pi)$ \\
      $\rho$ & 0.769 & $1$ & $-$ & $1$ & $(\pi,\pi),(\pi,\omega),(\gamma,\pi),(\gamma,\eta)$ \\
      $a_0$ & 0.983 & $0$ & $+$ & $1$ & $(\pi,\eta)$ \\
      $K^*$ & 0.894 & $1$ & $-$ & $\foh$ & $(\pi,\Lambda),(\pi,\Sigma),(\gamma,\Lambda),(\gamma,\Sigma)$ \\
      $K_1$ & 1.273 & $1$ & $+$ & $\foh$ & $(\gamma,\Lambda),(\gamma,\Sigma)$ \\
      $K_0^*$ & 1.412 & $0$ & $+$ & $\foh$ & $(\pi,\Lambda),(\pi,\Sigma)$ \\
      \hhline{======}
    \end{tabular}
  \end{center}
  \caption{Properties of all asymptotic particles and intermediate
    $t$-channel mesons entering the potential. For those particles,
    that appear in several charge states, averaged masses are
    used. For the mesons also all reaction channels, where the
    corresponding meson appears in a $t$-channel exchange, are given. 
    \label{tabparticleprops}} 
\end{table}
The interaction Lagrangians of these particles can be found in
Appendix \ref{appbornlagr}.

\subsubsection{\label{secrescontr}Resonance contributions}

For the spin-$\foh$ resonances, we follow the PS-PV convention used in
Refs. \cite{feusti98,feusti99}. For the positive-parity spin-$\foh$
resonances, PV coupling is used just as in the nucleon case. For
negative-parity spin-$\foh$ resonances, PS coupling is used since this
coupling has also been applied in other models for the $S_{11}(1535)$
on $\eta N$ photoproduction \cite{tiator,sauermann}. The $\omega N$
decay interactions are in analogy to the electromagnetic decays (see Ref.
\cite{gregi}) and are given in Appendix \ref{appres12lagr}. Note that
as a result of the problem of pinning down the corresponding resonance 
parameters reliably, $u$-channel contributions by hyperon resonances
in the $K\Lambda$ and $K\Sigma$ production are neglected as in Refs.
\cite{feusti98,feusti99}.

In combination with the conventional spin-$\fth$ couplings, e.g., for
$\Delta (1232) \ra \pi N$ (omitting isospin),
\bea
\mcl_{\Delta N\pi} = \frac{g_{\Delta N\pi}}{m_\pi} \bar u_\Delta^\mu
u_N \partial_\mu \pi \; ,
\label{lag32hadrpos}
\eea
the Rarita-Schwinger propagator $G^{\mu \nu} (q)$ also contributes
off-shell ($q^2 \neq m_R^2$) to spin-$\foh$ partial waves. To examine
the influence of the off-shell spin-$\foh$ contributions so-called
off-shell projectors have been introduced:
\bea
\Theta_{\mu \nu} (a) = 
g_{\mu \nu} - a \gamma_\mu \gamma_\nu \; ,
\label{offshellproj}
\eea
where $a$ is related to the commonly used off-shell parameter $z$
\cite{nath} by $a = (z + \sfoh )$. There have been theoretical
attempts to fix the value of $a$ \cite{peccei69,nath} and to thereby
remove the spin-$\foh$ contributions. However, in Ref. \cite{ben89} it has
been shown that these contributions are always present for any choice
of $a$. Furthermore, it has been argued that in an effective theory,
where the spin-$\foh$ spin-$\fth$ transition between composite
particles is described phenomenologically, these parameters should not
be fixed by a fundamental theory assuming pointlike particles, but
rather be determined by comparison to experimental data. This is also
confirmed by the fact that only a poor description of pion photoproduction
multipoles is possible when the values for $a$ given in Ref. \cite{nath}
are used for the $\Delta$ resonance \cite{ben89}. 

It has, furthermore, been shown \cite{pascatim} that for any choice of
the off-shell parameters, the ``conventional'' $\pi N \Delta$
interaction \refe{lag32hadrpos} leads to inconsistencies: Either the
constraints of the free theory are explicitly violated ($a \neq 1$)
\cite{nath} or it gives rise to the Johnson-Sudarshan-Velo-Zwanziger
problem \cite{jsvz} ($a = 1$). Pascalutsa and Timmermans
\cite{pascatim} have thus recently suggested an 
interaction that is invariant under gauge transformation of the
Rarita-Schwinger field ($u_R^\mu \ra u_R^\mu +\partial^\mu \epsilon$)
and consequently consistent with the free spin-$\fth$ theory. The
premise is that consistent interactions should not ``activate'' the
spurious spin-$\foh$ degrees of freedom, and therefore the full
interacting theory must obey similar symmetry requirements as the free 
theory. These interactions can be easily constructed by allowing only
couplings to the manifestly gauge invariant Rarita-Schwinger field
tensor,
\bea
U_R^{\mu \nu} = \partial^\mu u_R^\nu - \partial^\nu u_R^\mu \; ,
\eea
and its dual $\tilde U_R^{\mu \nu} = \foh \ve^{\mu \nu \alpha
  \beta} {U_R}_{\alpha \beta}$. The resulting amplitude
is therefore proportional to the spin-$\fth$ projector,
\bea
\mcp^{\mu \nu}_\fth (q) = 
g^{\mu \nu} - \frac{1}{3} \gamma^\mu \gamma^\nu - 
\frac{1}{3 q^2} 
( \slash q \gamma^\mu q^\nu + q^\mu \gamma^\nu \slash q) \; ,
\nonumber
\eea
as already anticipated by the {\it  ad hoc} prescription used in
Ref. \cite{willadel}. Pascalutsa has proposed in Ref. \cite{pascatim}
the following $\pi N \Delta$ interaction:
\bea
\mcl_{\pi N \Delta} = 
f_{\pi N \Delta} \bar{\tilde U}_R^{\mu \nu} \gamma_\mu \gamma_5 u_N
\partial_\nu \pi \; .
\label{lag32pasca}
\eea
Using this interaction, the net result is a Feynman amplitude that
resembles the conventional one, with the difference that the full
Rarita-Schwinger propagator $G^{\mu \nu}_\fth (q)$ is replaced by its
spin-$\fth$ part $- (\slash q - m)^{-1} \mcp^{\mu \nu}_\fth (q)$ and
the amplitude is multiplied by an overall $q^2$. Demanding on-shell
($q^2 = m_\Delta^2$) equivalence with the conventional interaction,
the coupling constant $f_{\pi N \Delta}$ can be identified to be 
$f_{\pi N \Delta} = g_{\pi N \Delta} / (m_\pi m_\Delta)$. 
This equivalence procedure can be generalized to any spin-$\fth$
vertex (in particular to the electromagnetic and vector meson decay
vertices given in Appendix \ref{appres32lagr}) by the replacement
\bea
\Gamma_\mu u_R^\mu \ra \Gamma_\mu \gamma_5 \gamma_\nu 
\tilde U_R^{\nu \mu} 
\label{conv32tocons32}
\eea
leading effectively to the substitution of the propagator $G_{\mu
\nu}$ and an additional overall factor of $q^2 / m_R^2$ in the Feynman 
amplitude. Here, $q$ denotes the four-momentum of the intermediate
resonance.

Pascalutsa has also shown \cite{pasca01} that using the
``inconsistent'' conventional couplings leading to $s$- and $u$-channel
contributions is equivalent at the $S$-matrix level to using the
``consistent'' (gauge-invariant) couplings plus additional contact
interactions. The advantage, however, of using consistent
couplings is that they allow for an easier analysis of separating
background and resonance contributions. This has also been confirmed
by Tang and Ellis \cite{tang} in the framework of an effective field 
theory. These authors have shown that the off-shell parameters are
redundant 
since their effects can be absorbed by contact interactions. In
addition, Pascalutsa and Tjon \cite{pascatjon} have demonstrated that
the gauge-invariant and the conventional $\pi N\Delta$ interaction
result in the same $\pi N$ threshold parameters once contact terms are
included and some coupling constants are readjusted. Pascalutsa
\cite{pasca01} has thus concluded that within an effective Lagrangian
approach, any linear spin-$\fth$ coupling is acceptable, even an
inconsistent one. The differences to the use of consistent
couplings plus contact terms are completely accounted for by a change
of coupling constants.

In our model, calculations with both spin-$\fth$ couplings are
performed to extract information on the importance of off-shell
contributions -- or, correspondingly, contact interactions -- from the
comparison with experimental data. I.e., for the pion-induced
reactions we present calculations where the additional spin-$\foh$
contributions are allowed in the spin-$\fth$ propagators and the
off-shell parameters are used as free parameters, and calculations
where these contributions are removed by the above prescription
\refe{conv32tocons32}. The remaining background contributions are
identical in both calculations, in particular the same $t$-channel
exchange diagrams are taken into account and no additional contact
diagrams are introduced when using the Pascalutsa couplings.

\subsection{Form factors}
\label{secformgauge}

To account for the internal structure of the mesons and baryons, as in
\cite{feusti98,feusti99}, the following form factors are introduced at
the vertices:
\bea
F_p (q^2,m^2) &=& \frac{\Lambda^4}{\Lambda^4 +(q^2-m^2)^2} \; ,
\label{formfacp} \\
F_t (q^2,m^2) &=& 
\frac{\Lambda^4 +\fof (q_t^2-m^2)^2}
{\Lambda^4 + \left(q^2-\foh (q_t^2+m^2) \right)^2} \; .
\label{formfact}
\eea
Here $q_t^2$ denotes the value of $q^2$ at the kinematical threshold
of the corresponding $s$, $u$, or $t$ channel. Guided by the results
of Refs. \cite{feusti98,feusti99} and to limit the number of free cutoff
parameters $\Lambda$, the following restrictions on the choice of
form factors and cutoff parameters are imposed on the present 
calculations:
\bi
\item{The same form factor shape [$F_p$ of Eq. \refe{formfacp}] and
    cutoff value $\Lambda_N$ is used at all nucleon-final-state
    vertices ($NN\pi $, $NN \eta$, $N\Lambda K$, $N\Sigma K$, and
    $NN\omega$) in the $s$ and $u$ channel.}
\item{The same form factor shape ($F_p$) is used
    at all baryon resonance vertices ($RN\gamma$, $RN\pi$, $RN\zeta$,
    $RN\eta$, $R\Lambda K$, $R\Sigma K$, and $RN\omega$), but it is
    distinguished between spin-$\foh$ and -$\fth$ resonances and
    between hadronic and the electromagnetic final state. This leads
    to four different cutoff values $\Lambda^h_\foh$,
    $\Lambda^\gamma_\foh$, $\Lambda^h_\fth$, and
    $\Lambda^\gamma_\fth$, where the second and fourth only contribute 
    in the global fits.}
\item{The same form factor shape [$F_p$ or $F_t$ of
    Eqs. \refe{formfacp} and \refe{formfact}] and cutoff value $\Lambda_t$
    is used at all baryon-baryon-meson $t$-channel vertices.}
\ei

\section{\label{secresultsintro}Description of Calculations}

From the Lagrangians introduced in Sec. \ref{secpot} and summarized
in Appendix \ref{applagr}, the spin dependent amplitudes
$\mcv^{fi}_{\lambda' \lambda} = \langle f | V | i \rangle$ are
calculated from the Feynman diagrams for the various reaction channels 
as described in Appendix \ref{appampli}. These spin dependent
amplitudes are then decomposed into helicity partial waves
$\mct_{\lambda' \lambda}^{IJ\pm}$ of good total isospin $I$, spin $J$,
and parity $P=\pm$ as discussed in Appendices \ref{apppwd} and
\ref{appiso}. 

For the determination of all parameters entering the
model, the calculation is compared to experimental data. To do so, the
$\pi N \ra \pi N$ partial waves (see Appendix \ref{apppartials}) and
the observables on all other reactions (see Appendix \ref{appobs}) are 
extracted from the helicity partial waves. This comparison is
performed via a $\chi^2$ minimization procedure, where the $\chi^2$
(per datum) is defined by
\bea
\chi^2 = \frac{1}{N} \sum\limits_{n=1}^N 
\left( \frac{x_c^n - x_e^n}{\Delta x_e^n} \right)^2 \; .
\eea
Here, $N$ is the total number of data points, $x_c^n$ ($x_e^n$) the
calculated (experimental) value and $\Delta x_e^n$ the experimental
error bar. For the pion-induced reactions, the implemented
experimental data are identical to the ones given in Ref.
\cite{gregi}. Altogether, more than 6800 data points are included in
the global and about 2400 in the purely hadronic fitting strategy,
which are binned into $96$ energy intervals; for each angle
differential observable we allow for up to $10-15$ data points per
energy bin. A summary of all references and more details on data base
weighing and error treatment are given in Ref. \cite{gregiphd}.

After having discussed all the ingredients of the model, the results
of the fitting procedure will be presented in the following
Sec. \ref{secpiindresults}. There, the results
from the fits to the pion-induced data (hadronic fits) are also
compared to those from the fits to pion- and 
photon-induced data (global fits). The extracted hadronic background
and resonance parameters are presented in Secs.
\ref{secresbacktcplg} and \ref{secresresparas}.

We have started the fitting procedure with an extension of the
preferred global fit parameter set SM-95-pt3 of Feuster and Mosel
\cite{feusti99}. The first step has been the inclusion of
the $K\Sigma$ and $\omega N$ data in a fit to the pion-induced
reaction data. In addition to the $t$-channel exchange processes
included in Refs. \cite{feusti98,feusti99}, we have taken into
account the exchange of the two scalar mesons $K_0^*(1430)$ and
$\sigma$ to improve the description of the associated strangeness
production and pion-nucleon elastic scattering, respectively, as
compared to Refs. \cite{feusti98,feusti99}. Furthermore, this
allows for more background contributions in the extended energy range
up to $\sqrt s = 2$ GeV. The $\sigma$ exchange is supposed to model
the correlated isoscalar-scalar two-pion exchange in $\pi N \ra \pi
N$. Since the direct coupling of the scalar $a_0$ meson to $\pi \eta$
($\mcl = - g_{a_0} m_{a_0} \pi \eta a_0$) was chosen in Refs.
\cite{feusti98,feusti99}, this coupling has 
also been used for the $K_0^*$ and the $\sigma$ meson in our first
calculation, thereby also accessing chiral symmetry breaking effects
as in Ref. \cite{goudsmit}; see Sec. \ref{secnuclchi}. At 
the same time, in this first calculation we have tried to minimize the
number of parameters and only varied a subset of all possible $\omega
N$ coupling constants, i.e., in the fitting process we have allowed for
two different couplings ($g_1$ and $g_2$) to $\omega N$ for those
resonances, that lie at or above the $\omega N$ threshold
[$P_{11}(1710)$, $P_{13}(1720)$, $P_{13}(1900)$,
$D_{13}(1950)$\footnote{The $D_{13}(1950)$ is denoted by
  $D_{13}(2080)$ by the PDG \cite{pdg}.}] and
one coupling ($g_1$) for the subthreshold resonance highest in mass:
$S_{11}(1650)$.

Since it has turned out in this calculation, that especially in the
$\omega N$ channel (and to some minor degree also in $K\Lambda$ and
$\eta N$ production) large background contributions, manifested by
large spin-$\fth$ off-shell parameters [cf. Eq. \refe{offshellproj}],
are needed, the subsequent calculations have been performed by also
allowing for more contributions from subthreshold resonances --- as,
e.g., $S_{11}(1535)\ra K\Lambda$ --- and coupling
possibilities\footnote{Since the $\omega N$ couplings of the
  $S_{11}(1650)$ have always turned out to be very small in the
  hadronic fits, finally only one coupling has been used in these
  fits.}. Note that in the coupled-channel model of Lutz {\it et
al.} \cite{lutz}, the authors have also found large subthreshold
contributions to $\gamma N/\pi N \ra \omega N$, in particular a
contribution assigned to the $D_{13}(1520)$. Recently, Titov and Lee
\cite{titov02}, Zhao \cite{zhao01}, and also Oh {\it et al.} \cite{oh01}
have extracted important 
$D_{13}(1520)$ and $S_{11}(1535)$ contributions in $\gamma N \ra 
\omega N$. Moreover, allowing for all possible contributions
is the only way to fully compare to predictions from quark models as, 
e.g., Ref. \cite{riska}, and to model all different helicity combinations of 
the $\omega N$ production mechanism [see Eqs. \refe{heliampli12} and
\refe{heliampli32}]. It is important to note that due to the
coupled-channel calculation, the couplings to one specific final state
are not only determined by the comparison to the experimental data of
this channel, but via rescattering also strongly constrained
by all other channels. Finally, upon the inclusion of the
photoproduction data in the global fitting analysis, the extracted
parameters can be further pinned down. 

Not unexpectedly, the inclusion of the chiral symmetry breaking $\sigma
\pi \pi$ coupling does not improve the description of $\pi N$ elastic
scattering significantly. Therefore, and to be in conformity with
chiral symmetry, all subsequent fits have been performed with the
chirally symmetric derivative $\sigma \pi \pi$ coupling 
[cf. Eq. \refe{lagrpin}]. The effects of the chiral symmetry breaking
coupling in comparison with the chiral symmetric one are discussed in
Sec. \ref{secsigmapasca}.

Feuster and Mosel \cite{feusti98,feusti99} have found
similarly good descriptions of experimental pion- and
photon-induced data on the final states $\gamma N$, $\pi N$, $2\pi N$, 
$\eta N$, and $K\Lambda$ up to 1.9 GeV, when either using the
form factor $F_p$ [Eq. \refe{formfacp}] or $F_t$ [Eq. \refe{formfact}]
for the $t$-channel meson 
exchanges. Since it is {\it a priori} not clear, whether these findings will
hold true for the extended energy region and model space, calculations
have been performed using both form factors. In addition, we have
checked the dependence of the results on the choice of the spin-$\fth$ 
resonance vertices (see Sec. \ref{secrescontr}) and the {\it a priori}
unknown $g_{\omega \rho \pi}$ coupling sign.

We choose the following notation for the labeling of the fits:
\bi
\item{``C'' or ``P'' denotes whether the conventional or Pascalutsa
    couplings are used for the spin-$\fth$ resonance vertices.}
\item{The following letter ``p'' or ``t'' denotes whether the
    form factor $F_p$ or $F_t$ [cf. Eqs. \refe{formfacp} and
    \refe{formfact} in Sec. \ref{secformgauge}] is used in the
    $t$-channel contributions.}
\item{The following symbol denotes whether the fit is a purely
    hadronic (``$\pi$'') or global (``$\gamma$'') fit.}
\item{The concluding symbol denotes the sign of the $g_{\omega \rho
      \pi}$ coupling.}
\item{For the chiral symmetry breaking calculation, a $\slash \chi$ is 
    inserted.}
\ei
The seven hadronic fits and four global fits, which have been performed,
can be summarized as follows:
\bi
\item{Using the conventional spin-$\fth$ vertices, four fits have been
    carried out allowing for both form factor shapes [$F_p$
    \refe{formfacp} or $F_t$ \refe{formfact}] in the $t$ channel and
    also both signs of the couplings of $g_{\rho \omega \pi}$:\\ 
    C-p-$\pi +$, C-p-$\pi -$, C-t-$\pi +$, C-t-$\pi -$.\\
    For the results of the last two fits see in particular Sec.
    \ref{secresbackform}.}
\item{One calculation has been performed with the chiral symmetry
    breaking direct $\sigma \vec \pi \vec \pi$ coupling (see Sec.
    \ref{secnuclchi}):\\ C-p-$\pi \slash \chi +$.\footnote{Some of the
      results of this calculation are published under G. Penner and
      U. Mosel, \JPR{C65}{055202}{2002}.}}
\item{Since in the conventional coupling fits, it has turned out that
    the $F_p$ $t$-channel form factor results in a better $\chi^2$
    result, only two fits using the Pascalutsa spin-$\fth$ vertices
    have been carried out:\\
    P-p-$\pi +$, P-p-$\pi -$.}
\item{For the global fits, we extended the best hadronic fits
    (C-p-$\pi \pm$, C-t-$\pi \pm$) to also include the photon-induced
    data:\\ 
    C-p-$\gamma +$, C-p-$\gamma -$, C-t-$\gamma +$, C-t-$\gamma -$.
    For the results of the last two fits see in particular Sec.
    \ref{secresbackform}.}
\ei

\section{Results on Pion-Induced Reactions}
\label{secpiindresults}

The extension of the Giessen model to also include a vector meson
final state requires some checks whether the new final state is
incorporated correctly. As pointed out in Ref. \cite{gregi} (see also
Appendix \ref{apppwd}), in
the presented partial-wave formalism this inclusion is straightforward 
by simply splitting up the $\omega N$ final state into its three
helicity states $\omega N_\fth$, $\omega N_\foh$, $\omega N_0$, where
the same helicity notation for $\omega N$ is used as given in 
Appendices \ref{appres12lagr} and \ref{appres32lagr}.
Thus effectively one has introduced three new final
states. The correct inclusion of these three final states has been
checked by simulating a single-channel problem, where just one
resonance, which couples to only one $\omega N$ helicity state, has
been initialized with the help of Eqs. \refe{heliampli12} and
\refe{heliampli32}, while all other final states are switched off. It
has been shown in Ref. \cite{feusti98} that the resulting
partial-wave $K$ matrix ,
\bea
\mck^{IJ\pm}_{\omega_\lambda \omega_\lambda} \sim 
\frac{-\sqrt s \Gamma_{\omega_\lambda} (s)}{s-m_R^2} \; ,
\label{kmatcheck}
\eea
leads via Eq. \refe{bsematinv} to a $\mct$ matrix that resembles a 
conventional relativistic Breit-Wigner. This artificial situation is
then similar to the low-energy $P_{33}$ $\pi N \ra \pi N$ partial
wave, which can be well approximated by a single resonance 
[$P_{33}(1232)$] only decaying and consequently contributing to $\pi N$.
Thus we have successfully checked that the partial-wave amplitude
$\mct^{IJ\pm}_{\omega_\lambda \omega_\lambda}$ resulting from the
single-helicity $\omega N$ situation has the correct width and energy
behavior and that all poles due to the resonance denominator in Eq. 
\refe{kmatcheck} cancel in the matrix inversion \refe{bsematinv}. 

The resulting $\chi^2$ values for all calculations performed are
presented in Table \ref{tabchisquares}. 
\begin{table}
  \begin{center}
    \begin{tabular}
      {l|r|r|r|r|r|r|r}
      \hhline{========}
       Fit & Total $\pi$ & $\chi^2_{\pi \pi}$ & $\chi^2_{\pi 2\pi}$ & 
       $\chi^2_{\pi \eta}$ & $\chi^2_{\pi \Lambda}$ & $\chi^2_{\pi \Sigma}$
       & $\chi^2_{\pi \omega}$ \\ 
       \hline
       C-p-$\pi +$ & 2.66 & 3.00 & 6.93 & 1.85 & 2.19 & 1.97 & 1.24 \\
       C-p-$\pi -$ & 2.69 & 2.76 & 6.86 & 1.84 & 2.40 & 2.36 & 1.12 \\
       P-p-$\pi +$ & 3.53 & 3.72 & 9.62 & 2.47 & 2.69 & 2.92 & 2.17 \\
       P-p-$\pi -$ & 3.60 & 3.96 & 8.49 & 2.50 & 3.31 & 2.79 & 2.03 \\
       C-p-$\pi \slash \chi +$ & 3.09 & 3.75 & 6.79 & 2.07 & 2.16 & 2.47 & 2.13 \\
       C-t-$\pi +$ & 3.09 & 3.32 & 7.46 & 2.06 & 2.48 & 2.42 & 3.48 \\
       C-t-$\pi -$ & 3.03 & 3.24 & 6.74 & 1.91 & 2.84 & 2.48 & 2.81 \\
       C-p-$\gamma +$ & 3.78 & 4.23 & 7.58 & 3.08 & 3.62 & 2.97 & 1.55 \\
       C-p-$\gamma -$ & 4.17 & 4.09 & 8.52 & 3.04 & 3.87 & 3.94 & 3.73 \\
       SM95-pt-3 & 6.09 & 5.26 & 18.35 & 2.96 & 4.33 & --- & --- \\
      \hhline{========}
    \end{tabular}
  \end{center}
  \caption{Resulting $\chi^2$ of the various
    fits. For comparison, we have also applied the preferred
    parameter set SM95-pt-3 of Ref. \cite{feusti99} to our
    extended and modified data base for energies up to $1.9$ GeV. For
    the $\chi^2$ results of the fits C-t-$\gamma\pm$, see text.
    \label{tabchisquares}}
\end{table}
Note that in contrast to Refs. \cite{feusti98,feusti99}, we have
included in the present calculation all experimental data up to the
upper end of the energy range, in particular also for all partial-
wave and multipole data up to $J=\fth$. A very good simultaneous
description of all pion-induced reactions is possible, even when the
photon-induced data are also considered. This
shows that the measured data for all reactions are indeed compatible
with each other, concerning the partial-wave decomposition and
unitarity effects. As a guideline for the quality of the present
calculation, we have also included a comparison with the preferred 
parameter set SM95-pt-3 of Ref. \cite{feusti99} applied to our extended
and modified data base. It is interesting to note that although this
comparison has only taken into account data up to 1.9 GeV for the
final states $\gamma N$, $\pi N$, $2\pi N$, $\eta N$, and $K \Lambda$,
the present best global calculation C-p-$\gamma +$ results in a better 
description in almost all channels; only for $\pi N \ra \eta N$ the
$\chi^2$ of Ref. \cite{feusti99} is slightly better. This is due to the
fact that for example for the understanding of $K\Lambda$ production, 
the coupled-channel effects due to the final states $K\Sigma$
and $\omega N$ have to be included. This is discussed in Sec.
\ref{secrespl} below; see also the discussion on $K\Lambda$
photoproduction in PMII \cite{pm2}.

The results for the hadronic fits in Table \ref{tabchisquares} also
reveal that while $\omega N$ production seems to be rather
independent of the sign of $g_{\omega \rho \pi}$, the effect of
sign switching becomes obvious in the $K\Lambda$ and $K\Sigma$
results, showing that both reactions are very sensitive to rescattering
effects due to the $\omega N$ channel. Only the global fitting
procedure gives a significant preference of the positive sign 
for $g_{\omega \rho  \pi}$ in the pion-induced $\omega N$ 
production. It is also interesting that while in Ref.
\cite{feusti98} similar results have been found using either one of
the form factors $F_t$ and $F_p$ for the $t$-channel meson exchanges,
the extended data base and model space shows a clear preference of
using the form factor $F_p$ for all vertices, i.e., also for the
$t$-channel meson exchange. Especially in the global fitting
procedure, not even a fair description of the experimental data has
been possible. This is discussed in detail in Sec.
\ref{secresbackform}.

Therefore we do not display the results of the fits C-t-$\pi
\pm$/C-t-$\gamma \pm$ in the following; furthermore, for reasons of
clarity, we restrict ourselves in this section to displaying the 
pion-induced results for the best global fit C-p-$\gamma +$, the best
hadronic fit C-p-$\pi +$, and the calculation using the Pascalutsa
spin-$\fth$ vertices P-p-$\pi +$. Only in those cases, where important
differences are found, also the other calculations are discussed. 

In the subsequent sections, we start with a discussion of the
influence of the treatment of the $\sigma$ meson and the spin-$\fth$
vertices on the pion-induced results. Then the different channels are
discussed separately and the section ends with the presentation of the
background and resonance properties.

\subsection{$\sigma$ meson, chiral symmetry, and spin-$\fth$ vertices} 
\label{secsigmapasca}

\begin{figure}
  \begin{center}
    \parbox{16cm}{
      \parbox{75mm}{\includegraphics[width=75mm]{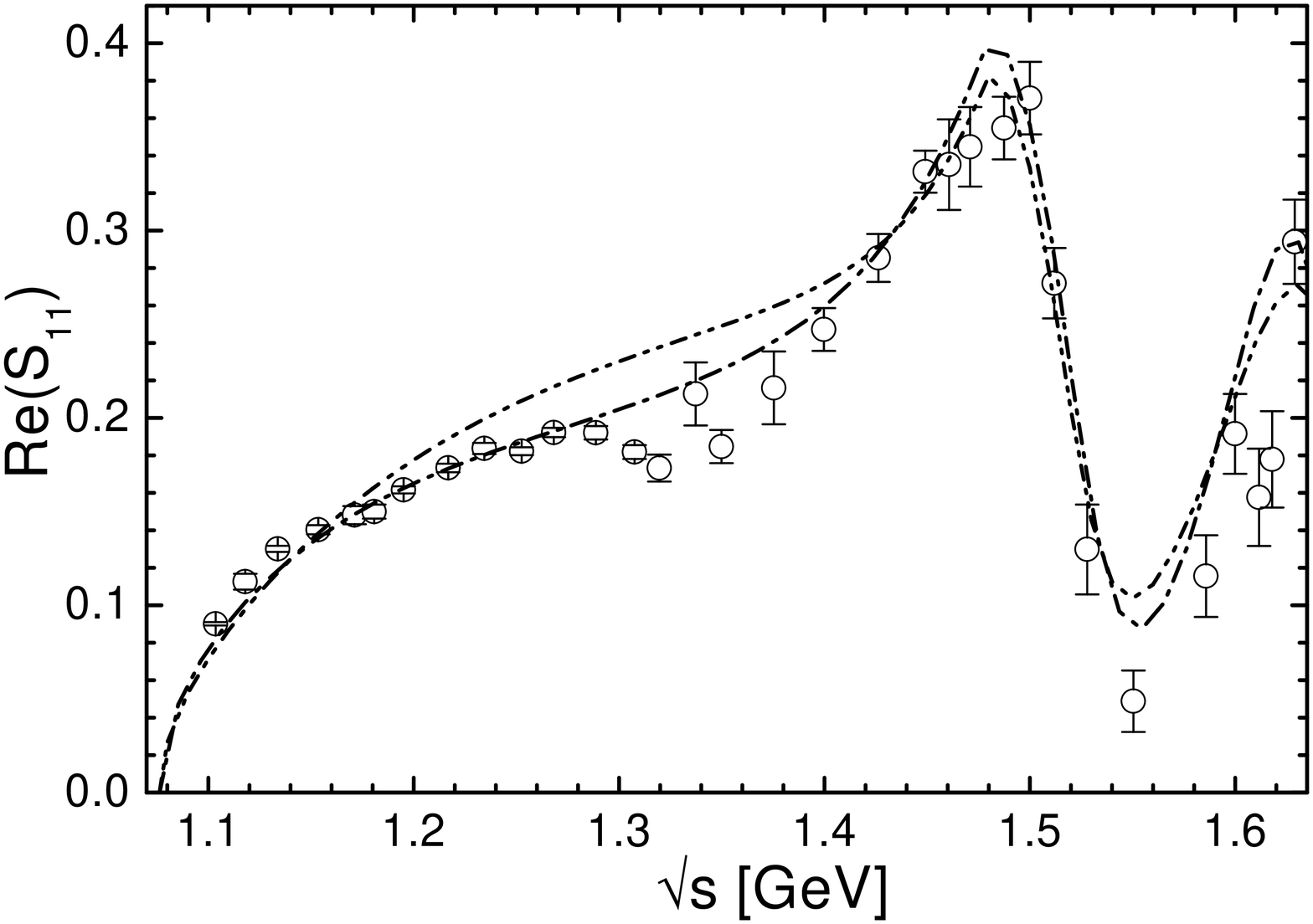}}
      \parbox{75mm}{\includegraphics[width=75mm]{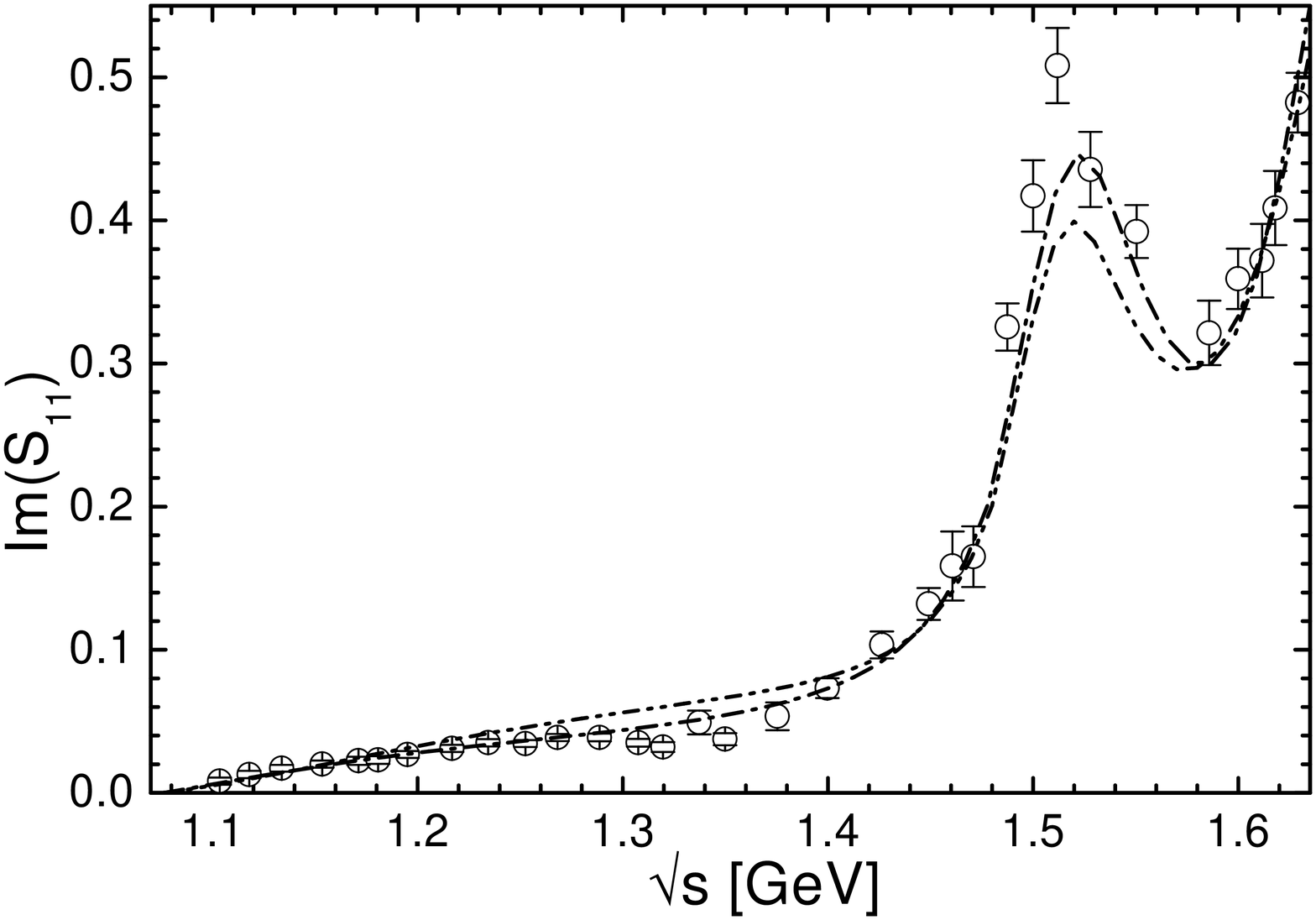}}
      }
    \caption{Effect of the chirally symmetric (calculation
      C-p-$\pi+$: dash-dotted) as compared to the chiral symmetry
      breaking (calculation C-p-$\pi \slash \chi +$:
      dash-double-dotted) $\sigma \pi \pi$ coupling in the $S_{11}$
      $\pi N$ elastic 
      partial wave. \textit{Left:} real part; \textit{right:}
      imaginary part. Data are from Ref. \cite{SM00}.
      \label{figs11chiral}} 
  \end{center}
\end{figure}

As compared to the calculation of Refs. \cite{feusti98,feusti99}
we have added a $\sigma$ meson $t$-channel exchange. In Sec.
\ref{secnuclchi} it has been pointed out that the inclusion of a
$\sigma$ meson is not necessary from the viewpoint of chiral symmetry,
when pseudovector $\pi NN$ coupling is used. However, the $\sigma$
meson can still be used to simulate the correlated two-pion
scalar-isoscalar exchange, but conformity with chiral symmetry then 
requires a derivative $\sigma \pi \pi$ coupling. The preference of a
chirally symmetric coupling has become obvious, when we have switched
from the chiral symmetry breaking coupling (calculation C-p-$\pi \slash
\chi +$) to the chirally symmetric derivative coupling (calculation
C-p-$\pi +$): Even without any refitting the $\chi^2$ in the $\pi
N$ partial waves improves by about $10\%$. This improvement
comes especially from the threshold region in the $S_{11}$ (and also
$P_{13}$) partial wave, see Fig. \ref{figs11chiral}, and
even extends up to the energy region of the second resonance ($\sqrt s
\approx 1.65$ GeV). 

\begin{figure}
  \begin{center}
    \parbox{75mm}{\includegraphics[width=75mm]{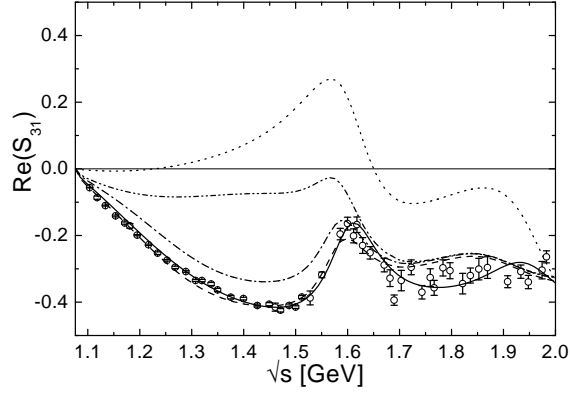}}
    \caption{Effect of the $\sigma$ meson exchange on the real part of 
      the $S_{31}$
      partial wave in $\pi N$ scattering. P-p-$\pi+$ (solid line),
      P-p-$\pi+$ without $\sigma$ (dotted), C-p-$\pi+$ (dashed),
      C-p-$\pi+$ without $\sigma$ (dash-dot), C-p-$\pi+$ without
      $P_{33}(1232)$ (dash-double-dotted). Data are from Ref. \cite{SM00}.
      \label{figsigma}} 
  \end{center}
\end{figure}

The importance of the inclusion of a chirally symmetric $\sigma$ meson 
becomes especially obvious in the calculations, where the Pascalutsa
spin-$\fth$ vertices (cf. Sec. \ref{secrescontr}) are used. It turns
out in the present model that the use of the chirally symmetric
coupling is mandatory: With the nonderivative coupling, not even a
fit to low-energy (up to $1.4$ GeV) $\pi N$ scattering has been
possible. In Refs. \cite{feusti98,feusti99}, where the $\sigma$
meson was not included, it was shown that in particular the $\pi N$
$S_{31}$ partial wave can hardly be described when the 
spin-$\foh$ off-shell contributions of the $P_{33}(1232)$ were
neglected. In the present calculations, however, we find that the
inclusion of a chirally symmetric $\sigma$ meson exchange with a
derivative $\sigma \pi \pi$ coupling allows the description of 
low-energy $\pi N$ elastic scattering even without this off-shell
contributions, i.e., using the Pascalutsa prescription for the
spin-$\fth$ vertices. From
Fig. \ref{figsigma} 
it is obvious that a good description of the $S_{31}$ partial
wave is indeed possible when the Pascalutsa couplings are used. At the
same time it turns out that the $\sigma$ meson as a background
contribution is enhanced as compared to when the conventional
spin-$\fth$ couplings are used. This is not only manifested in the
increase of the $\sigma$ couplings (see Table \ref{tabborncplgs}
below), but also the $t$-channel cutoff parameter $\Lambda_t$ (see
Table \ref{tabcutoff} below) increases by a factor of $2$, meaning
that the missing spin-$\foh$ off-shell background contributions of the 
spin-$\fth$ resonances are
compensated by larger $t$-channel diagram contributions in the lower
partial waves of all reaction channels. The resemblance of the
calculations P-p-$\pi+$ without the $\sigma$ meson and C-p-$\pi+$
without the $P_{33}(1232)$ resonance also asserts the finding of
Pascalutsa \cite{pasca01} and Pascalutsa and Tjon \cite{pascatjon}
that the two prescriptions for the spin-$\fth$ vertices become
equivalent when additional background contributions are included,
i.e., when the spin-$\foh$ off-shell contributions are reshuffled into
other contributions. Similar observations concerning the importance of
the inclusion of a $\sigma$ meson have also been made in the full BSE
$\pi N \ra \pi N$ model of Lahiff and Afnan \cite{lahiff}. These
authors have also allowed for the inclusion and neglect of the $P_{33}
(1232)$ spin-$\foh$ off-shell contributions by using conventional and
Pascalutsa $\pi N \Delta$ couplings. A ten times smaller $g_{\sigma
  NN} g_{\sigma \pi \pi}$ value in the conventional as compared to the
Pascalutsa case was found. At the same time, the cutoff value of the
$\sigma$ form factor in the conventional case has been much softer thus
reducing the $\sigma$ contribution even further.

\subsection{$\pi N \ra \pi N$}
\label{secrespp}

\begin{figure}
  \begin{center}
    \parbox{16cm}{
      \parbox{16cm}{\includegraphics[width=16cm]{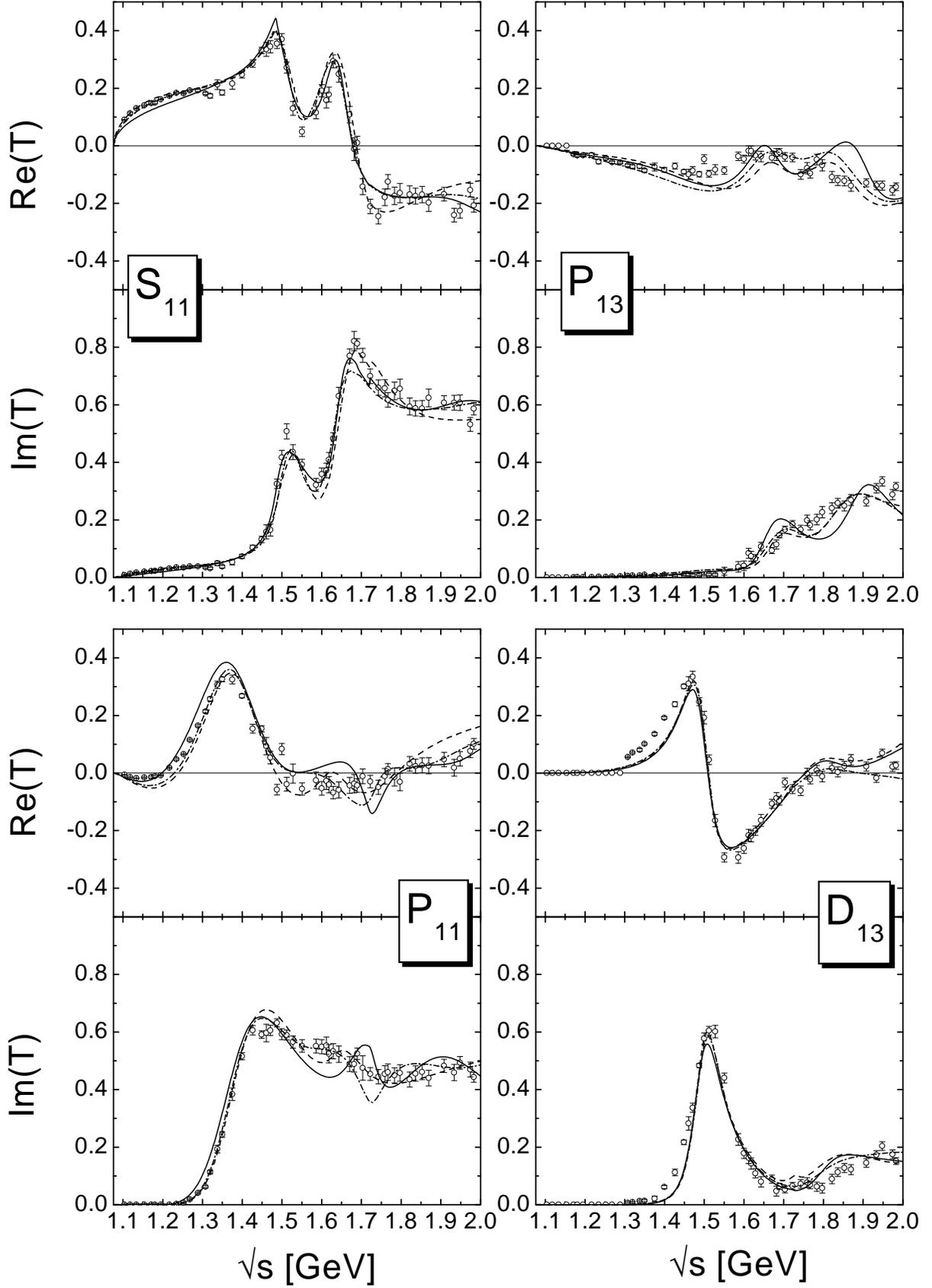}}
      }
    \caption{$\pi N \ra \pi N$ partial waves for $I=\foh$. Calculation 
      C-p-$\gamma +$: solid line, C-p-$\pi +$: dotted line, P-p-$\pi
      +$: dashed line. Data are from Ref. \cite{SM00}.
      \label{figppi12}}
  \end{center}
\end{figure}

\begin{figure}
  \begin{center}
    \parbox{16cm}{
      \parbox{16cm}{\includegraphics[width=16cm]{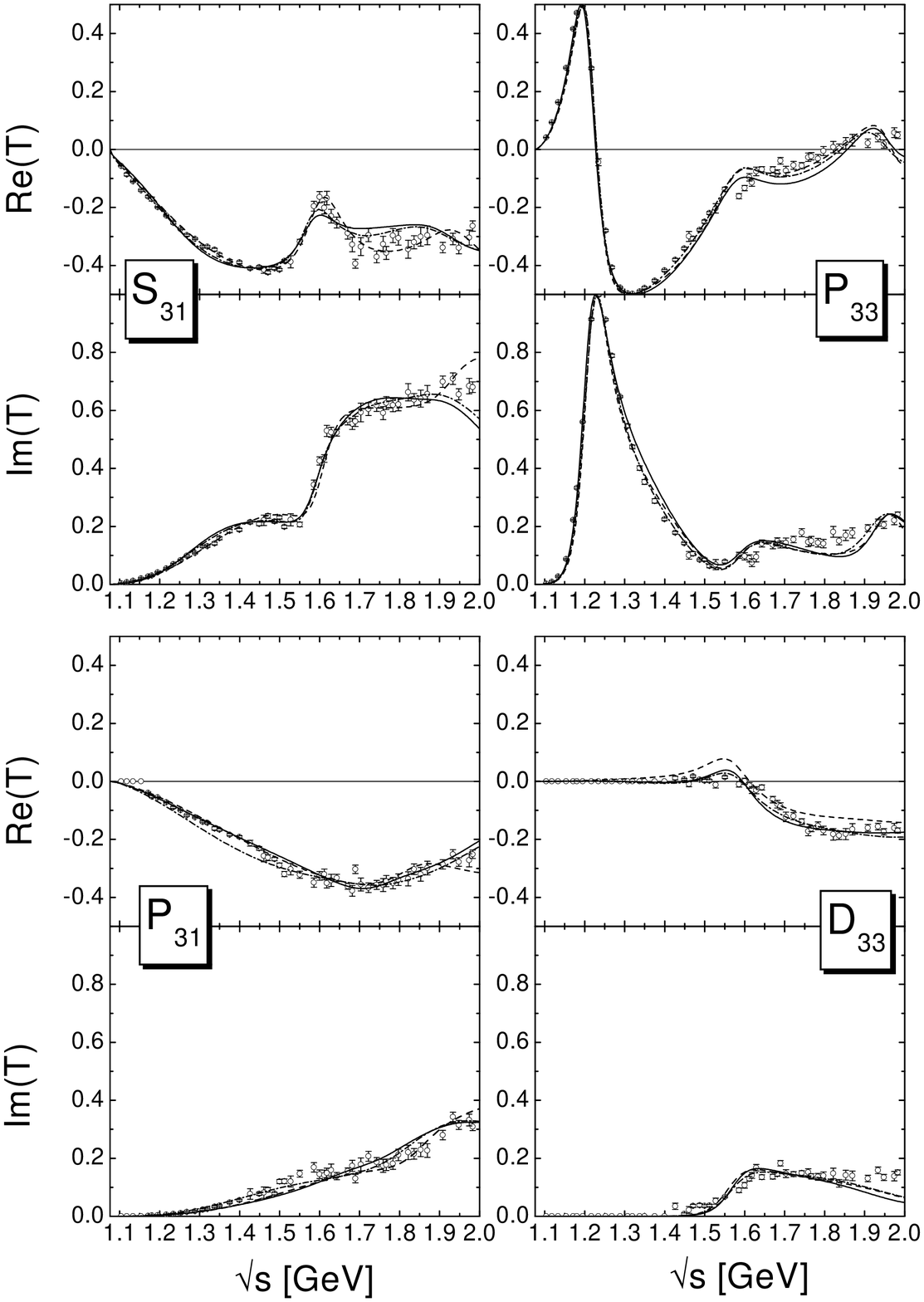}}
      }
    \caption{$\pi N \ra \pi N$ partial waves for $I=\fth$. Notation
      and data as in Fig. \ref{figppi12}.
      \label{figppi32}}
  \end{center}
\end{figure}

The resulting descriptions of the $\pi N$ elastic scattering partial
waves are shown in Figs. \ref{figppi12} and \ref{figppi32} 
in comparison with the continously updated single-energy partial 
wave \cite{SM00} analyses of the Virginia Polytechnic Institute (VPI)
group, which greatly simplifies 
the analysis of experimental data within the coupled-channel
formalism. Note that for those energies, where the single-energy
solutions have not been available, the gaps have been filled with the 
energy-dependent solution of the VPI group. 
In most partial waves, the hadronic calculations using the Pascalutsa 
(P-p-$\pi +$) and conventional (C-p-$\pi +$) spin-$\fth$ vertices are
very similar and equally well reproduce the $\pi N \ra \pi N$
single-energy data points of \cite{SM00}. The largest differences are
found in the 
\bi
\item{$P_{11}$ wave around the $P_{11}(1710)$ resonance. Since there
    is no prominent structure in the $\pi N$ elastic scattering data, the
    width of this resonance is difficult to fix resulting in the
    different structures in Fig. \ref{figppi12}. This also explains
    why the $P_{11}(1710)$ mass as given by the references in the
    Particle Data Group (PDG)
    review \cite{pdg} ranges from about 1.69 to 1.77 GeV.}
\item{$S_{11}$ wave around the $S_{11}(1650)$ resonance. Due to the
    missing off-shell contributions a more pronounced resonance
    behavior is needed in the Pascalutsa calculation to be able to
    describe the high-energy tails of the real and imaginary part.}
\item{$S_{31}$ wave above 1.7 GeV. In this partial wave, it has turned
    out that adding a second resonance [besides the $S_{31}(1620)$]
    around 1.98 GeV improves the $\chi^2$ considerably in the Pascalutsa
    calculation. However, the same does not hold true for the other
    calculations, which consequently show less structure in the 
    high-energy tail. See also Sec. \ref{secresresparas} below.}
\item{$D_{13}$ wave above 1.8 GeV. In this partial wave, it has also
    turned out that adding a third resonance between 1.7 and 1.8 GeV, 
    improves the $\chi^2$ considerably in the Pascalutsa
    calculation. Since the resulting resonance is rather narrow
    ($\Gamma_{tot} \approx 55$ MeV), the difference to the other
    calculations remains small and is only visible in the imaginary
    part between 1.7 and 1.8 GeV. See also Section
    \ref{secresresparas} below.}
\ei
The calculation with the chiral symmetry breaking $\sigma$
contribution is not shown in Figs. \ref{figppi12} and \ref{figppi32}
since it is very similar to the calculation C-p-$\pi +$; the main
differences are contained in the low-energy tails of the spin-$\foh$
partial waves and especially in the $S_{11}$ wave, see
Fig. \ref{figsigma} above.

For the extension of the model up to 2 GeV it turns out to be
essential to add a resonance in the $P_{13}$, $P_{31}$, and $P_{33}$
partial waves as compared to Refs. \cite{feusti98,feusti99}. This is in line 
with Manley and Saleski \cite{manley92}, who found additional states
around 1.88, 1.75, and 2.01 GeV, respectively. Without these
resonances, those three partial waves cannot completely be described 
above 1.8 GeV in our model; see also Refs. 
\cite{feusti98,feusti99}. However, in the $P_{I3}$ waves, the new
resonances are at the boundary of the energy 
range of the present model. This means that their properties cannot be
extracted with certainty, but in both partial waves there is a clear
indication for an additional contribution. See also Sec.
\ref{secresresparas} below.

The most striking differences between the global and the purely
hadronic fits can be seen in the low-energy tails of the $S_{11}$ and
$P_{11}$ waves, which in the latter case is accompanied by an increase
of the mass and widths of the $P_{11}(1440)$. While in the hadronic
calculations the threshold behavior of all $J=\foh$ partial waves is
nicely reproduced, which also leads to $\rho NN$ couplings in line
with the KSRF relation (see Sec. \ref{secresbackborn} below), in
the global calculation this description is inferior. The reason for
this behavior can be found in the necessity of the reduction of the
nucleon form factor cutoff $\Lambda_N$ in the global fits due the
$E_{0+}^{p/n}$ multipoles, see also the discussion on pion
photoproduction in PMII \cite{pm2}. Thereby the
low-energy interference pattern in $\pi N$ scattering between the
$\rho$ meson and the nucleon is misbalanced and deteriorates in
comparison with the hadronic fits. Moreover, the resonant structure
due to the $P_{13}(1900)$ in the $P_{13}$ wave turns out to be more
pronounced in the global fits as compared to the hadronic
calculations. This is a consequence of the necessity of an enhanced
$P_{13}$ contribution in the $\omega N$ production mechanism, see
Sec. \ref{secrespo}. In the isospin-$\fth$ partial
waves, there is hardly any difference between the hadronic and the
global fit results. The reason is that the $I=\fth$ resonances only
contribute to pion and $K\Sigma$ photoproduction, and are hence not
submitted to that many additional constraints of the photoproduction
data as the isospin-$\foh$ resonances.

For a detailed discussion of the individual resonance contributions to 
the partial waves and the discrepancies in the $D_{13}$ partial wave
below $1.45$ GeV, see Sec. \ref{secresresparas} below.

\subsection{$\pi N \ra 2 \pi N$}
\label{secresp2}

\begin{figure}
  \begin{center}
    \parbox{16cm}{
      \parbox{16cm}{\includegraphics[width=16cm]{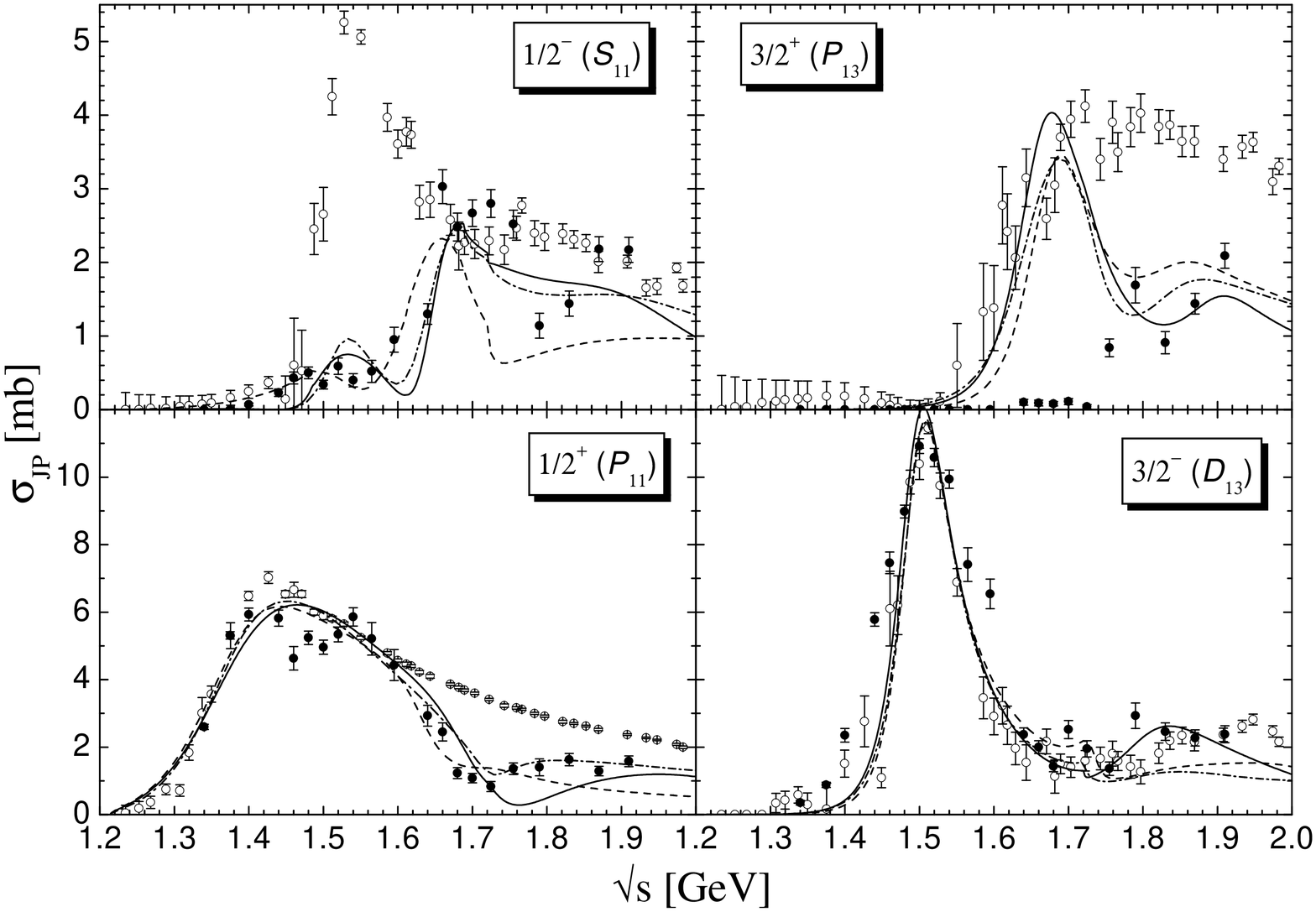}}
      \parbox{16cm}{\includegraphics[width=16cm]{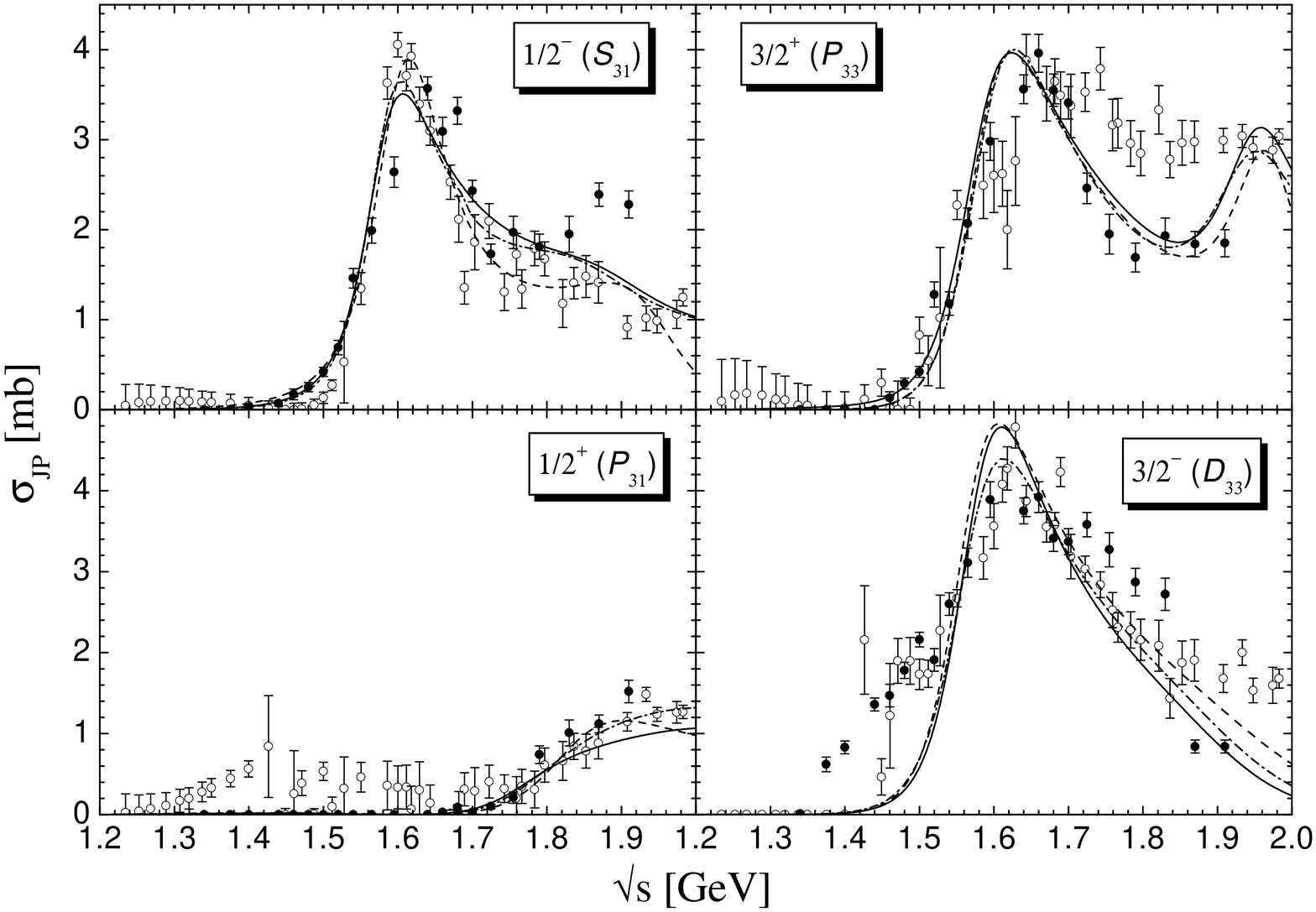}}
      }
    \caption{$\pi N \ra 2\pi N$ partial-wave ($J^P$) cross sections
      for $I=\foh$ (\textit{upper panel}) and $I=\fth$ (\textit{lower
        panel}). The solid dots ($\bullet$) are taken from
      Ref. \cite{manley84}, the open 
      dots ($\circ$) are the inelastic $\pi N \ra \pi N$ partial-wave
      cross sections extracted from the VPI analysis
      \cite{SM00}. Notation as in Fig. \ref{figppi12}.
      \label{figp2both}}
  \end{center}
\end{figure}

\begin{figure}
  \begin{center}
    \parbox{16cm}{
      \parbox{16cm}{\includegraphics[width=16cm]{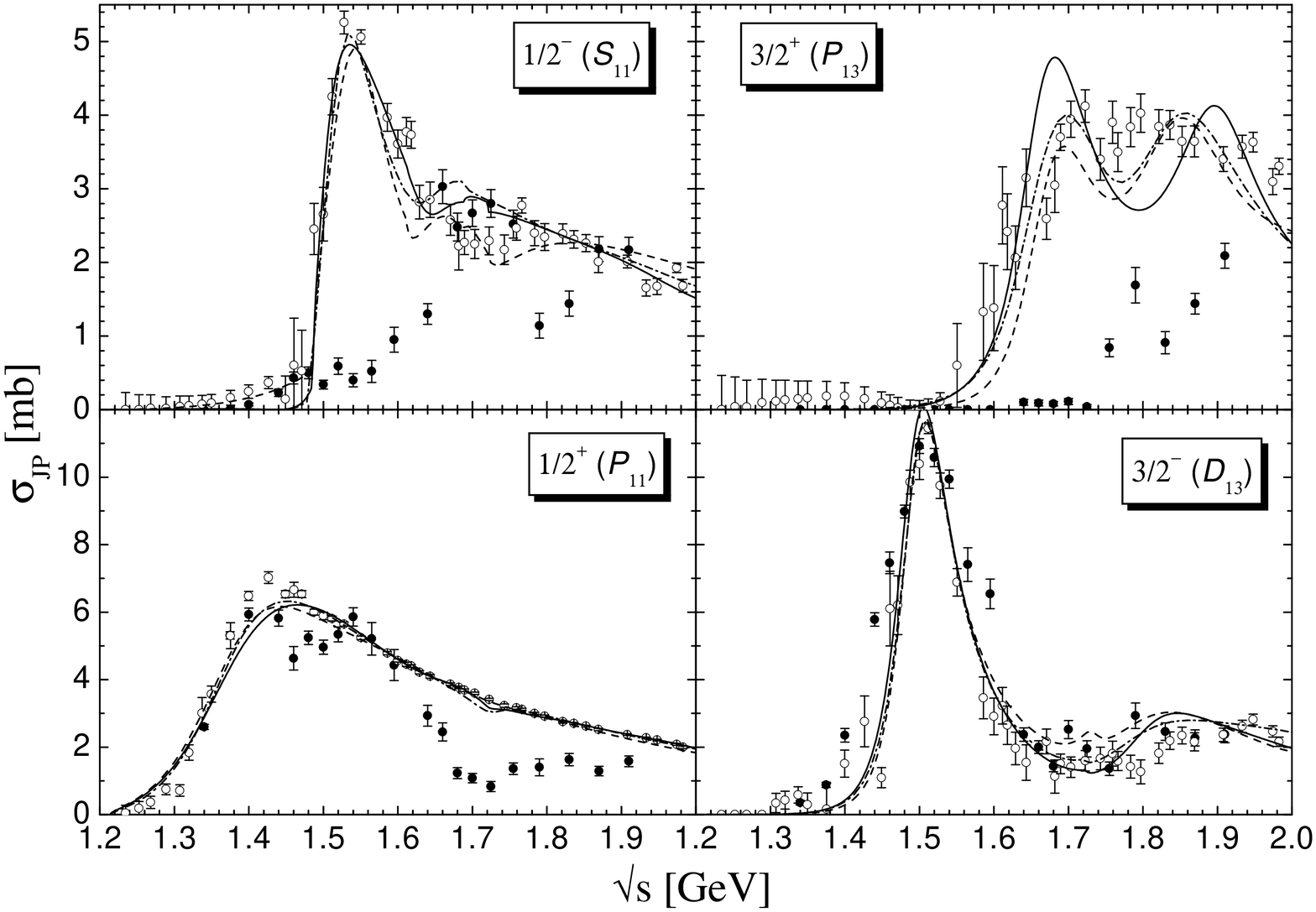}}
      \parbox{16cm}{\includegraphics[width=16cm]{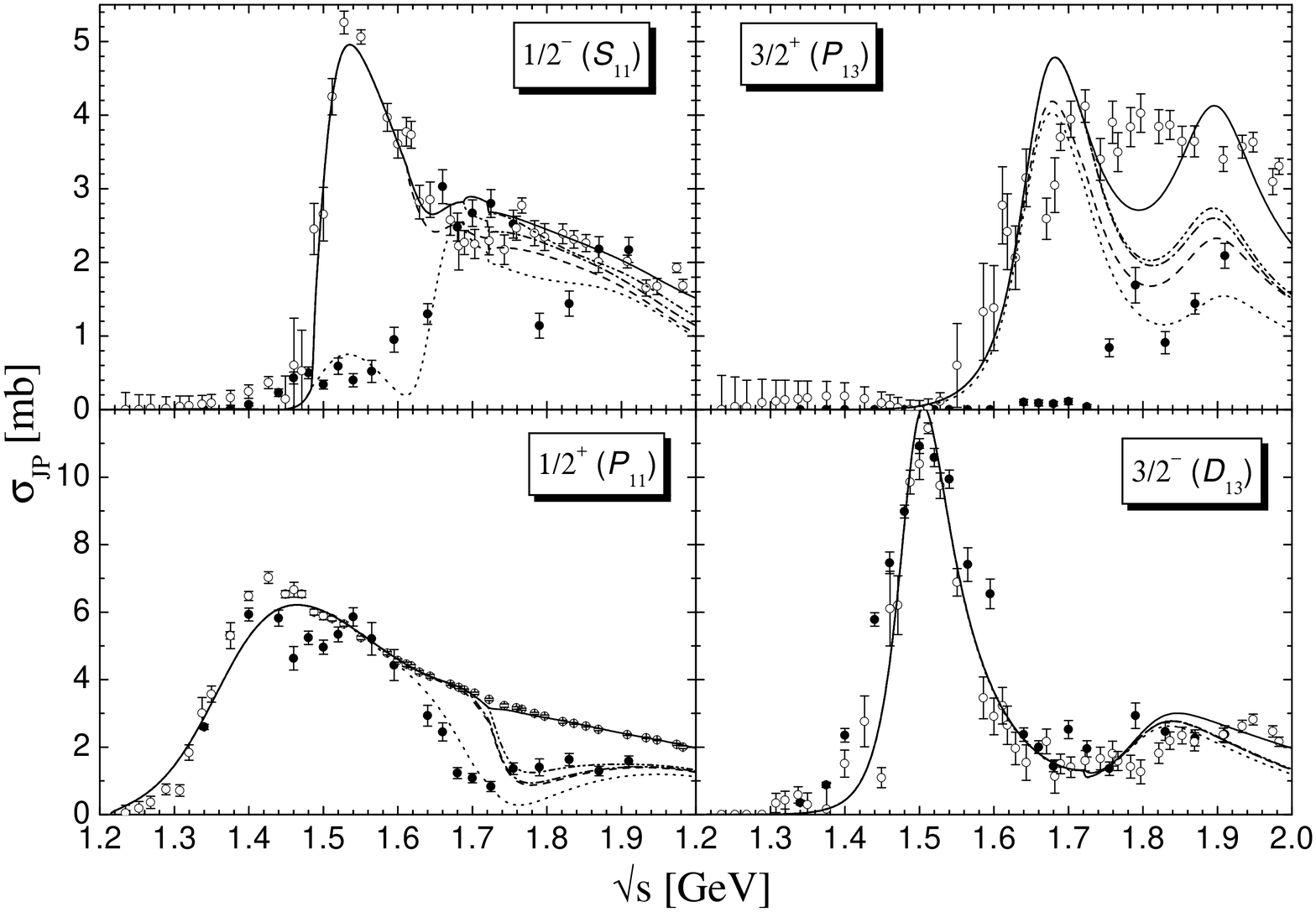}}
      }
    \caption{Inelastic partial-wave cross sections of $\pi N \ra \pi N$
      for $I=\foh$. Data as in Fig. \ref{figp2both}. \textit{Upper
        panel:} Notation as in Fig. \ref{figppi12}. \textit{Lower
        panel:} Decomposition of the inelasticities for calculation
      C-p-$\gamma +$. Partial-wave cross section of $2\pi N$: dotted,
      $+\eta N$: dashed, $+K\Lambda$: dash-dotted,  $+K\Sigma$:
      dash-double-dotted, total ($+\omega N$): solid line.
      \label{figppini12}}
  \end{center}
\end{figure}

Manley {\it et al.} \cite{manley84} have performed a partial-wave analysis
of pion-induced two-pion production on the nucleon taking into account
the two-pion isobar states $\pi \Delta$, $\rho N$, $\sigma N$, and
$\pi N^*(1440)$. Since in our model only one effective two-pion state 
($\zeta N$) is included, where $\zeta$ is an artificial
isovector-scalar meson, it is not possible to compare our calculation 
to the partial waves extracted in Ref. \cite{manley84} for the individual
$2\pi N$ final states. To get a handle on the strength of the $2\pi N$
flux in the various partial waves, we use as experimental input the
$\pi N \ra 2\pi N$ partial-wave cross sections defined by
\bea
\sigma^{IJP} = \frac{4 \pi}{\avec k^2} 
\sum_{\lambda ,\lambda'} (J + \sfoh)
\left| \mct_{\lambda' \lambda}^{IJP} \right|^2
\nonumber
\eea
that were also extracted in Ref. \cite{manley84}. These cross sections
correspond to the sum of all individual $2\pi N$ fluxes for one
partial wave, thus representing the total $2\pi N$ inelasticity. As a
consequence of modeling the $2\pi N$ state by a two-body state within
our model, one cannot expect that all details of these data can be
described within the model. In particular, the threshold and
phasespace behavior is different from the individual three-body final
states. However, even with the assumption that the $\zeta$ meson only
couples to resonances (see Sec. \ref{secnuclchi}, the $2\pi N$ flux
is  well reproduced in most partial waves up to $J=\fth$; see
Fig. \ref{figp2both}.
This indicates that the pion-induced $2\pi N$ production is indeed
dominated by baryon resonances. Since the $2\pi N$ final state
clearly dominates all partial-wave inelasticities besides $S_{11}$,
$P_{11}$, and $P_{13}$ (see below), cf. Fig. \ref{figp2both}, the
qualitative description of this channel is mandatory in a unitary
model. The various calculations
for the $2\pi N$ partial-wave cross sections are very similar in all
partial waves, with the exception of the $S_{11}$ wave. There, the
Pascalutsa calculation results in a largely decreased $S_{11}$ $2\pi
N$ production above 1.7 GeV, below the $2\pi N$ production
data. Although the $S_{11}$ $\omega N$ partial-wave cross section is
increased simultaneously by about 0.5 mb as compared to the
conventional calculations, the resulting total inelasticity is still
reduced, see Fig. \ref{figppini12}. 
All calculations show a kink structure in the $S_{11}$ and the
$D_{13}$ $2\pi N$ flux at the $K\Sigma$ and the $\omega N$ thresholds, 
respectively, indicating that $2\pi N$ flux is moved to the
corresponding channels. 

The largest changes in the $2\pi N$ production upon inclusion of the
photoproduction data can be observed in the $P_{11}$ and $D_{13}$
waves above the $\omega N$ threshold. The inclusion of the very
precise preliminary $\omega N$ photoproduction data of the SAPHIR
Collaboration \cite{barthom} requires that inelastic contributions
are moved from $2\pi N$ to $\omega N$ in the $P_{11}$ wave and vice
versa in the $D_{13}$ case. This can also be seen in the dramatic
change of the total $\pi N \ra \omega N$ cross section behavior when
the photoproduction data are included, see Fig. \ref{figpotot} below.
Otherwise, similarly to the $\pi N \ra \pi N$ case, also the $2\pi N$
production is only slightly changed by the inclusion of the
photoproduction data. A small, but interesting change can, however, be
observed in the high energy tail of the $P_{31}$ and $P_{33}$ waves,
which can be traced back to the shift of inelasticity caused by
$K\Sigma$ from $P_{33}$ in the hadronic calculations to $P_{31}$ in
the global calculations; see also Sec. \ref{secresps}.

The only obvious discrepancy between the calculated $2\pi N$ 
partial-wave cross sections and the Manley {\it et al.} \cite{manley84} data
is given in the $P_{13}$ partial wave. In the energy region between
1.55 and 1.72 GeV the inelasticity increases up to 4 mb in line with
the calculated $2\pi N$ cross section, while the measured $2\pi N$ 
cross section is still zero. At the same time the total cross sections
from all other open inelastic channels ($\eta N$, $K \Lambda$, and $K
\Sigma$) add up to significantly less than $4$ mb. This indicates that
either the extracted $2 \pi N$ partial-wave cross section is not
correct in the $P_{13}$ partial wave or another inelastic channel
(i.e., an additional $3 \pi N$ channel) gives noticeable contributions
to this partial wave. The same problem with the $P_{13}$ inelasticity
has also been observed in a resonance parametrization of $\pi N \ra
\pi N$ and $\pi N \ra 2 \pi N$ by Manley  and Saleski
\cite{manley92}. Since this is the only partial wave where such a
large discrepancy is observed, no additional final state is introduced 
in the present model, but instead, we have largely increased the error
bars of the $2\pi N$ data points in this energy region. However, it
would be desirable to account for $3\pi N$ contributions in future
investigations by the inclusion of, e.g., a $\rho \Delta$ final
state. This might also clarify whether there is a missing ($3\pi N$)
contribution in the $P_{33}$ wave above 1.7 GeV, see
Fig. \ref{figp2both} and Sec. \ref{secresresparas} below.  So far,
no analysis has given such a contribution.

In addition, there is the same problem as in $\pi N$ scattering with
the description of the rise of the $2\pi N$ production in the $D_{I3}$
waves, i.e. in the $D_{13}$ wave below $1.45$ GeV and in the $D_{33}$
wave below $1.55$ GeV, see Fig. \ref{figp2both}. This effect is
probably due to the effective description of the $2\pi N$ state in the
present model; see the detailed discussion in  Sec.
\ref{secresresparas} below.

It is interesting to note that the inelasticities of $\pi N \ra \pi N$
scattering only enter the fitting procedure indirectly, since the real
and imaginary part of the partial waves are the input for the
calculations. Therefore the very good description of the partial-wave
inelastic $\pi N$ cross sections in all calculations, see the upper
panel in Fig. \ref{figppini12}, is an outcome of summing up the 
partial-wave cross sections of all other $\pi N$-induced
channels. Note that the inelasticities for the $I=\fth$ partial waves
are not shown for the different calculations, since due to the
smallness of the $K\Sigma$ contributions, the results are almost
identical to the $2\pi N$ partial-wave cross sections. From
Figs. \ref{figp2both} and \ref{figppini12} we can thus deduce that
not only is the PWD of all inelastic channels on 
safe grounds, but also that all important channels for the considered
energy region are included. At the same time, this shows that the
experimental data on the various reactions are indeed compatible with
each other, in particular no significant discrepancy between the
measured $\pi N$ inelasticity and the sum of all partial-wave cross
sections is observed. The only exceptions are the aforementioned
indications for missing ($3\pi N$) contributions in the $P_{I3}$
waves.

Note also that the inclusion of the photoproduction data only
slightly changes the total inelasticities of the individual partial
waves. The only noticeable differences between the hadronic and global 
calculation is a decrease of the $S_{11}$ inelasticity between $1.6$
and $1.7$ GeV, and an increase in the $P_{13}$ inelasticity around the 
$P_{13}(1720)$. 

In the lower panel of Fig. \ref{figppini12}, the decomposition of the
$\pi N$ inelasticity of the best global fit C-p-$\gamma +$ is
shown. It can be deduced that the $\pi N$ 
inelasticities are made up in all partial waves mainly by the 
$2\pi N$ channel. This also allows us to deduce that the Manley $2\pi N$ 
data \cite{manley84} are in line with the $\pi N$ inelasticitis of the 
VPI analysis \cite{SM00}. The only contradictions can be observed in
the $D_{13}$ wave at 1.6, 1.7, and 1.8 GeV, in the $S_{31}$ wave above 
1.85 GeV and the $D_{33}$ wave between 1.7 and 1.85 GeV.

Besides the $2\pi N$ channel, there are in all partial waves important
contributions to the inelasticities from other channels. Thus the
necessity of the inclusion of a large set of final states in a
coupled-channel calculation can be seen in various partial waves:
\bi
\item{In the $S_{11}$ wave there is the well known $\eta N$
    contribution around the $S_{11}(1535)$. Note that the $\eta N$
    inelasticity also exhibits a second hump, which 
    is due to the interference between the $S_{11}(1535)$ and the
    $S_{11}(1650)$ resonances, although the latter only has a very 
    small $\eta N$ width. See also Sec. \ref{secrespe}.}
\item{In the $P_{11}$ wave there is also an important contribution by
    the large $\eta N$ and $\omega N$ widths of the $P_{11}(1710)$
    resonance. This contrasts previous analyses
    \cite{feusti99,manley92}, where this contribution has been
    assigned to the $K\Lambda$ channel.}
\item{The $P_{13}$ wave contains important contributions from
    $\eta N$ and $\omega N$ as well, where the first one stems from
    the $P_{13}(1900)$ resonance, while the latter one consists of
    important contributions from both $P_{13}$ resonances.}
\item{The $D_{13}$ wave is also fed by a smoothly increasing $\omega
    N$ contribution.}
\ei

The other final states, i.e., the associated strangeness channels
$K\Lambda$ and $K\Sigma$, are only of minor importance for the $\pi N$ 
inelasticities. While both give visible contributions in the $S_{11}$
wave, $K\Lambda$ also shows up in the $P_{13}$ and $K\Sigma$ in the
$P_{11}$ wave.

\subsection{$\pi N \ra \eta N$}
\label{secrespe}

\begin{figure}
  \begin{center}
    \parbox{16cm}{
      \parbox{75mm}{\includegraphics[width=75mm]{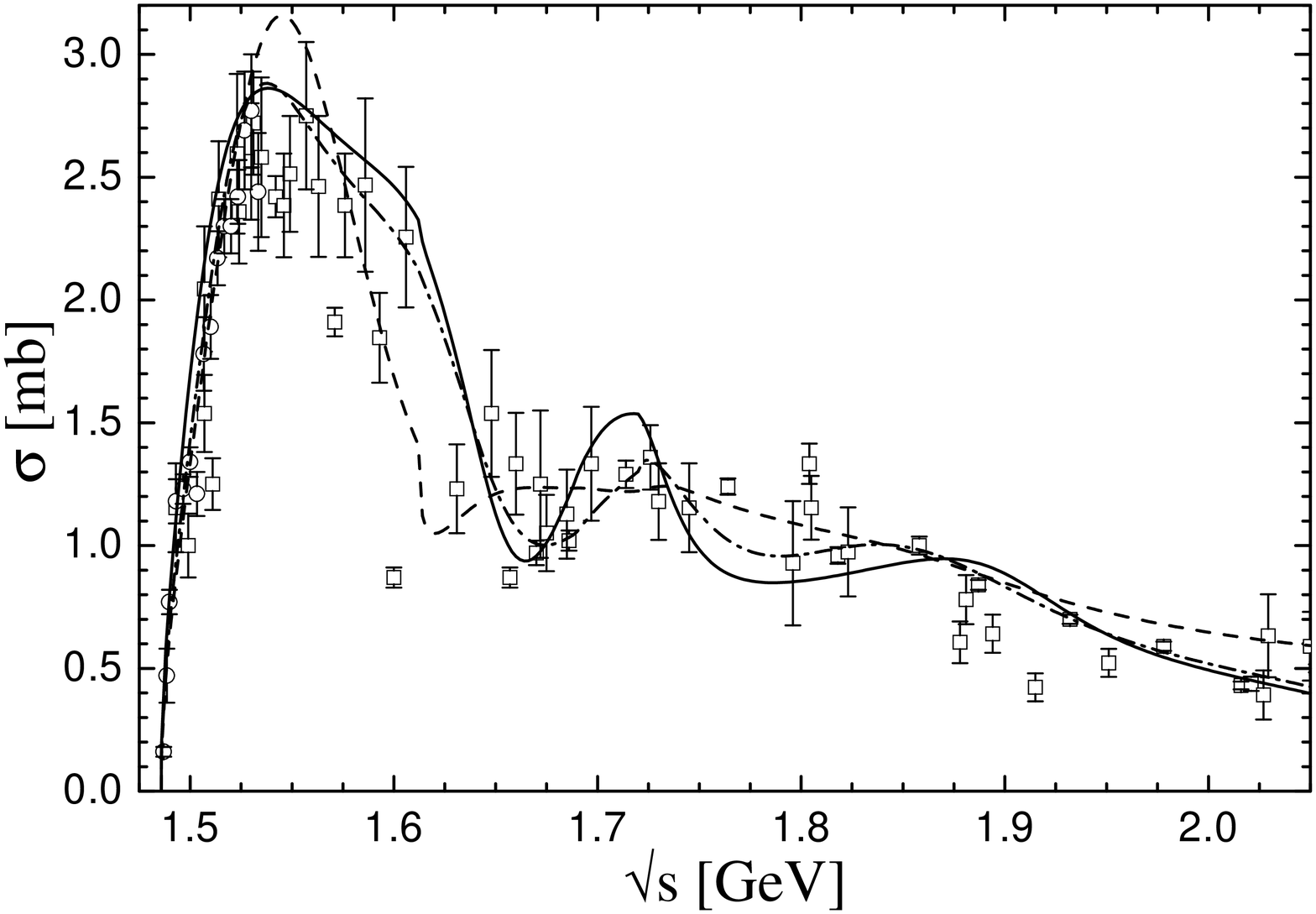}}
      \parbox{75mm}{\includegraphics[width=75mm]{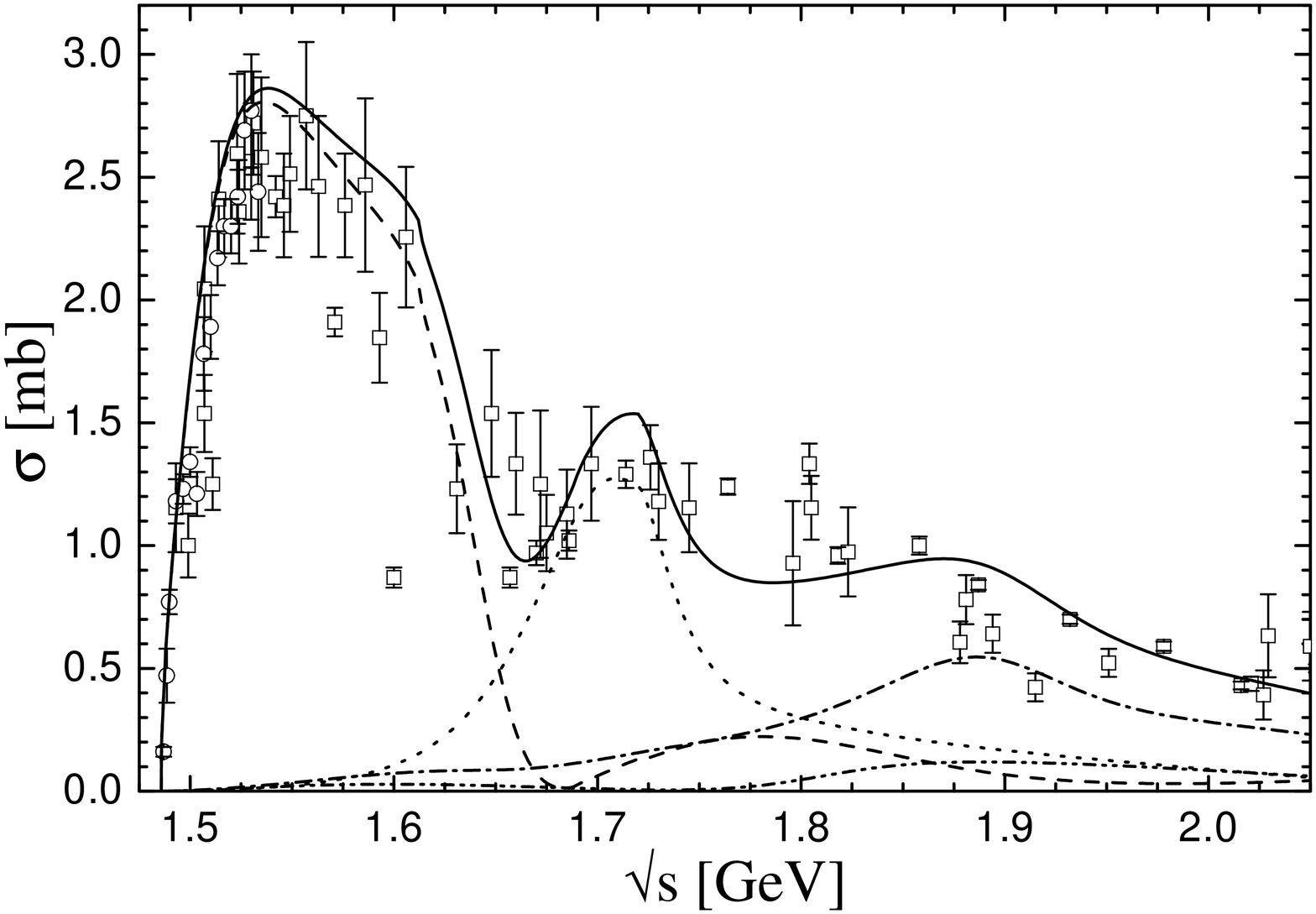}}
      }
    \caption{$\pi^- p \ra \eta n$ total cross section. The new
      threshold data from Ref. \cite{morrison} are denoted by $\circ$, all 
      other data (see Ref. \cite{feusti98})
      by $\Box$. \textit{Left:} Results from the different
      calculations. Notation as in
      Fig. \ref{figppi12}. \textit{Right:} 
      Partial-wave decomposition of the total cross
      section for the calculation C-p-$\gamma
      +$. $J^P=\foh^-(S_{11})$: dashed line; $\foh^+(P_{11})$: 
      dotted line; $\fth^+(P_{13})$: dash-dotted line;
      $\fth^-(D_{13})$: dash-double-dotted line (in brackets the $\pi
      N$ notation is given). The sum of all partial waves is given by
      the solid line.
      \label{figpetot}}
  \end{center}
\end{figure}

In the first coupled-channel effective Lagrangian model on $\eta N$ 
production by Sauermann {\it et al.} \cite{sauermann}, this channel has been 
described by a pure $S_{11}$ mechanism for energies up to $\sqrt s =
1.75$ GeV. As Fig. \ref{figpetot} shows,
the $\pi N \ra \eta N$ reaction is indeed dominantly composed of the
$S_{11}$ contribution due to the $S_{11}(1535)$, however, only for
energies up to $\approx 1.65$ GeV. Due to its large $\eta N$ width the 
$P_{11}(1710)$ dominates in the following energy window up to $1.8$
GeV, while for the highest energies, the $P_{13}(1900)$ resonance is
strongest. The double hump structure in the $S_{11}$ contribution is
due to the destructive interference between the $S_{11}(1535)$ and
$S_{11}(1650)$ resonances, even though the latter one has a much
smaller $\eta N$ decay ratio. This interference pattern exhibits
maximal destructive interference at the $S_{11}(1650)$ resonance
position, while above 1.7 GeV the $S_{11}$ contribution is
resurrected.

\begin{figure}
  \begin{center}
    \parbox{16cm}{
      \parbox{16cm}{\includegraphics[width=16cm]{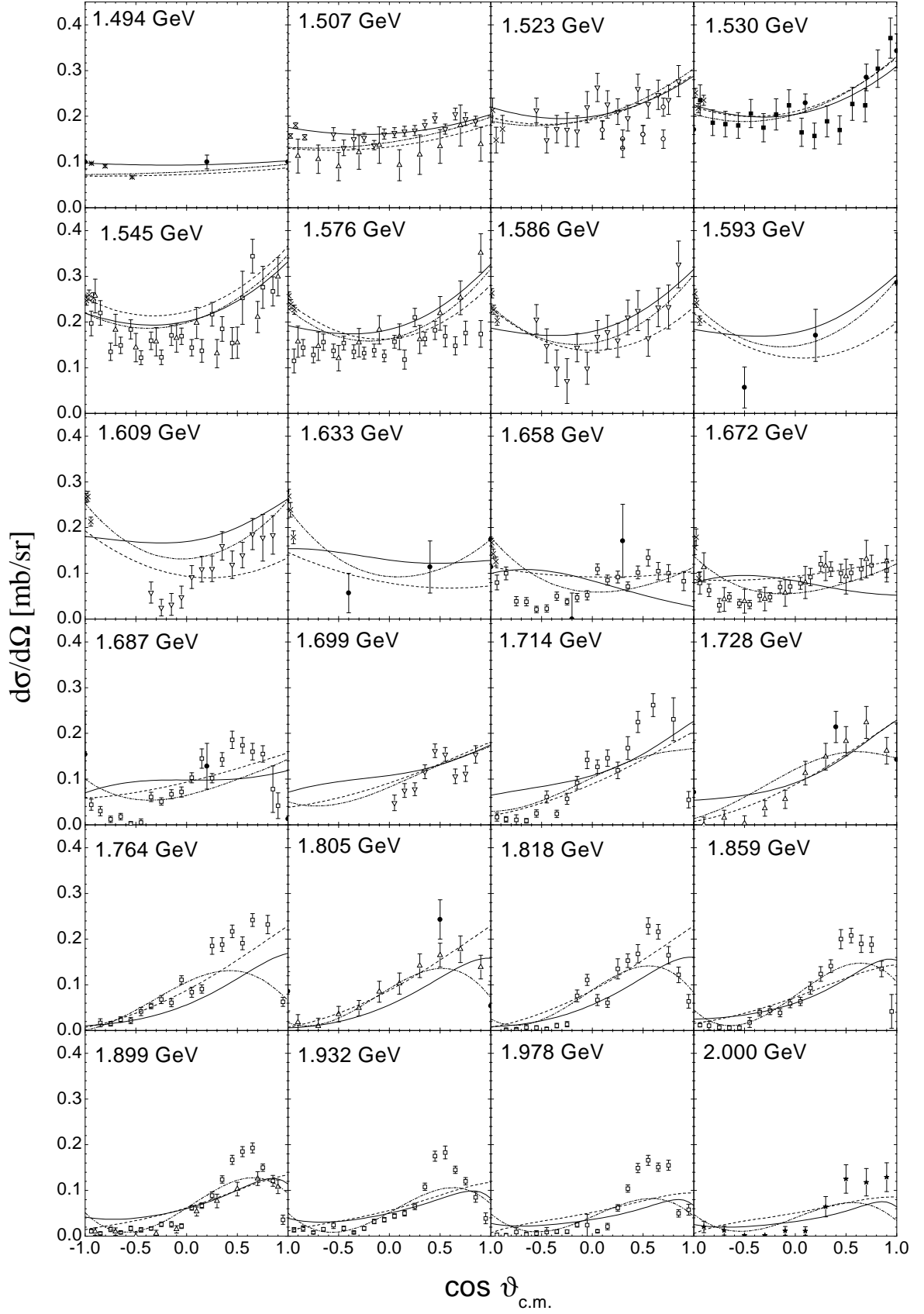}}
      }
    \caption{$\pi^- p \ra \eta n$ angle-differential cross section.
      For the data references, see Refs. \cite{feusti98} and \cite{gregiphd}. 
      Notation as in Fig. \ref{figppi12}.
      \label{figpedif}}
  \end{center}
\end{figure}

The importance of the $P_{11}(1710)$ contribution has also been found
in the resonance parametrization of $\pi N \ra \pi N$ for $I=\foh$ and 
$\pi N \ra \eta N$ by Batini\'c {\it et al.} \cite{batinic}, who extracted a
total width for this resonance of about 120 MeV and an $\eta N$ decay
ratio of almost 90\%. However, in contrast to the results of these
authors, we also find in the present calculation important
contributions of the $P_{13}(1900)$ at higher energies. These
contributions are in line with the observed differential cross section
at higher energies, see Fig. \ref{figpedif}.
However, some deviations in the differential cross section behavior
between calculation and experimental data are observed and the angular
structure cannot be fully described. But one has to note, that 
at higher energies, there are almost only experimental data available
from Brown {\it et al.} \cite{brown} ($\Box$ in Fig. \ref{figpedif}),
which enter with enlarged error 
bars due to problems with the momentum calibration in the experiment,
see Refs. \cite{batinic,feusti98}. Hence these discrepancies hardly
influence the fitting procedure and the resulting $\chi^2$ is still
rather good. Since at energies above 1.8 GeV, there are almost only
data available from Brown {\it et al.} \cite{brown}, a reliable decomposition
in this region can only be achieved after the inclusion of the $\eta
N$-photoproduction data.

In this reaction channel, large differences between the Pascalutsa
and conventional calculations are observed. This is related to the
visible differences in the $S_{11}$ $\pi N \ra \pi N$ partial wave,
since this partial wave constitutes the largest contribution in the
$\eta N$ production mechanism. An obvious difference is that the
Pascalutsa calculation results in less angular structure of the 
angle-differential cross section at higher energies, however,
influencing the resulting $\chi^2$ only to a minor degree, see
above. On the other side, the inclusion of the photoproduction data
hardly changes the total cross section behavior. Only the
$P_{11}(1710)$ contribution is slightly emphasized, which also leads
to the observed differences in the differential cross
section. Moreover, the $\omega N$ threshold effect in the $P_{11}$
wave can be clearly observed in calculation C-p-$\gamma +$ and
C-p-$\pi +$.

\subsection{$\pi N \ra K\Lambda$}
\label{secrespl}

\begin{figure}
  \begin{center}
    \parbox{16cm}{
      \parbox{75mm}{\includegraphics[width=75mm]{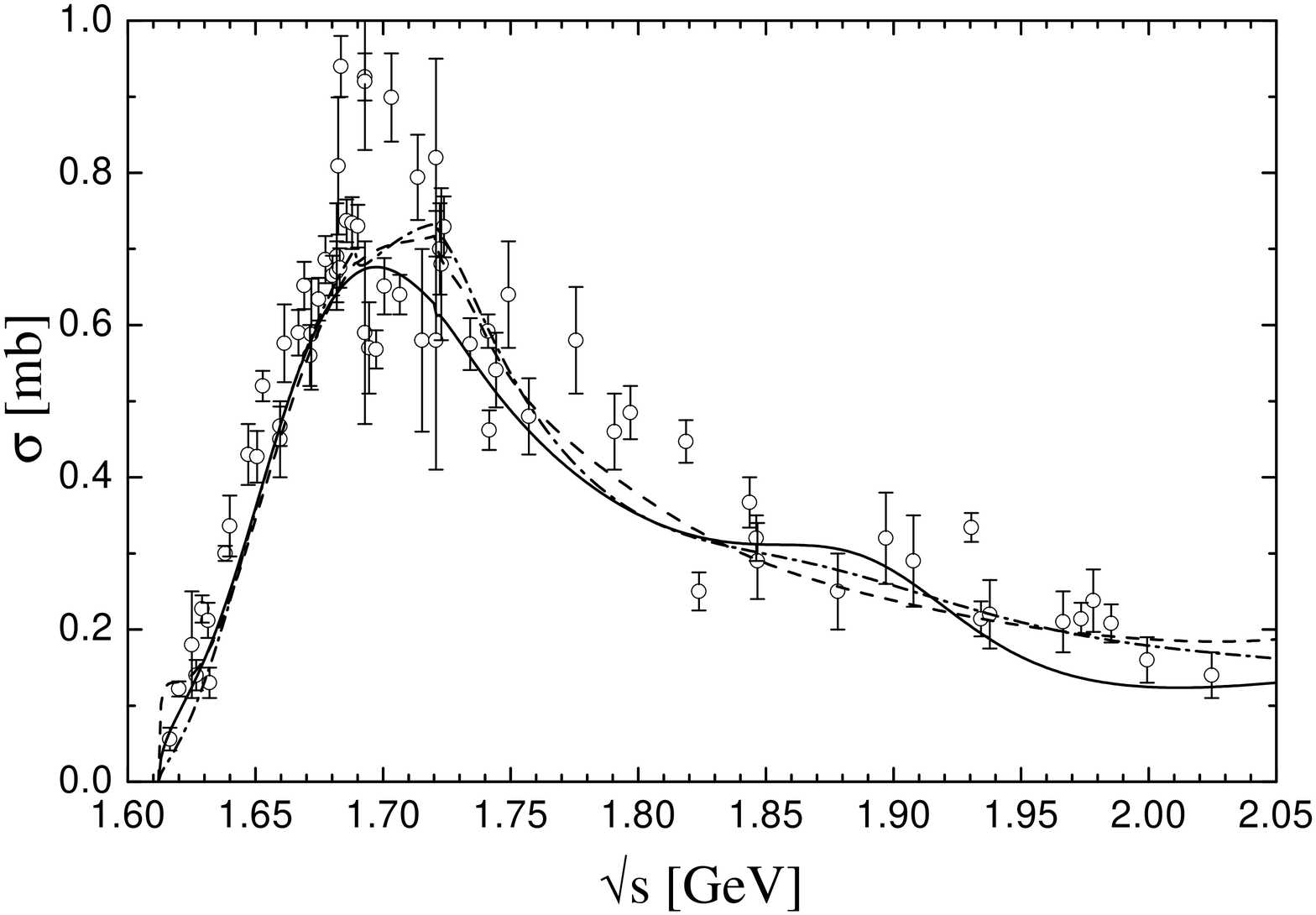}}
      \parbox{75mm}{\includegraphics[width=75mm]{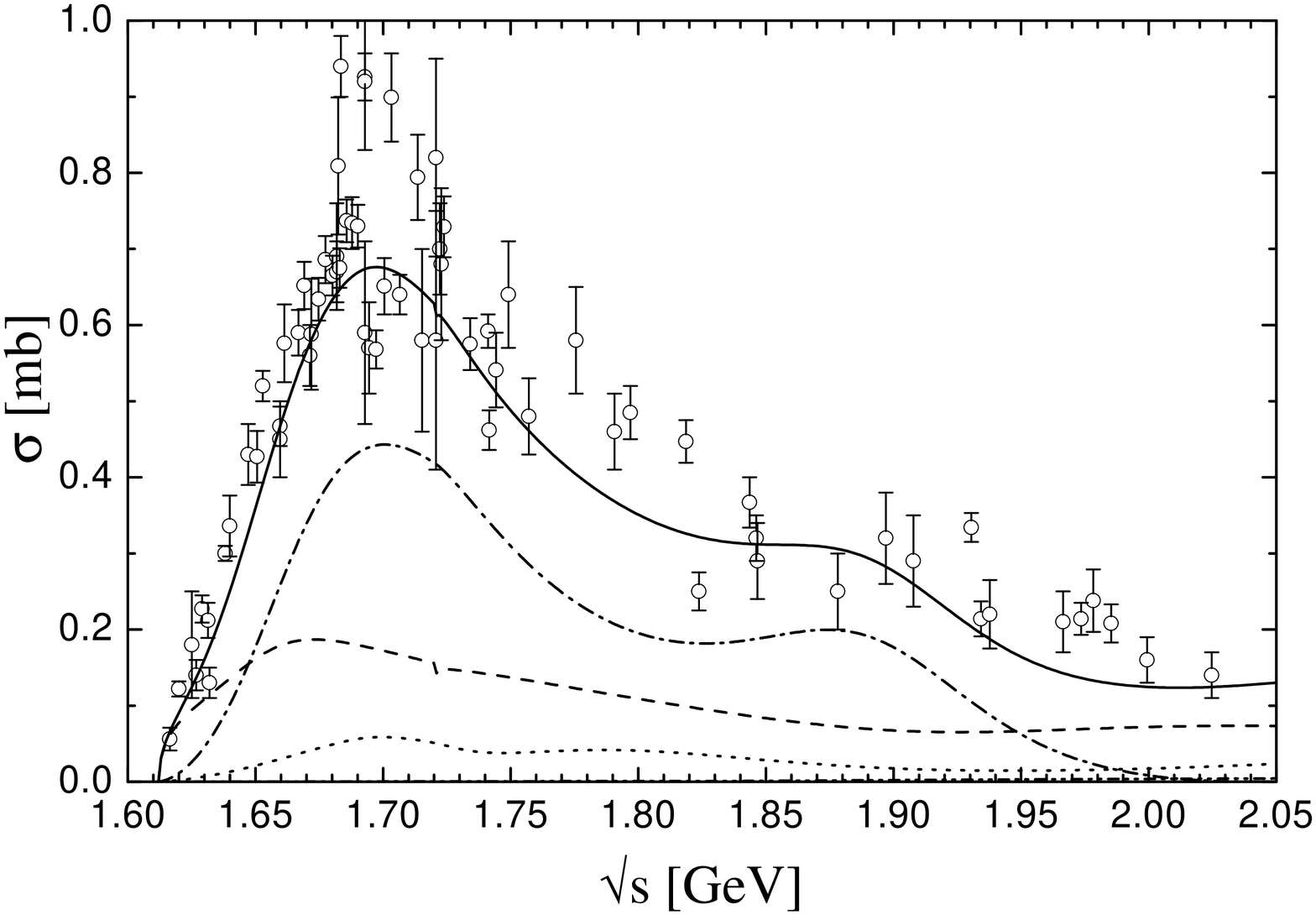}}
      }
    \caption{$\pi^- p \ra K^0 \Lambda$ total cross section. For the
      data references, see Ref. \cite{feusti98}.
      \textit{Left:} Results of the different calculations. Line code
      as in Fig. \ref{figppi12}. \textit{Right:} Partial-wave
      decomposition of the total cross section. Notation as in
      Fig. \ref{figpetot}. 
      \label{figpltot}}
  \end{center}
\end{figure}

$K\Lambda$ production turns out to be a channel which is very
sensitive to rescattering effects. The inclusion of the 
$K\Sigma$ and $\omega N$ final states strongly alters the total
cross section in this 
reaction, 
especially in the hadronic calculations, see Fig. \ref{figpltot}. 
In both of the displayed hadronic calculations,
the $K\Sigma$ channel leads to a kink in the $S_{11}$
partial wave, which has already been observed in the coupled-channel
chiral SU(3) model of Ref. \cite{nobby} including only $S$ and $P$ waves,
while the $\omega N$ channel strongly influences the $P$ waves. 
\begin{figure}
  \begin{center}
    \parbox{16cm}{
      \parbox{16cm}{\includegraphics[width=16cm]{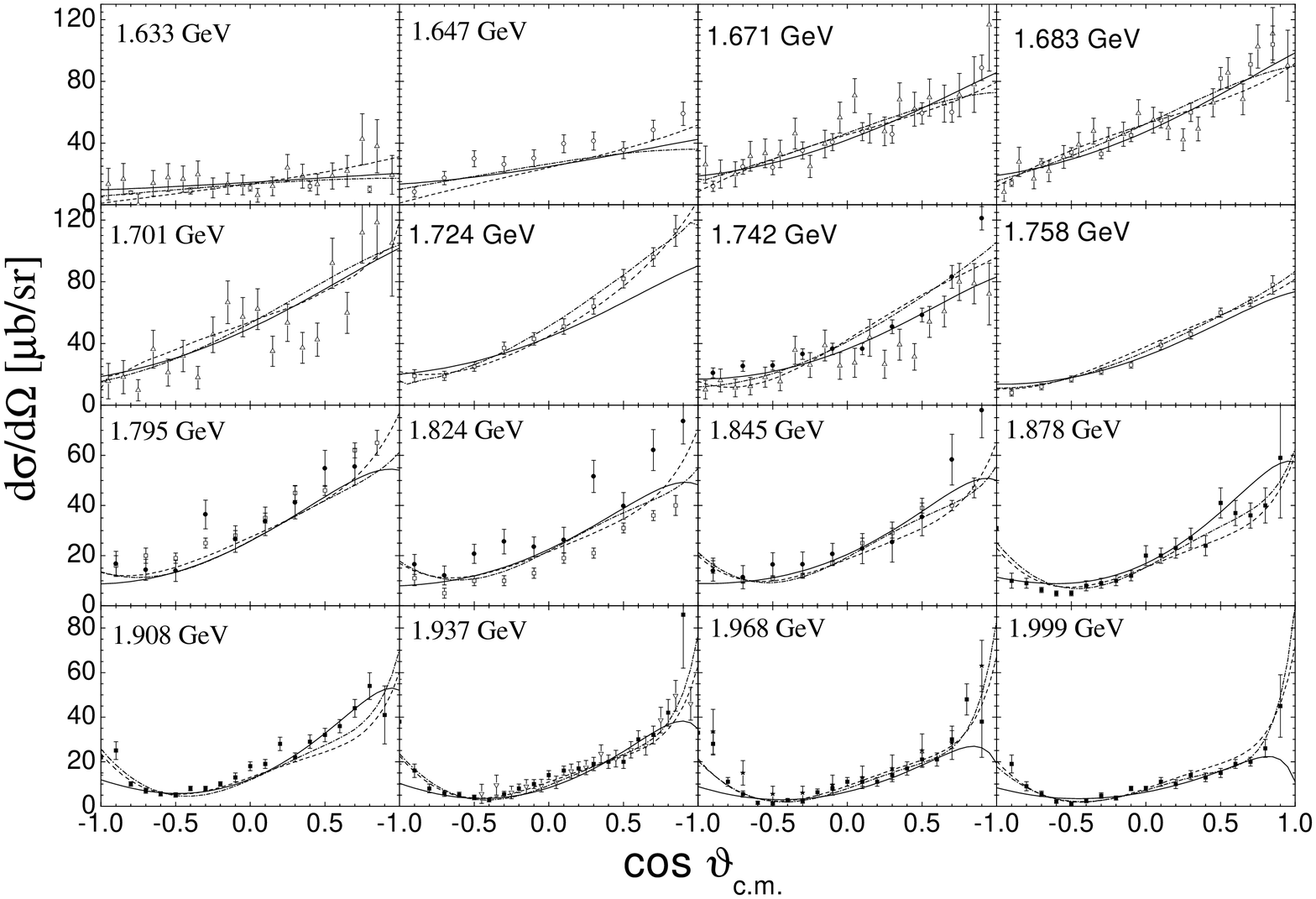}}
      \parbox{16cm}{\includegraphics[width=16cm]{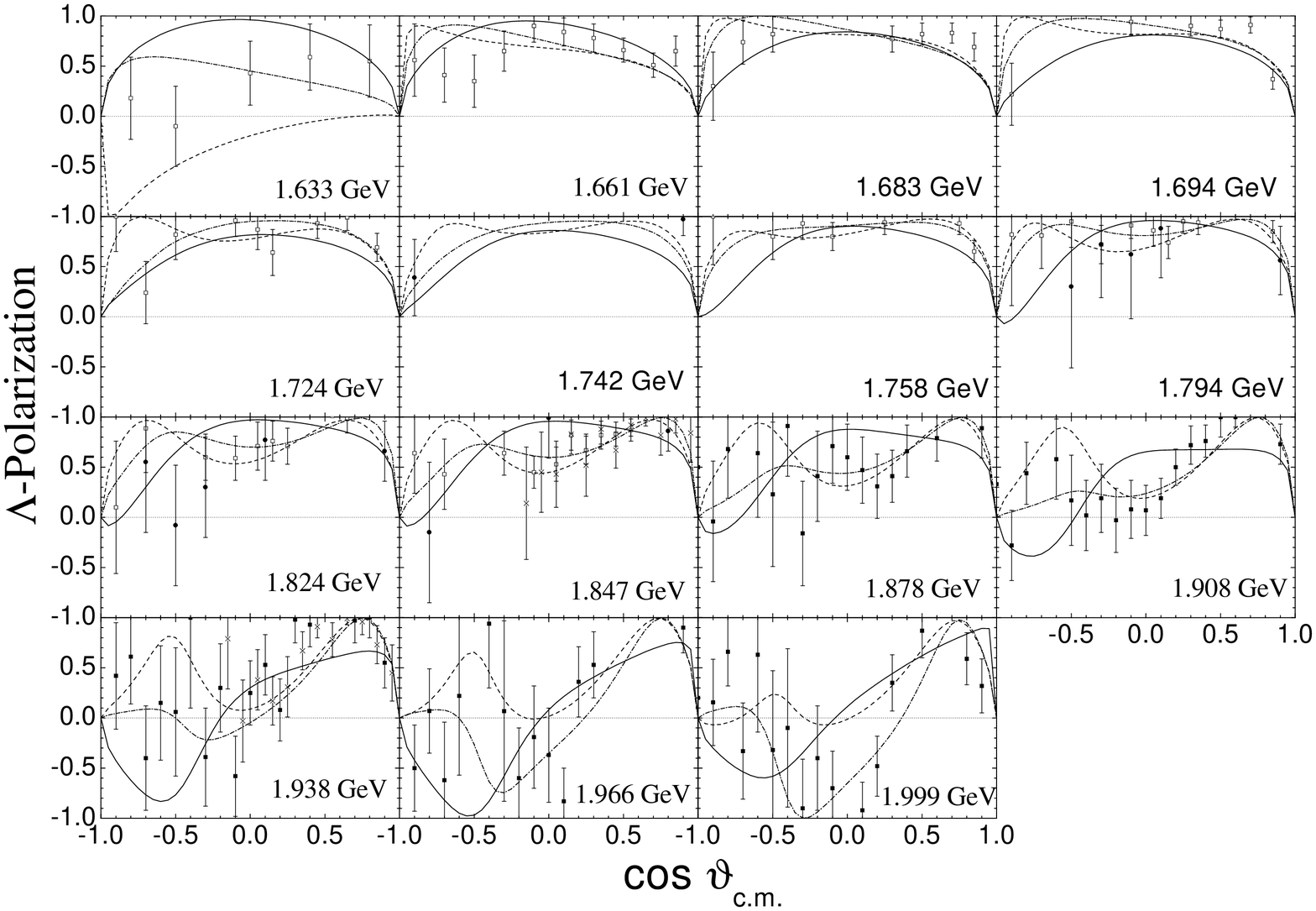}}
      }
    \caption{$\pi^- p \ra K^0 \Lambda$ angle- differential cross
      sections (\textit{upper panel}) and polarization measurements
      (\textit{lower panel}). 
      For the data references, see Refs. \cite{feusti98,gregiphd}.
      Notation as in Fig. \ref{figppi12}. 
      \label{figpldifpol}}
  \end{center}
\end{figure}
The inclusion of these coupled-channel effects and of the
$P_{13}(1900)$ resonance are major improvements as compared to
Refs. \cite{feusti98,feusti99}. There, these mechannisms 
were not included and thus the $K\Lambda$ channel was not
subjected to any threshold effect and the peaking behavior around 1.7
GeV had to be fully described by the $P_{11}(1710)$ resonance. In the
extended model space, this resonancelike behavior is mainly caused by
the $P_{13}(1720)$ resonance, but also influenced by the opening of
these two channels. 

The $S$ wave behavior in the Pascalutsa calculation P-p-$\pi +$
differs above 1.65 GeV from that in the conventional calculation
C-p-$\pi +$ (see Sec. \ref{secrespp} and Fig. \ref{figppi12}). The
largest differences between these calculations can thus be observed in
the $S_{11}$ wave contribution, which is more
pronounced in the Pascalutsa calculation giving rise to a slightly
different behavior at the lowest energies and at the $K\Sigma$
threshold. The coupled-channel effects become less obvious once the
photoproduction data are included. In the global calculation
C-p-$\gamma +$ the $S_{11}$ and $P_{13}$ waves are only slightly
influenced by the $\omega N$ threshold, while the $K\Sigma$ threshold
effect has completely vanished. Note that the $P_{13}$ wave dominates
over almost the complete considered energy region. The second most
important part comes from the $S_{11}$ staying almost constant in the
upper energy range, while close to threshold, a slight peak caused by
the $S_{11}(1650)$ is visible.

Although the new $P_{13}(1900)$ only has a small $K\Lambda$ width, it
improves the description of the reaction significantly due to
rescattering, similarly to the $S_{11}(1650)$ resonance in $\pi N \ra
\eta N$. Thus the $P_{13}(1900)$ gives rise to a good description
of the angle differential observables, while in Ref. \cite{feusti98} only
contributions from the $S_{11}(1650)$ and $P_{11}(1710)$ resonances
were found. The improvement becomes most visible in the high energy
region, where the full angular structure of the cross section 
and polarization of the $K\Lambda$ channel can be described, see
Fig. \ref{figpldifpol}. Especially for a description of the upward
bending behavior of the differential cross section at
backward angles at the highest energies, the inclusion of the
$P_{13}(1900)$ turns out to be important. Note that due to the change
of the $K_0^*$ coupling (cf. Table \ref{tabborncplgs}), the extreme
forward peaking behavior of the hadronic calculations is not visible
any more in the global calculation.

The polarization data hardly influence the determination of the
parameters due to the large error bars, see
Fig. \ref{figpldifpol}. However, all calculations give a good
description of the angular and energy dependent structure, in
particular the pure positive polarization for lower energies and the
change to negative values for the backward angles at higher energies.

\subsection{$\pi N \ra K\Sigma$}
\label{secresps}

\begin{figure}
  \begin{center}
    \parbox{16cm}{
      \parbox{75mm}{\includegraphics[width=75mm]{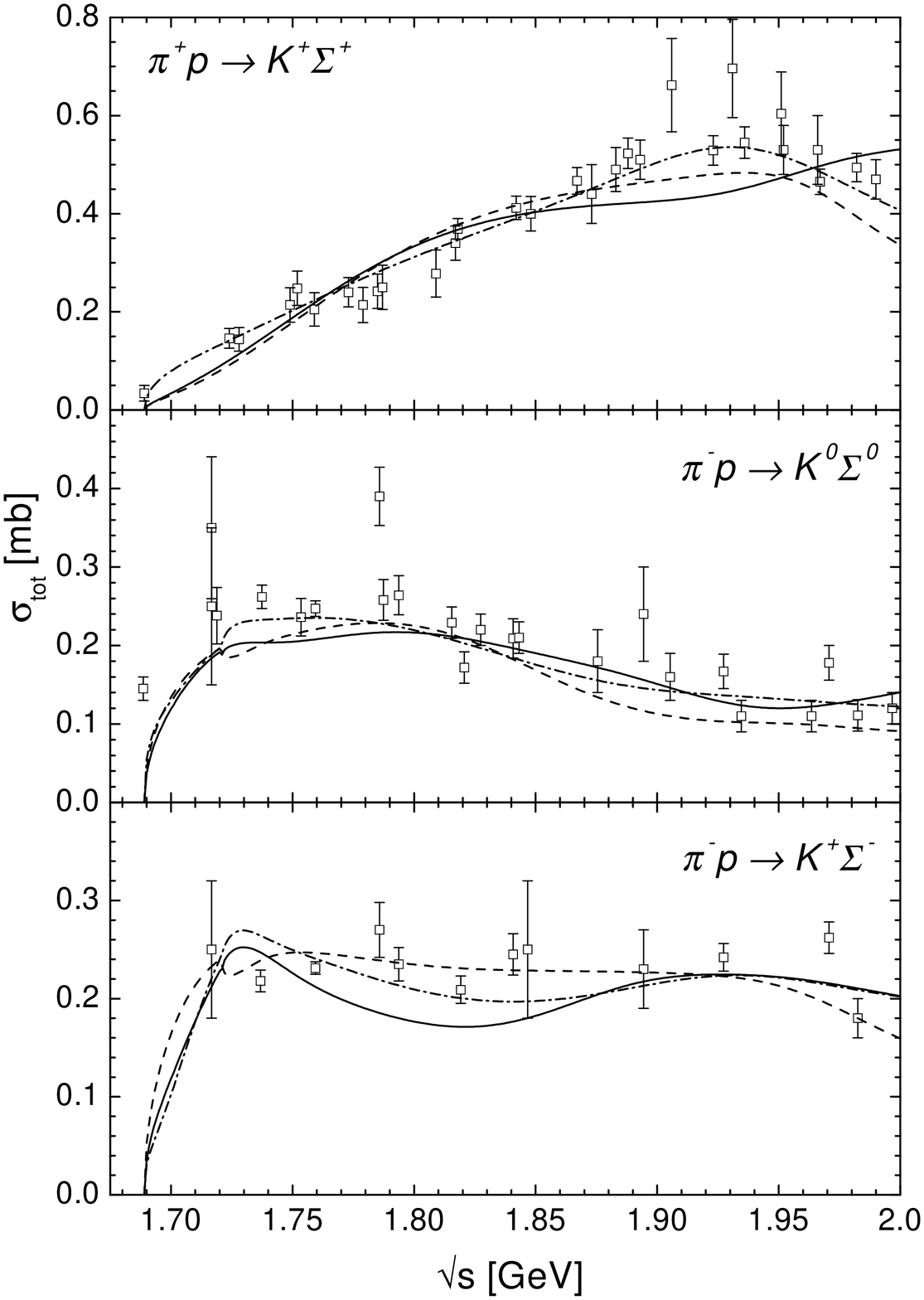}}
      \parbox{75mm}{\includegraphics[width=75mm]{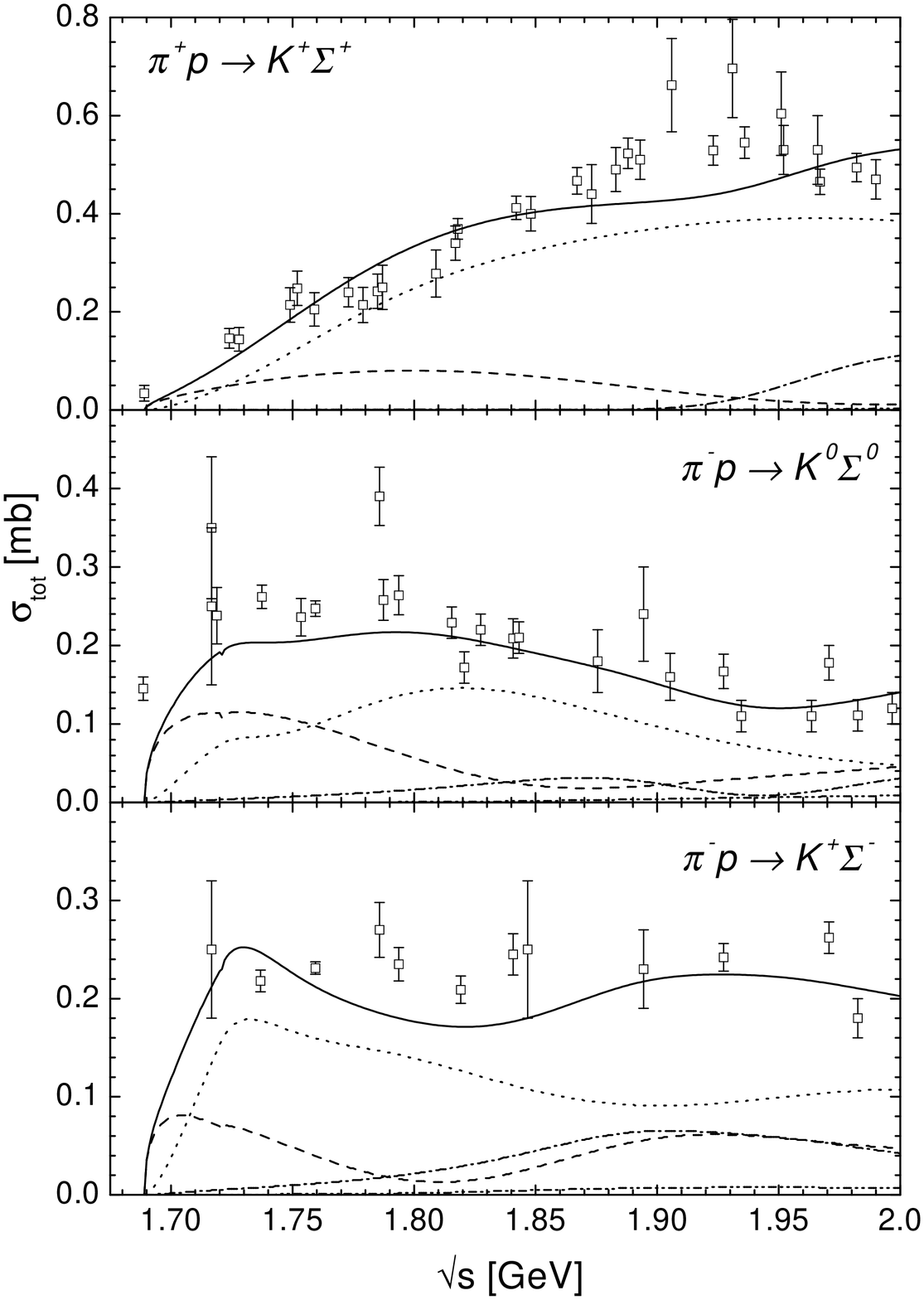}}
      }
    \caption{$\pi N \ra K\Sigma$ total cross sections for the
      different charge reactions. Notation as in Fig. \ref{figppi12}. 
      For the data references, see Refs. \cite{gregi,gregiphd}.
      \textit{Left:} Results of the different calculations. Notation
      as in Fig. \ref{figppi12}. \textit{Right:} Partial-wave
      decomposition of the total cross section for the calculation
      C-p-$\gamma +$. $J^P=\foh^-(S_{I1})$: dashed line;
      $\foh^+(P_{I1})$: dotted line; $\fth^+(P_{I3})$: dash-dotted
      line; $\fth^-(D_{I3})$: dash-double-dotted line. The sum of all 
      partial waves is given by the solid line.
      \label{figpstot}}
  \end{center}
\end{figure}

Due to the isospin structure of the $K\Sigma$ final state, the $\pi N
\ra K\Sigma$ channel is similar to $\pi N$ elastic
scattering. The reaction process is determined by two isospin
amplitudes ($I=\foh$ and $I=\fth$), while data have been taken for the
three charge reactions $\pi^+ p \ra K^+ \Sigma^+$, $\pi^-
p \ra K^0 \Sigma^0$, and $\pi^- p \ra K^+ \Sigma^-$. Since the first
reaction is purely $I=\fth$, it allows a stringent test of the
$I=\fth$ (resonance) contributions in the present model, while the
other two are a mixture of $I=\foh$ and $I=\fth$ contributions [see
Eqs. \refe{amplii32spec}]. Within our model it is possible to describe
all three charge reactions with approximately the same quality, see
Table \ref{tabchisquareps}, 
\begin{table}
  \begin{center}
    \begin{tabular}
      {l|c|c|c|c}
      \hhline{=====}
       Fit & Total $\chi^2_{\pi \Sigma}$ & 
       $\chi^2 (\pi^-p \ra K^0 \Sigma^0)$ &
       $\chi^2 (\pi^-p \ra K^+ \Sigma^-)$ &
       $\chi^2 (\pi^+p \ra K^+ \Sigma^+)$ \\
       \hline
       C-p-$\pi +$ & 1.97 & 2.14 & 1.85 & 1.97 \\
       C-p-$\pi -$ & 2.37 & 3.08 & 1.86 & 1.96 \\
       P-p-$\pi +$ & 2.93 & 3.34 & 1.67 & 3.01 \\
       P-p-$\pi -$ & 2.80 & 3.04 & 1.90 & 2.91 \\
       C-p-$\pi \slash \chi+$ & 2.48 & 2.63 & 2.29 & 2.42 \\
       C-t-$\pi +$ & 2.42 & 3.18 & 1.61 & 2.05 \\
       C-t-$\pi -$ & 2.48 & 3.67 & 1.92 & 1.66 \\
       C-p-$\gamma +$ & 2.97 & 2.76 & 2.06 & 3.45 \\
       C-p-$\gamma -$ & 3.94 & 4.06 & 4.90 & 3.53 \\
      \hhline{=====}
    \end{tabular}
  \end{center}
  \caption{Resulting $\chi^2$ of the various
    fits for the three different charge reactions of $\pi N \ra
    K\Sigma$.\label{tabchisquareps}}
\end{table}
corroborating the isospin decomposition of the $K\Sigma$ channel in
the present calculation. From the total cross section behavior, shown
in Fig. \ref{figpstot},
one deduces, that the threshold behavior of the reactions with
$I=\foh$ contributions is influenced by a strong $S_{11}$ wave,
arising from the $S_{11}(1650)$ just below the $K\Sigma$
threshold, and $P_{I1}$-wave dominance for increasing energies, which
stem from the $P_{31}(1750)$ and in particular the
$P_{11}(1710)$. However, the $P_{13}(1900)$ is also visible in the
$K^+\Sigma^-$ channel. In the pure $I=\fth$ reaction the $S$ wave
importance is largely reduced, and the $P$ waves dominating over the
complete energy range. Note that the $J^P=\fth^-$ waves do not give
any noticeable contribution to the cross sections, see also below. 
In the hadronic reactions it turns out that the main
contribution to the $I=\fth$ channel comes from the
$P_{33}(1920)$, however, the inclusion of the photoproduction data
moves this strength over to the $P_{31}(1750)$; see also Sec.
\ref{secresp2} above. A similar observation is made in the $I=\foh$
sector, where strength is also moved over from the $P_{13}$ to the
$P_{11}$ waves and the latter one is realized in a large
$P_{11}(1710)$ $K\Sigma$ width.

\begin{figure}
  \begin{center}
    \parbox{16cm}{
      \parbox{16cm}{\includegraphics[width=16cm]{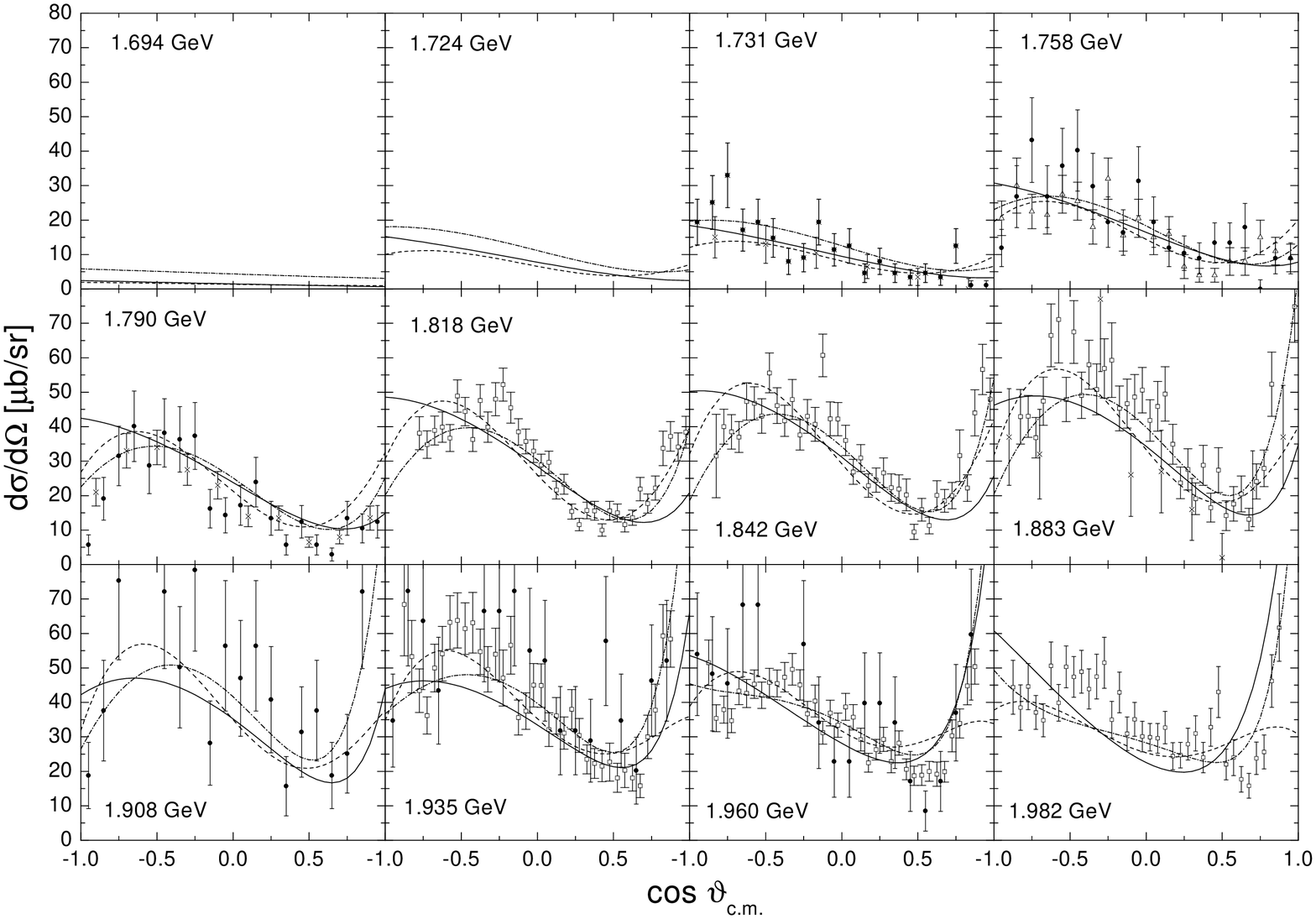}}
      \parbox{16cm}{\includegraphics[width=16cm]{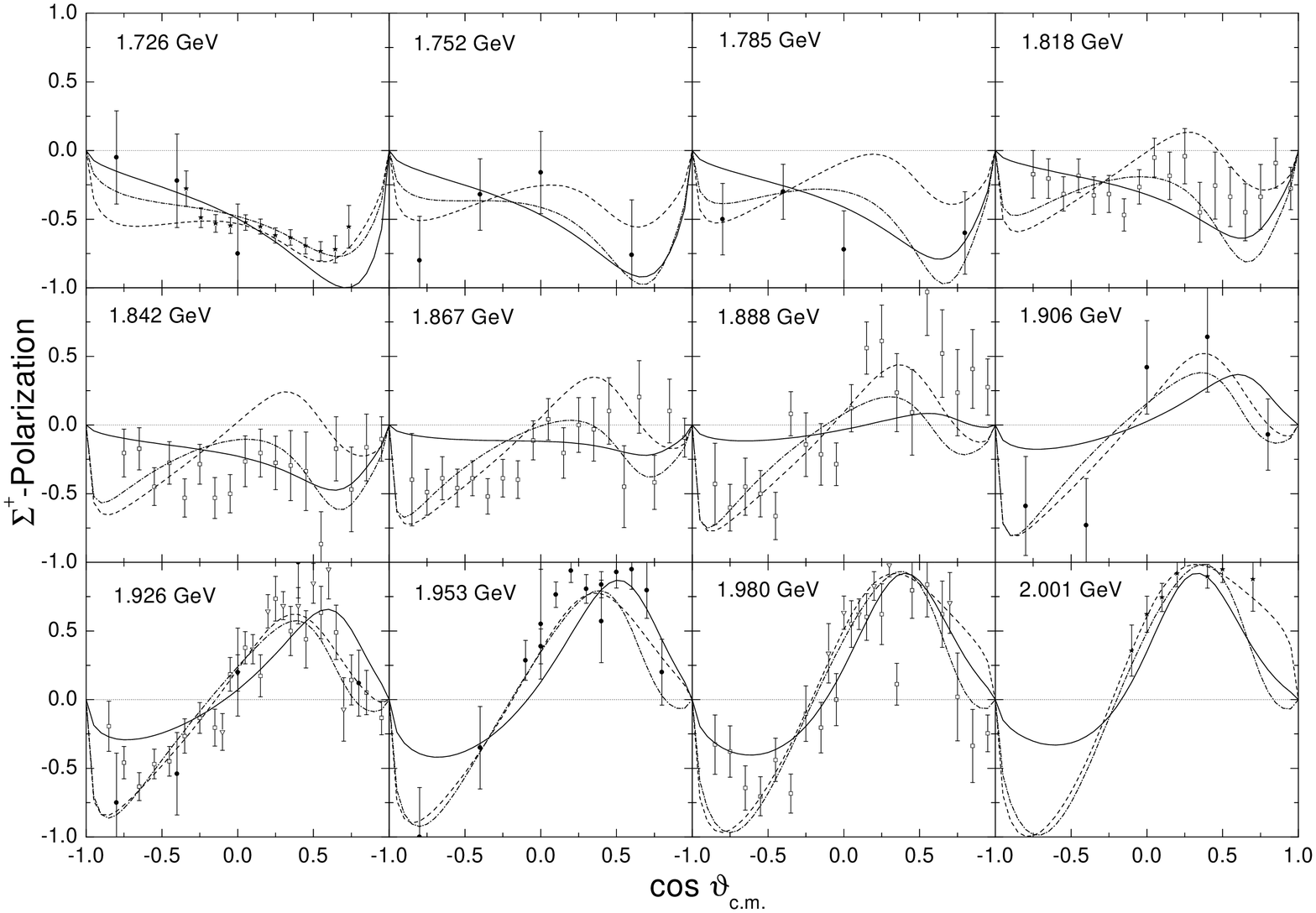}}
      }
    \caption{$\pi^+ p \ra K^+ \Sigma^+$ differential cross sections
      (\textit{upper panel}) and $\Sigma^+$-polarization
      measurements (\textit{lower panel}). Notation as in
      Fig. \ref{figppi12}. 
      For the data references, see Refs. \cite{gregi,gregiphd}.
      \label{figpspdifpol}}
  \end{center}
\end{figure}

\begin{figure}
  \begin{center}
    \parbox{16cm}{
      \parbox{16cm}{\includegraphics[width=16cm]{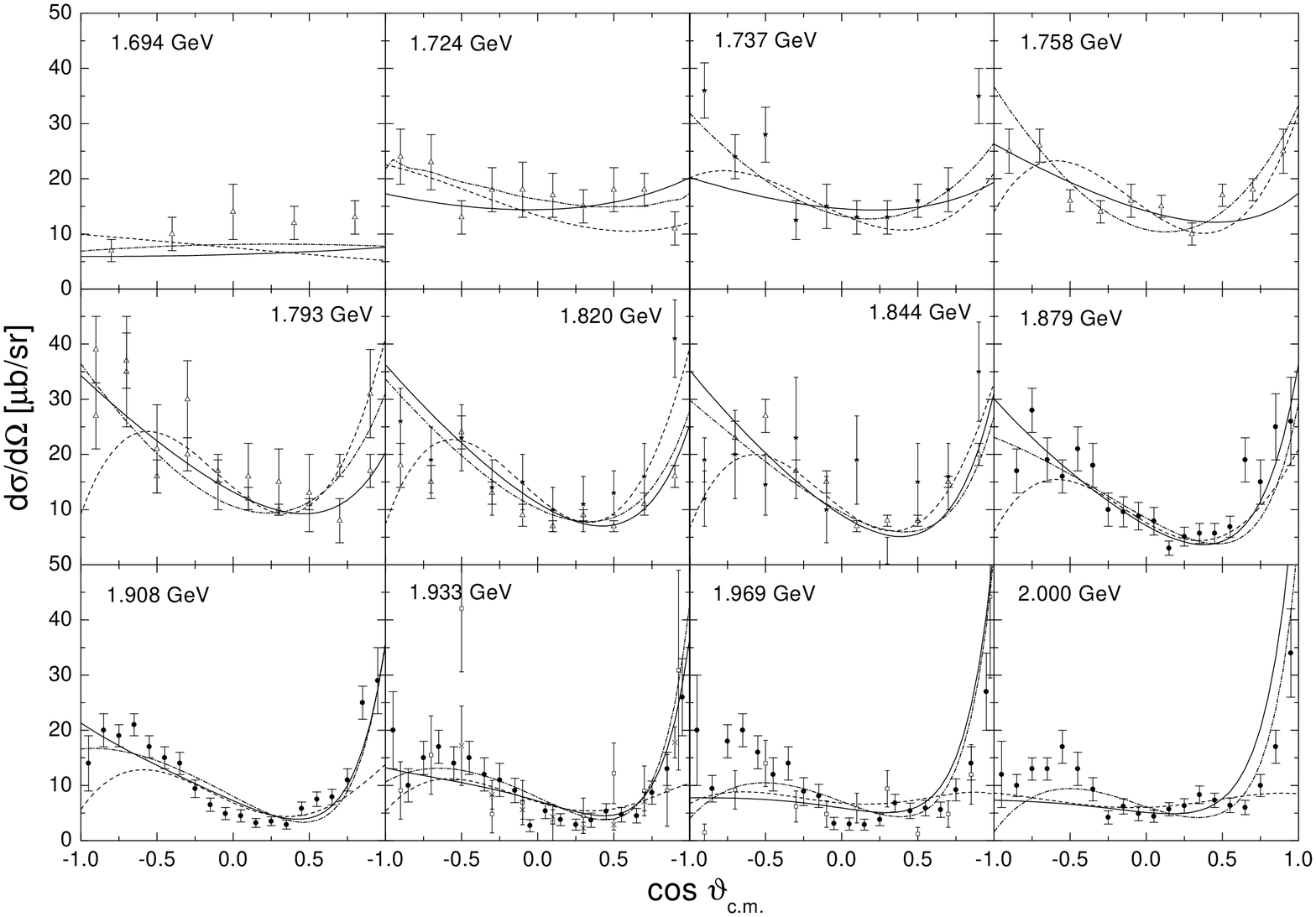}}
      \parbox{16cm}{\includegraphics[width=16cm]{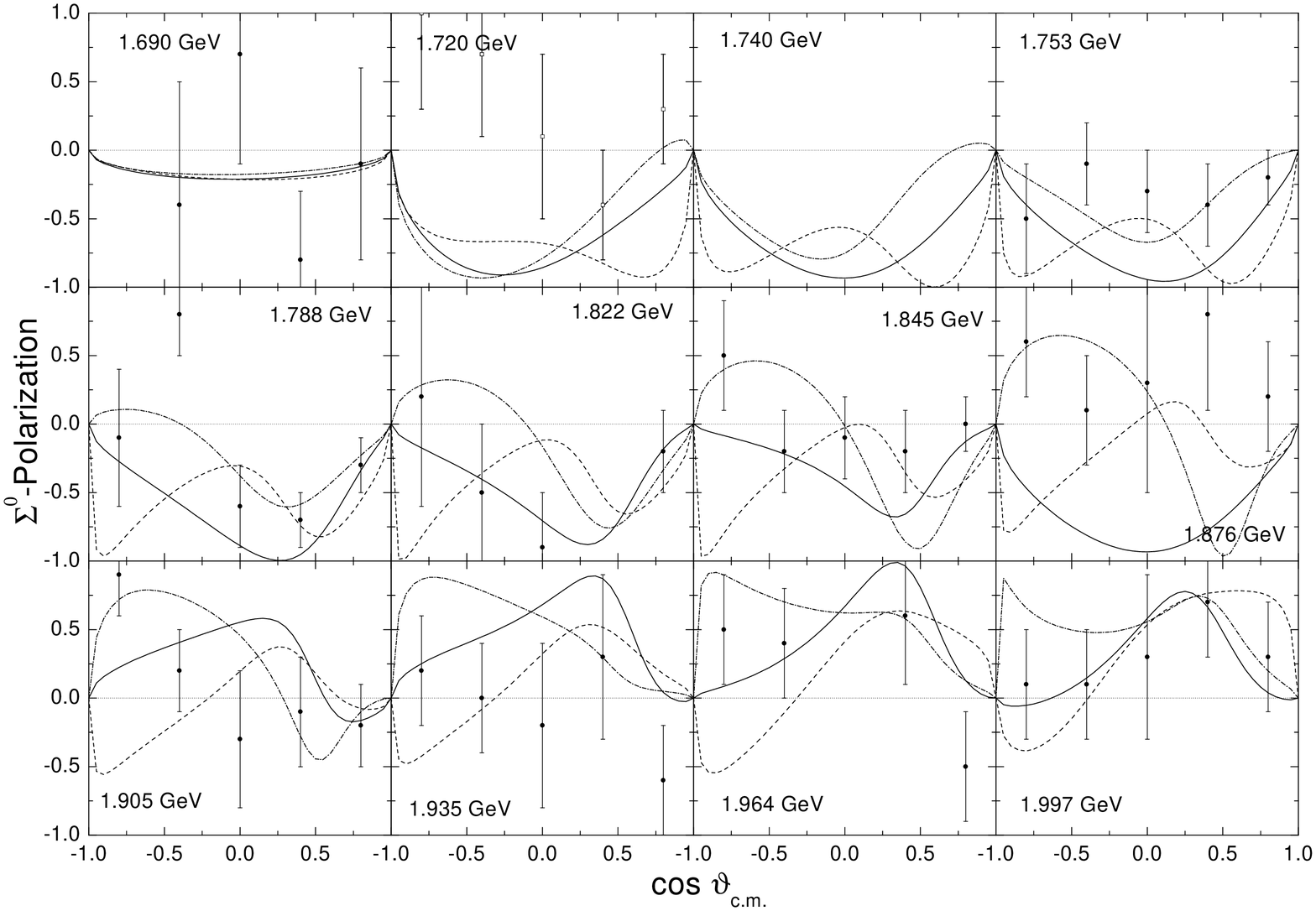}}
      }
    \caption{$\pi^- p \ra K^0 \Sigma^0$ differential cross sections
      (\textit{upper panel}) and $\Sigma^0$-polarization
      measurements (\textit{lower panel}). Notation as in
      Fig. \ref{figppi12}. 
      For the data references, see Refs. \cite{gregi,gregiphd}.
      \label{figps0difpol}}
  \end{center}
\end{figure}

\begin{figure}
  \begin{center}
    \parbox{16cm}{
      \parbox{16cm}{\includegraphics[width=16cm]{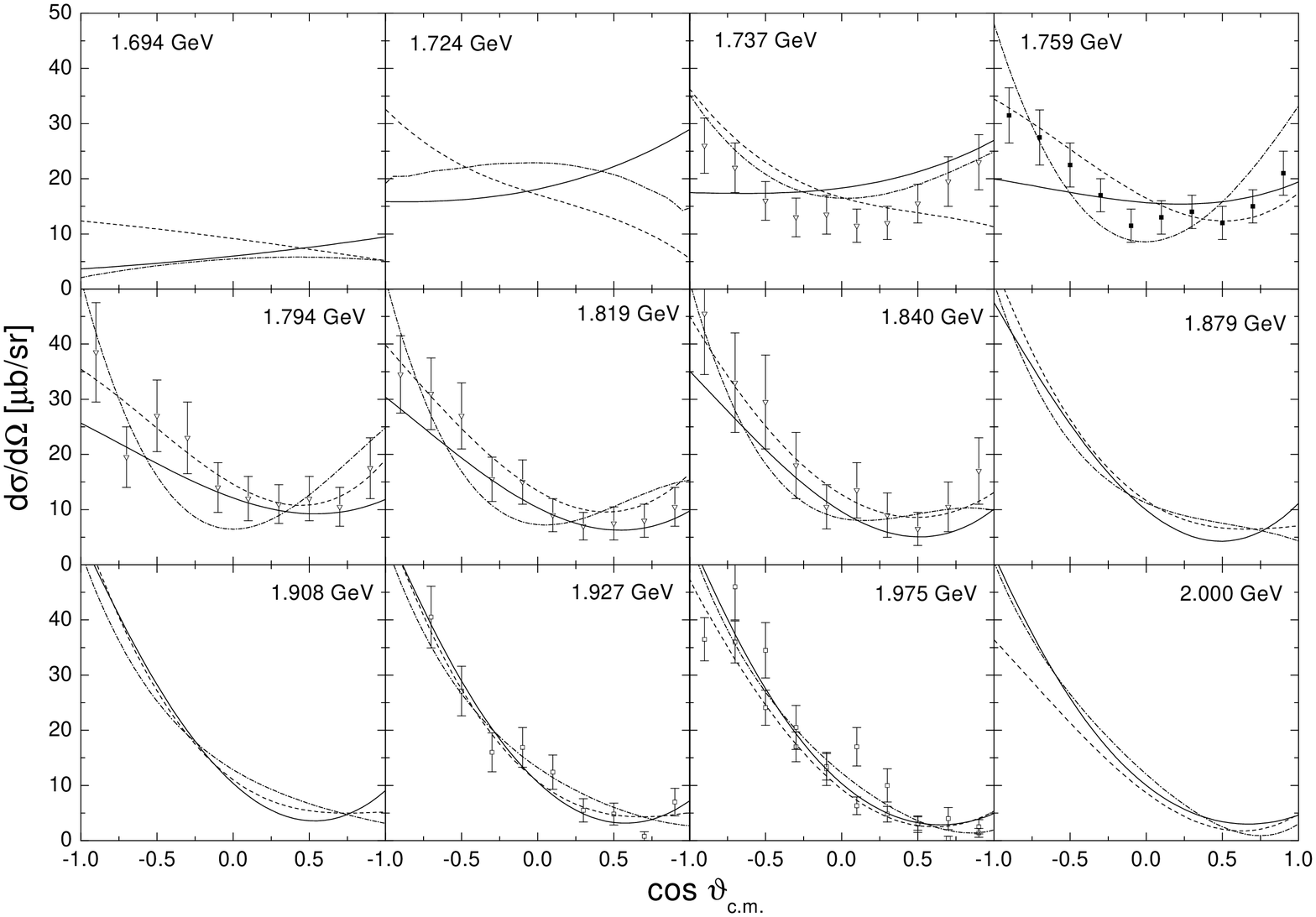}}
      }
    \caption{$\pi^- p \ra K^+ \Sigma^-$ angle-differential cross
      section. Notation as in Fig. \ref{figppi12}.
      For the data references, see Refs. \cite{gregi,gregiphd}.
      \label{figpsmdif}}
  \end{center}
\end{figure}

These contributions result in a very good description of the
differential cross sections and polarization measurements for all
three reactions, see Figs. \ref{figpspdifpol} -- \ref{figpsmdif}.
As pointed out above, the three reaction channels, which are built up
by only two isospin amplitudes, allow for strong constraints on the
partial-wave decomposition of the $K\Sigma$ production. Within our
model the full angular structure of all three charge reactions can be
well described, while in the SU(3) model of Ref. \cite{nobby} problems have 
been observed with the description of the backward peaking behavior of
the angle differential $\pi^- p \ra K^+ \Sigma^-$ cross section at
higher energies. This large difference to the other two charge
reactions, who both show a forward peaking behavior in this energy
range, can, however, be easily explained with the help of the
$t$-channel meson contributions of $K^+$ and $K_0^*$. Since both are
$I=\foh$ particles, they can only contribute to $\pi^- p \ra K^0
\Sigma^0$ and $\pi^+ p \ra K^+ \Sigma^+$, but not to $K^+ \Sigma^-$
production, which consequently tends to small values at forward
angles. The lack of $t$-channel contributions also explains the good
result of the calculation C-t-$\pi +$ for $\pi^- p \ra K^+ \Sigma^-$,
where the form factor $F_t$ has been used, although this form factor
leads in general to worse results (see Tables \ref{tabchisquares} and 
\ref{tabchisquareps}). On the other hand, the very good result of
C-t-$\pi -$ for $\pi^+ p \ra K^+ \Sigma^+$ has to be compensated by
a much worse $\pi^- p \ra K^0 \Sigma^0$ result.

This is also related to the observed difference between the
Pascalutsa and the conventional calculations in the differential cross 
section of $K\Sigma$ production at higher energies. The large forward
peaking behavior for higher energies in 
the $K^+\Sigma^+$ and $K^0\Sigma^0$ production cannot be described in
the Pascalutsa calculation. Due to the lack of the spin-$\fth$ offshell
contributions, in this calculation a larger cutoff value $\Lambda_t$ is
extracted, thus giving rise to more background contributions over the
complete angle and energy range. At the same time, a description of
the forward peaking behavior at high energies requires large couplings
to the $t$-channel mesons, but in the Pascalutsa calculations this
would spoil the agreement at backward angles and lower
energies. Consequently, the most striking differences between the
Pascalutsa and conventional calculations are found in the high-energy
region. For more details on the $t$-channel form factors and couplings,
see the discussion in Secs. \ref{secresbackform} and
\ref{secresbacktcplg}.

While the polarization measurements for $\pi^- p \ra K^0 \Sigma^0$
hardly influence the parameter extraction due to the large error bars, 
the measurements for $\pi^+ p \ra K^+ \Sigma^+$ largely constrain the
$I=\fth$ contributions, see Figs. \ref{figpspdifpol} and
\ref{figps0difpol}. The change of negative to positive polarization
values at forward angles with increasing energy, peaking around $\cos
\vt \approx 0.4$ is nicely described as a result of the $P_{33}(1920)$
contribution, confirming the strong necessity of $K\Sigma$ flux in the 
$P_{33}$ partial wave at higher energies. Note further that although
the contribution of the $D_{33}(1700)$ to the total cross section is
negligible (cf. Fig. \ref{figpstot}), it leads to the negative hump at 
$\cos \vt \approx 0.7$ in the $\Sigma^+$ polarization close to
threshold, thus affirming
the necessity of subthreshold contributions. Polarization
measurements of comparable quality for the reactions with
isospin-$\foh$ contributions would be very interesting for testing
the importance of the various resonance contributions, since due to
the large error bars, the different calculations for the polarization
measurement in $\pi^- p \ra K^0 \Sigma^0$ result in a quite different
behavior. The only common characteristic of the different calculations
in the $K^0 \Sigma^0$ polarization is caused by the $D_{33}(1700)$ and
$P_{33}(1920)$ resonances, enforcing the change from negative
polarization values at low energies to positive values at high
energies in the forward region.

\subsection{$\pi N \ra \omega N$}
\label{secrespo}

\begin{figure}
  \begin{center}
    \parbox{16cm}{
      \parbox{75mm}{\includegraphics[width=75mm]{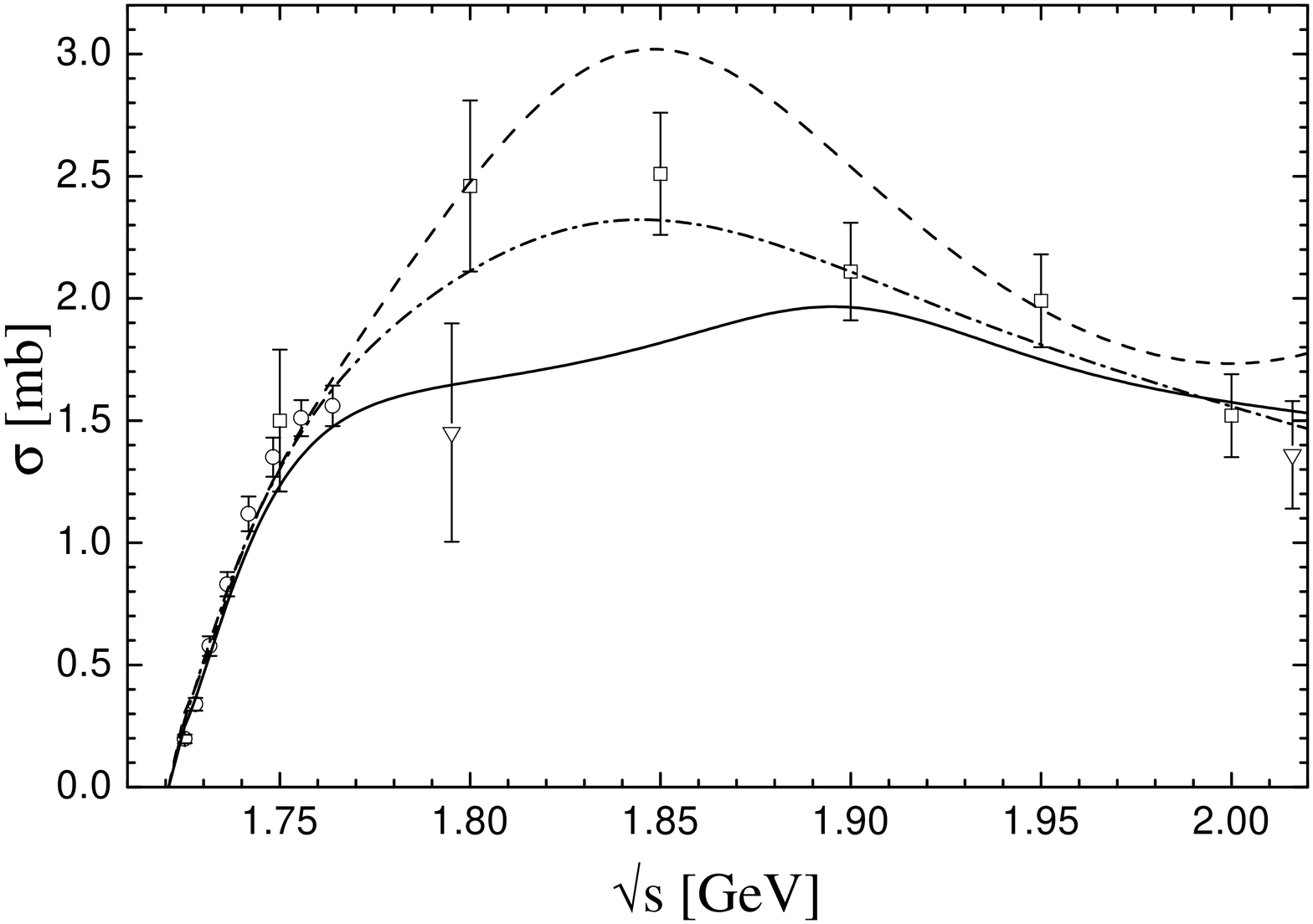}}
      \parbox{75mm}{\includegraphics[width=75mm]{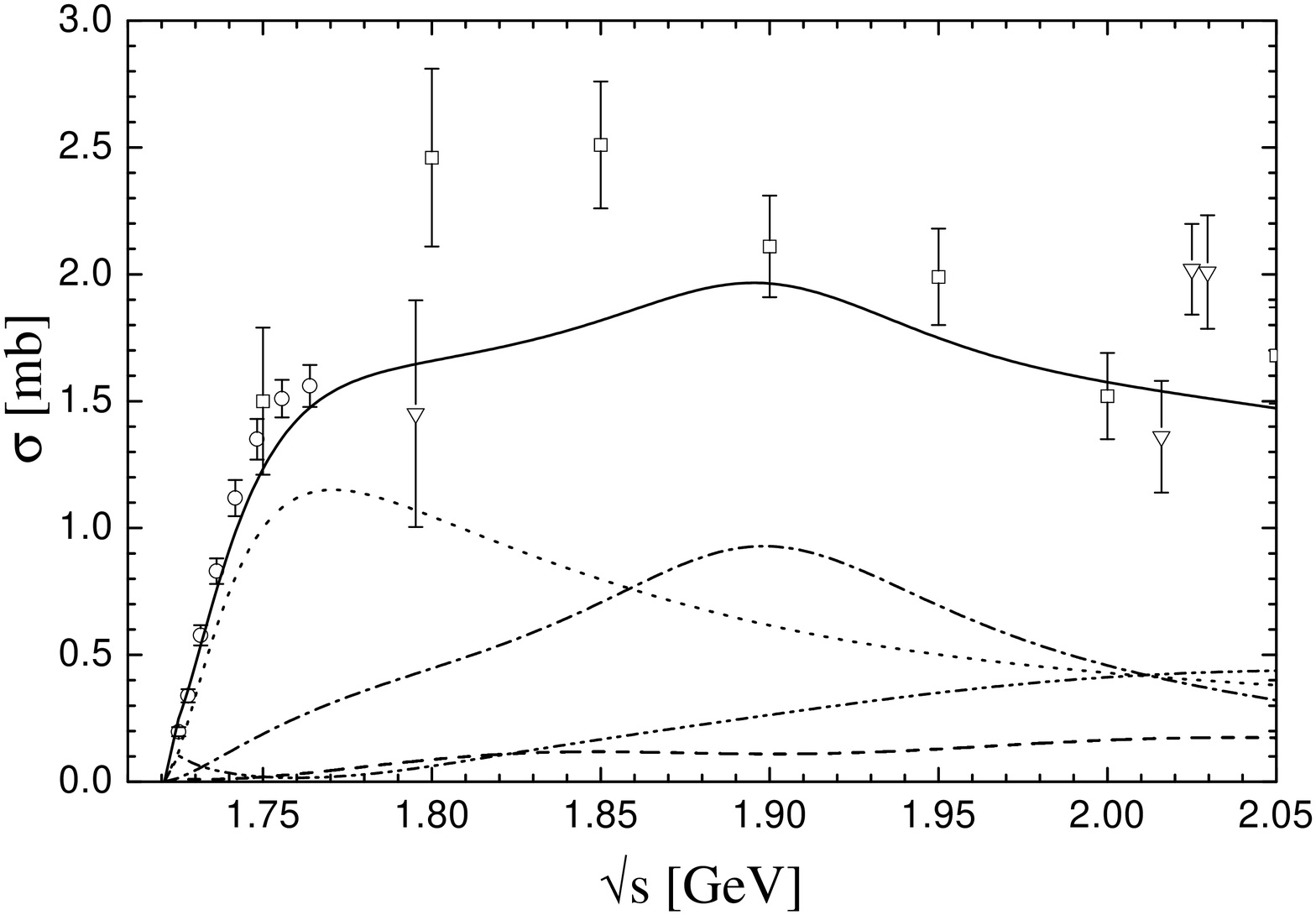}}
      }
    \caption{$\pi^- p \ra \omega n$ total cross section. 
      For the data references, see Refs. \cite{gregi,gregiphd}.
      \textit{Left:} Results of different calculations. Line code as in
      Fig. \ref{figppi12}. \textit{Right:} Partial-wave decomposition
      of the total cross section. $J^P=\foh^-$: dashed line; $\foh^+$:
      dotted; $\fth^+$: dash-dotted; $\fth^-$: dash-double-dotted.
      \label{figpotot}}
  \end{center}
\end{figure}

As can be seen from Fig. \ref{figpotot} the $\omega N$ channel, 
which strongly influences all other reactions, cannot be completely 
fixed by using the pion-induced data alone. While in the hadronic 
calculations C-p-$\pi +$ and P-p-$\pi +$, the total cross section is
dominated by a $J^P=\fth^-$ wave, resonating below 1.85 GeV and
accompanied by a strong $\fth^+$ wave, this picture is 
changed once the much more precise $\omega N$ photoproduction data
from the SAPHIR Collaboration \cite{barthom} are included. In the
global calculation, the $\foh^+$ and $\fth^+$ waves dominate up to 
energies of 2 GeV. The $P_{11}(1710)$ leads to the peaking in the
$\foh^+$ wave around 1.76 GeV, while the $P_{13}(1900)$ gives 
rise to the peaking behavior of the $\fth^+$ contribution around 1.9
GeV, see Fig. \ref{figpotot}. This decomposition leads to a slower
increase of the total cross section at energies above 1.745 GeV; a
property which is also indicated by the precise Karami total cross
section data \cite{karami}. This is in contrast to our findings in
Ref. \cite{gregi}, where a dominant
$\fth^-$ contribution has been extracted because the more precise
photoproduction data have not been considered simultaneously.   
The comparison of this result with the coupled-channel model of Lutz
{\it et al.} \cite{lutz} is especially interesting, because there, $\pi N
\ra \omega N$ is described by a pure $\fth^-$ production
mechanism. This is due to the fact that in the 
model of Ref. \cite{lutz} no $P$ wave contributions are included. These
authors' findings seem to lead to an overestimation of the $\pi N$
inelasticity in the $\fth^-$ ($D_{13}$) channel, which just starts
overshooting the experimental data at the $\omega N$
threshold. Unfortunately, they do not compare their calculation to the
angle-differential Karami cross section \cite{karami}, which would
allow for a further evaluation of the quality of their
calculation. There has also been a single-channel analysis on $\pi N
\ra \omega N$ by Titov {\it et 
al.} \cite{titov}\footnote{Note that Ref. \cite{titov} has not used 
  the correct experimental data, but followed the claim of
  Ref. \cite{hanhart99}; see Refs. \cite{gregi,gregidata}.}.
These authors have extracted dominant contributions from the
subthreshold $S_{11}(1535)$, $S_{11}(1650)$, and $P_{11}(1440)$
resonances, which only give minor contributions in the present
calculation. These authors also neglected the $P_{11}(1710)$ and
resonances beyond the $P_{13}(1720)$, both of which turn out to be
most important in the present calculation. 

This once again shows the
necessity of the inclusion of photoproduction data for a reliable
analysis of resonance properties, especially in channels (as the
$\omega N$ production), where only few precise pion-induced data are
available.

\begin{figure}
  \begin{center}
    \parbox{16cm}{
      \parbox{16cm}{\includegraphics[width=16cm]{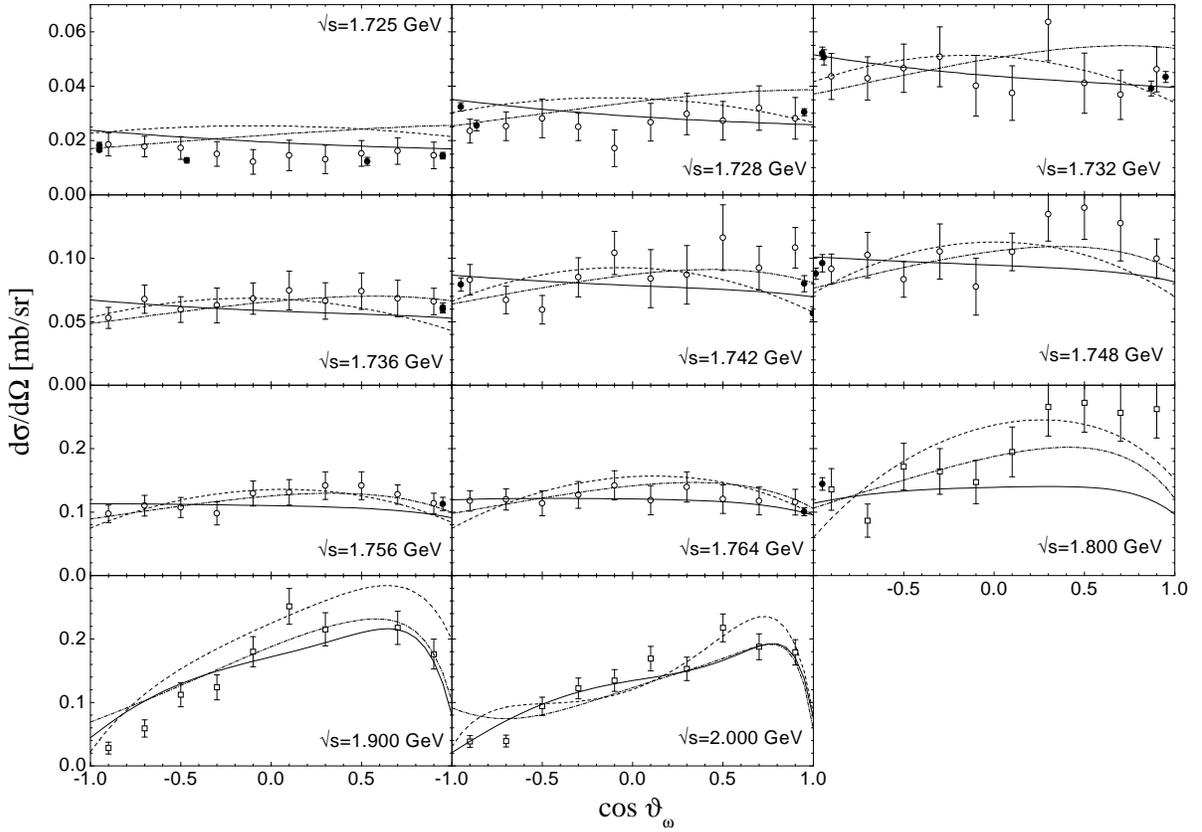}}
      }
    \caption{$\pi^- p \ra \omega n$ angle-differential cross
      section. Line code as in Fig. \ref{figppi12}. For the data
      references, see Refs. \cite{gregi,gregiphd}.
      \label{figpodif}}
  \end{center}
\end{figure}

The differential cross section shows an almost flat behavior close to
threshold, see Fig. \ref{figpodif}, even for the global calculation
dominated by $P$ waves.
To get a handle on the angle-differential structure of the cross
section for higher energies ($\sqrt s \geq 1.8$ GeV) we have used the
corrected cosine event distributions given in Ref. \cite{danburg} to
also extract differential cross sections with the help of the given
total cross sections. While the differential cross section at forward
angles is almost constant above 1.8 GeV, the backward cross section
decreases. These data points strongly constrain the nucleon
$u$-channel contribution thereby restricting the $\omega NN$ coupling
constants, and the downbending behavior is best described by the
global fit. At these energies also the forward peaking behavior
becomes visible which is due to the $t$-channel $\rho$ meson
exchange. This contribution is also the reason why the 
forward peaking behavior is more pronounced in the Pascalutsa
calculation. Although the extracted $\rho NN$ coupling is smaller than 
in the other calculations, the cutoff value $\Lambda_t$ (cf. Tables
\ref{tabborncplgs} and \ref{tabcutoff} below) is much
larger than in the other calculations resulting in an effectively
larger contribution, see also the discussion in Secs.
\ref{secresbackform} and \ref{secresbacktcplg} below.

It should also be noted that the $\omega N$ parameters are
not constrained by the $\omega N$ data points alone but also greatly
influenced by the $\pi N$ inelasticities and cusp effects 
appearing in $\eta N$, $K\Lambda$, and $K\Sigma$ production due to the
$\omega N$ threshold opening. Therefore the extracted partial-wave
decomposition of $\pi N \ra \omega N$ is on safe grounds, since all
other channels and in particular the $\pi N \ra \pi N$ partial waves
and inelasticities and the pion-induced $2\pi N$ production are well 
described in the energy region above the $\omega N$
threshold. However, more precise cross section measurements at
energies above $1.76$ GeV and polarization measurements of the $\pi N
\ra \omega N$ production would be the perfect tool to corroborate the
present findings.

\section{Extracted Hadronic Parameters}

\subsection{Background contributions and $t$-channel form factors}
\label{secresback}

The values of all Born and $t$-channel coupling constants, which have
been varied during the calculation, are listed in Table
\ref{tabborncplgs}.
\begin{table}[t]
  \begin{center}
    \begin{tabular}
      {l|r|l|r|l|r|l|r}
      \hhline{========}
      $g$ & value & $g$ & value & $g$ & value & $g$ & value \\
      \hhline{========}
      $g_{NN\pi}$      &  12.85 & $g_{NN\sigma} \cdot g_{\sigma \pi \pi}$ & $ 22.92$ & $g_{NN\rho}$       &   4.53 & $\kappa_{NN\rho}$     &   1.47 \\
      &  12.75 & & $ 25.14$ & &   4.40 & &   1.41 \\
      &  12.77 & & $ 26.88$ & &   5.59 & &   1.51 \\
      &  12.80 & & $ 39.16$ & &   2.71 & &   1.16 \\
      &  13.01 & & $ 13.66$ & &   2.21 & &   1.30 \\
      \hline
      $g_{NN\eta}$     &   0.10 & $g_{NNa_0}$ & $-70.60$ & $g_{NN\omega}$     &   3.94 & $\kappa_{NN\omega}$        &  $-0.94$ \\
      &   0.12 & & $-45.82$ & &   3.87 & &   0.17 \\
      &   0.06 & &   39.56  & &   4.06 & &   0.48 \\
      &   0.07 & &  $-2.98$ & &   3.90 & &   0.59 \\
      &   0.29 & &  $ 8.60$ & &   3.94 & &  -0.90 \\
      \hline
      $g_{N\Lambda K}$ & $-12.20$ & $g_{N\Lambda K_0^*}$ &  52.54 & $g_{N\Lambda K^*}$ & $-27.61$ & $\kappa_{N\Lambda K^*}$ &  $-0.50$ \\
      & $-12.88$ & & $  2.32$ & & $-28.29$ & &  $-0.55$ \\
      & $-18.48$ & & $-25.56$ & & $-27.85$ & &  $-0.36$ \\
      & $-14.35$ & & $  2.36$ & & $  3.10$ & &  $ 0.01$ \\
      & $-11.53$ & & $-11.58$ & & $ -5.86$ & &  $-0.39$ \\
      \hline
      $g_{N\Sigma K}$  &   2.48 & $g_{N\Sigma K_0^*}$ & $-52.30$ & $g_{N\Sigma K^*}$  &   4.33 & $\kappa_{N\Sigma K^*}$  &  $-0.86$ \\
      &   1.56 & & $-54.44$ & &    3.88  & &  $-0.98$ \\
      &  15.39 & & $ 65.28$ & &    2.29  & &  $ 0.40$ \\
      &  12.44 & & $ -2.14$ & &  $-4.22$ & &  $-0.33$ \\
      &   2.50 & & $ 11.06$ & &  $ 0.71$ & &  $-0.11$ \\
      \hhline{========}
    \end{tabular}
  \end{center}
  \caption{Nucleon and $t$-channel couplings. First line: C-p-$\gamma
    +$; second line: C-p-$\gamma -$; third line: C-p-$\pi +$; fourth line:
    P-p-$\pi +$; fifth line: C-t-$\pi +$. The values for the $K_1$ meson
    are given for the global calculations C-p-$\gamma +$ and
    C-p-$\gamma -$. 
    \label{tabborncplgs}}
\end{table}
Note that no other background parameters are used in the
calculations, emphasizing the reduced freedom of the background in our
model as compared to analyses driven by resonance models (see, e.g.,
Ref. \cite{vrana}).

\subsubsection{Born couplings}
\label{secresbackborn}

Our values of $g_{\pi NN}$ are consistently lower than the values
extracted by other groups, for example the value of $g_{\pi NN} =
13.13$ from the VPI group \cite{SM00}. However, one has to keep in
mind that the present calculation considers a large energy region
using only one $\pi NN$ coupling constant, thereby putting large
constraints through all production channels on this coupling and the
threshold region only plays a minor role. For example in the global
fits, the $\pi NN$ coupling is especially influenced by the
$t$-channel pion exchange mechanism of $\omega N$ photoproduction,
which is due to the restriction of using only one cutoff value
$\Lambda_t$ for all $t$-channel diagrams. 

For the other couplings of the nucleon to the pseudoscalar final state 
mesons, the situation in the pion-induced reactions is
different. As found in previous analyses
\cite{sauermann,feusti98,feusti99} the $\eta NN$ coupling turns out to 
be very small and the precise value thus hardly influences the
$\chi^2$ of $\eta N$ production. Also in $\pi N \ra
K\Lambda / K\Sigma$, the Born couplings are only of minor importance
due to the large offshellness of the nucleon and the associated large
reduction of its contributions by the hadronic form factor. For
example, a doubling of the $KN\Lambda/\Sigma$ coupling constants
keeping all other contributions fixed leads to a worsening in $\chi^2$
for $\pi^- p \ra K^0 \Sigma^0/K^+ \Sigma^-$ of only about 10\%, and
for $\pi^- p \ra K^0 \Lambda$ of about 15\%. This also explains, why
the $NK\Sigma$ coupling extracted from the pion-induced data alone,
always ends up to be large compared to SU(3) expectations. However,
the situation changes drastically when the photoproduction data is
included. As a result of gauge invariance, the importance of the Born
diagrams is enhanced in the photoproduction reactions and allows to
determine the Born couplings more reliably. The resulting relations
between the Born couplings of our best global fit are actually close
to SU(3) relations with $\alpha_{FD} = F/(F+D) \in [0.25;0.41]$ (see, e.g., 
Ref. \cite{dumbrajs}), which is around the value of $\alpha_{FD} \approx
0.35$ predicted by the Cabibbo-theory of weak interactions and the
Goldberger-Treiman relation \cite{dumbrajs}.

As has already been pointed out in Ref. \cite{gregi}, the $\omega NN$ 
coupling constants have more influence on 
the angular dependent behavior of the pion-induced reaction process
than the $NK\Lambda$ and $NK\Sigma$ couplings and can therefore be
better fixed already in the hadronic fits, see Table
\ref{tabborncplgs}. This is a result of the
nucleon $u$-channel contribution, which strongly influences the
behavior of the angle-differential cross section in the backward
direction at higher energies, and explains why the resulting
values for this coupling are very similar in all calculations. Note
that a value $g_{\omega NN} \approx 4$ is extracted 
in our calculations, even though the same nucleon cutoff 
$\Lambda_N\approx 1$ GeV (see Table \ref{tabcutoff}) is used for
all final states, which is in contrast to the results found in
single-energy analysis (see, e.g., Ref. \cite{titov}).

\subsubsection{$t$-channel form factors}
\label{secresbackform}

It is interesting to compare our value of $g_{\omega NN} \sim 4$ with,
e.g., the value of $15.9$ which has been extracted in the Bonn-model
for nucleon-nucleon scattering \cite{machleidt}. In nucleon-nucleon
scattering, the $\omega$ only contributes via $t$-channel exchange and 
thus its coupling is always modified by a form factor. The actual shape 
of the form factor and the kinematic region are thus of great
importance for the applicability of the extracted coupling.

We have examined the influence of the form factor shape by performing
calculations with two different form factors $F_p$ \refe{formfacp} and
$F_t$ \refe{formfact} for the $t$-channel exchanges. In Ref. 
\cite{feusti98} no significant differences in the resulting quality of
the fits have been found, when either of the two form factors has been
used and consequently, in Ref. \cite{feusti99} only calculations using
$F_t$ have been performed. However, as Table \ref{tabchisquares}
shows, this result is not valid any more for the extended channel
space and kinematic region of the present model. The calculations
C-t-$\pi \pm$, which use $F_t$ instead of $F_p$ as in C-p-$\pi \pm$,
result in an overall description, which is worse by more than 10\%,
with the largest differences in the $\pi N \ra \omega N$
reaction. This reaction differs from $\eta N$, $K\Lambda$, and
$K\Sigma$, which have comparable $\chi^2$, in that respect, that in
the $t$ channel the $\rho$ meson is exchanged. Since this exchange
also contributes to $\pi N$ elastic scattering, the combination of
coupling and form factor for the $NN\rho$ vertex is tested in two
different reactions and thus in a wide kinematic region. As a result
of the larger data base for $\pi N$ elastic scattering, the value of
$g_{\rho NN}$ is adjusted to this reaction and there is no freedom
left for $\pi N \ra \omega N$. Since the calculations using $F_p$ can
describe both reactions simultaneously, the form factor shape $F_p$
seems to be applicable to a wider kinematic region than $F_t$. Note 
that this finding is even fortified when we look at the global
fits. There, no satisfying description of the experimental data using
$F_t$ has been possible, see PMII \cite{pm2}. This comes about
because of the quite different $q^2$ dependent behavior of the two
form factors $F_p$ and $F_t$ below the pole mass and in the low
$|t|=|q^2|$ region.

\subsubsection{$t$-channel couplings}
\label{secresbacktcplg}

Having performed calculations with two different $t$-channel
form factor shapes allows us to compare those couplings, which only
contribute to $t$-channel processes. As can be seen from Table
\ref{tabborncplgs}, large differences in these couplings are found
comparing the calculations with the conventional spin-$\fth$
couplings, with the Pascalutsa couplings, and with the use of $F_t$
instead of $F_p$ in the $t$ channel, while in the two global fits
C-p-$\gamma \pm$, differing only by the sign of $g_{\omega \rho \pi}$, 
the couplings are almost identical. The reduction of the $t$-channel
couplings when $F_t$ is used is not surprising, since the form factor
shape \refe{formfact} leads to less damping than $F_p$
\refe{formfacp}. In the case of the Pascalutsa calculations, the need
for background contributions also in lower partial waves is enhanced,
thereby leading to larger cutoff values $\Lambda_t$, see Table
\ref{tabcutoff}.
\begin{table}[t]
  \begin{center}
    \begin{tabular}
      {r|r|r|r}
      \hhline{====}
      $\Lambda_N$ [GeV] & 
      $\Lambda^h_\foh$ [GeV] & 
      $\Lambda^h_\fth$ [GeV] & 
      $\Lambda^h_t$ [GeV] \\
      \hhline{====}
      0.96 & 4.00 & 0.97 &  0.70 \\
      0.96 & 4.30 & 0.96 &  0.70 \\
      1.16 & 3.64 & 1.04 &  0.70 \\
      1.17 & 4.30 & 1.02 &  1.80 \\
      1.11 & 3.80 & 1.00 &  0.70 \\
      \hhline{====}
    \end{tabular}
  \end{center}
  \caption{Cutoff values for the form factors. First line: C-p-$\gamma
    +$; second line: C-p-$\gamma -$; third line: C-p-$\pi +$; fourth
    line: P-p-$\pi +$; fifth line: C-t-$\pi +$. The upper
    index $h$ denotes that the value is applied to a
    hadronic vertex, while the lower one denotes
    the particle going off-shell, i.e., $N$: nucleon; $\foh$:
    spin-$\foh$ resonance; $\fth$: spin-$\fth$ resonance; $t$:
    $t$-channel meson.
    \label{tabcutoff}}
\end{table}
At the same time, the corresponding couplings have to be reduced to
prevent an overshooting at forward angles and higher energies as in
$\pi N \ra K\Sigma$, see Sec. \ref{secresps} above. Comparing the
last three lines in Table \ref{tabcutoff}, where basically three
different background models have been used, one still finds that the
off-shell behavior of the nucleon and resonance contributions are
similarly damped, thus leading to similar resonant structures in the
three calculations C-p-$\pi +$, P-p-$\pi +$, and C-t-$\pi +$.

Thus our analysis shows that coupling constants extracted from $t$-channel 
processes strongly depend on the chosen cutoff function and cutoff
value. As in the $\pi N \ra \omega N$ reaction, this can in particular
lead to the effect that a calculation with a smaller $t$-channel
coupling (P-p-$\pi +$) results in larger $t$-channel contributions
than a calculation with a smaller coupling (C-p-$\pi +$), see
Fig. \ref{figpodif} above. Only when those couplings are also tested
close to the on-shell point or a wide kinematic range, the
applicability of the couplings and form factors is subjected to more
stringent test and the extracted values and form factor shapes become
meaningful. In the present model, this holds true for $NN\rho$ and 
$NN\sigma$ in $\pi N$ elastic scattering,  and the $NN\omega$,
$NN\pi$, and $NN\eta$ couplings, where the latter three appear
simultaneously in $s$-, $u$-, and $t$-channel processes.

Hence couplings as $g_{\omega NN}$ from, e.g., the Bonn-model
\cite{machleidt}, can only be interpreted in combination with the
cutoff used {\it and} in the kinematic region where it has been
applied to. This point has also been examined by Pearce and Jennings
\cite{pearce}. These authors have shown that the use of form factors as 
ours as compared to the one in the Bonn potential leads to large
differences in the off-shell behavior of the effective couplings. 

A similar consideration as for the $\pi NN$ coupling has also to be
applied to the $\rho NN$ coupling. Due to the fitting of the complete
energy region from threshold up to 2 GeV, the resulting $\rho NN$
coupling represents an averaged coupling which can deviate 
from values extracted in a restricted kinematic regime. Furthermore,
the $\rho NN$ coupling is also influenced by $\pi$ and $\eta$ 
photoproduction and also pion-induced $\omega$ production. Thus it is 
{\it a priori} not clear how well the resulting coupling reproduces the
KSRF relation. As pointed out in Sec. \ref{secnuclchi}, the 
KSRF relation, which relates the $\rho$ $t$-channel exchange to the
Weinberg-Tomazawa contact term, requires a coupling of $g_{\rho NN} =
2.84$. At first sight, it seems from Table \ref{tabborncplgs} 
that only in the calculations when the Pascalutsa spin-$\fth$
couplings is used is this relation fulfilled. However, the only
meaningful quantity entering the calculations is the product of
form factor and coupling constant. Evaluating $F_p$ for $\Lambda_t =
1.804$ ($0.705$) as in calculation P-p-$\pi +$ (C-p-$\pi +$) for
$q^2=0$ shows that $g_{\rho NN}^{eff} = g_{\rho NN} \cdot F_p (q^2 =
0) = 2.62$ ($2.31$) at threshold; thus both calculation result in a
similar effective coupling close to the KSRF value. Although the
$\rho$ tensor coupling $\kappa_\rho \approx 1.6$ turns out to be small
compared to the empirical VMD value of $3.71$, it points in the
direction of the value recently extracted in a model based on a gauge
formalism including $\rho$ mesons, baryons, and pionic loop
contributions \cite{jido}.

It is interesting to note that the $\rho NN$ coupling constant is
decreased in the global fits as compared to the purely hadronic fits,
thus deviating from the KSRF relation. The reason for this
behavior is related to the cutoff value $\Lambda_N$ of the nucleon
form factor. It is well known that the $\rho$ and nucleon contributions 
interfere in low-energy $\pi N$ elastic scattering. Since the pion
photoproduction multipoles $E^{p/n}_{0+}$ (see the discussion on pion
photoproduction in PMII \cite{pm2})
demand a reduced nucleon contribution at higher energies,
$\Lambda_N$ is decreased from $1.15$ GeV for the hadronic fits to
$0.95$ GeV for the global fits, thereby damping this contribution. At
the same time, this also affects the interference between $\rho$ and
nucleon at lower energies, leading to the necessity of simultaneously
reducing the $\rho NN$ coupling. Nevertheless, the same interference
as in the hadronic fits cannot be achieved and the low-energy tails of 
the $S_{11}$ and $P_{11}$ are not as well described; see
Fig. \ref{figppi12} above. 

As we have pointed out above, chosing the chirally symmetric $\sigma
\pi \pi$ coupling leads to consistently better results in $\pi N$
elastic scattering, even in the intermediate energy region. Our final
results always require a positive $g_{\sigma NN} g_{\sigma \pi \pi}$
value as in Pearce and Jennings \cite{pearce}\footnote{Note that
  Pearce and Jennings \cite{pearce} found a very large $\sigma$
  coupling of $g_{\sigma NN} g_{\sigma \pi \pi} \approx 1800$.}, which
means that the $\sigma$ contribution is attractive in the $S$ waves
and repulsive in the $P$ waves. 
The actual value of the $\sigma$ coupling strongly depends on the
choice of the spin-$\fth$ couplings. When the Pascalutsa couplings are
used, we always find a larger value for this coupling, thereby
indicating the need for stronger background contributions in $\pi N$
elastic scattering; see Sec. \ref{secsigmapasca} above. 

The other $t$-channel couplings ($a_0$, $K^*$, $K_0^*$),
in particular those of the scalar mesons $a_0$ and $K_0^*$, 
turn out to be large in almost all calculations. However, since
the value of $t$ is rather negative and thus the $t$-channel meson far
off-shell, the effective contribution is strongly damped by the
form factor in the corresponding processes. For $K\Lambda$ and
$K\Sigma$ production, we have included two $t$-channel processes in
the pion- ($K^*$ and $K_0^*$) and two in the photon-induced ($K^*$ and 
$K_1$) reactions. In the purely hadronic fits, the differentiation
between the $K_0^*$ and $K^*$ meson is difficult; in the global fits,
however, the freedom of the relative importance of the mesons is
reduced, since the $K^*$ contributes to both the hadro- and the
photoproduction reactions.

In the case of using the Pascalutsa spin-$\fth$ couplings, the
$t$-channel couplings differ significantly from the values of the
other calculations. This is because the missing spin-$\foh$ off-shell
contributions of the spin-$\fth$ resonances have to be compensated by
other background, i.e., $t$-channel, contributions and thus the
extracted cutoff value for the $t$-channel processes $\Lambda_t$
becomes much larger. This also means that the $t$-channel
contributions are not only important in the extreme forward region
(low $|t|$), but rather for the complete $\cos \vt$
range. Consequently, very large $t$-channel couplings for $a_0$,
$K^*$, and $K_0^*$ would not be in line with the angle-differential
observables and thus the couplings are reduced; see also the
discussion about $K\Sigma$ production in Sec. \ref{secresps}.

\subsection{Scattering lengths}

The scattering lengths and effective ranges extracted from the present
analysis are in general agreement with the values obtained by other
groups, see Table \ref{tabscattlength}.
\begin{table}[t]
  \begin{center}
    \begin{tabular}
      {l|l|r|r|c}
      \hhline{=====}
      \multicolumn{2}{c|}{} & present & Lutz {\it et al.} \cite{lutz} &
      Others \\
      \hhline{=====}
      $\pi N$ & $a^\foh$ & 0.197 & & $0.246^a$ \\
              & $r^\foh$ & 0.660 & & \\
              & $a^\fth$ & $-0.117$ & & $-0.130^a$ \\
              & $r^\fth$ & 18.33 & & \\
      \hline
      $\eta N$ & $a^\foh$ & $0.991+\mi 0.347$ & $0.43+\mi0.21$ & $0.710(30)+\mi 0.263(23)^b$ \\
               & $r^\foh$ & $-2.081-\mi 0.812$ & & \\
      \hline
      $K\Lambda$ & $a^\foh$ & $-0.154+\mi 0.084$ & $0.26+\mi 0.10$ & $-0.148+\mi 0.165^c$ \\
                 & $r^\foh$ & $-3.021+\mi 0.187$ & & \\
      \hline
      $K\Sigma$ & $a^\foh$ & $-0.270+\mi 0.172$ & $-0.15+\mi 0.09$ & $-0.363+\mi 0.112^c$ \\
                & $r^\foh$ & $-4.032+\mi 2.064$ & & \\
                & $a^\fth$ & $-0.011+\mi 0.005$ & $-0.13+\mi 0.04$ & $-0.126+\mi 0.046^c$ \\
                & $r^\fth$ & $34.79-\mi 3.561$ & & \\
      \hline
      $\omega N$ & $\bar a^\foh (J=\foh)$ & $-1.093+\mi 0.958$ & $-0.45+\mi 0.31$ & \\
                 & $\bar r^\foh (J=\foh)$ & $-0.001+\mi 7.765$ & & \\
                 & $\bar a^\foh (J=\fth)$ & $-0.228+\mi 0.621$ & $-0.43+\mi 0.15$ & \\
                 & $\bar r^\foh (J=\foh)$ & $13.31-\mi 17.11$ & & \\
                 & $\bar a^\foh$ & $-0.516+\mi 0.733$ & $-0.44+\mi 0.20$ & $1.6+\mi 0.30^d$ \\
                 & $\bar r^\foh$ & $ 8.873-\mi 8.820$ & & \\
      \hhline{=====}
    \end{tabular}
  \end{center}
  \caption{Scattering length (in fm) from the present analysis in
    comparison with other calculations. The upper index denotes the
    isospin. $^a$: Ref. \cite{SM00}. $^b$: Ref. \cite{batinic}. $^c$:
    Ref. \cite{inoue}. $^d$: Ref. \cite{klingl99}.
    \label{tabscattlength}}
\end{table}
For the vectormeson state $\omega N$ we follow the notation of Lutz
{\it et al.} \cite{lutz} for the extraction of the scattering length:
\bea
\bar a^\foh = \sfot \bar a^\foh (J=\sfoh) + \sftt \bar a^\foh (J=\sfth) 
\eea
and similarly for $\bar r^\foh$. The upper index denotes the
isospin. The $\omega N$ helicity state combinations contributing at
threshold are \cite{lutz}
\bea
|\omega N; J=\sfoh \rangle &=& |\omega N, \sfoh; J=\sfoh \rangle +
\smallfrac{1}{\sqt} |\omega N, +0; J=\sfoh \rangle \; ,
\nonumber \\
|\omega N; J=\sfth \rangle &=& |\omega N, \sfth; J=\sfth \rangle +
\smallfrac{1}{\sqth} |\omega N, \sfoh; J=\sfth \rangle +
\sqrt{\smallfrac{2}{3}} |\omega N, +0; J=\sfoh \rangle \; .
\eea
The extracted scattering lengths, however, have to be taken with care,
since the present analysis does not concentrate on the threshold
regions of the reactions, but aims on a description of a large energy
range. This can result in significant differences to well known
values, as, e.g., in the $\pi N$ elastic scattering, see the
discussion in Secs. \ref{secrespp} and \ref{secresback}. Furthermore, 
in particular in the $\omega N$ case, more polarization measurements
are needed for a reliable determination of the exact decomposition of
the production mechanism close to threshold, see Sec.
\ref{secrespo} and also the discussion on $\omega$ photoproduction in
PMII \cite{pm2}.

\subsection{Resonances}
\label{secresresparas}

In the extension of the energy range and final state space, the
inclusion of more resonances as compared to Feuster and Mosel 
\cite{feusti98,feusti99} has become necessary. We find striking
evidence for three more resonances, which are of vital importance for
a satisfying description of all 
experimental data below 2 GeV: a $P_{31}(1750)$, a $P_{13}(1900)$, and
a $P_{33}(1920)$, which are only rated by the PDG \cite{pdg} by one,
two, and three stars, respectively. Omitting one of these
resonances, the calculations result in a considerably worse total
$\chi^2$ by more than $15\%$. We can furthermore corroborate the
findings of Feuster and Mosel \cite{feusti98,feusti99} that there is
a strong need for a $D_{13}$ resonance in the energy range between 1.9 
and 2 GeV.

In the global calculations, the properties of almost all considered
resonances can be very well fixed (see Tables \ref{tabrespropsouri32} 
$-$ \ref{tabresaparas}), even the couplings of the subthreshold
resonances are practically identical for C-p-$\gamma +$ and
C-p-$\gamma -$.
\begin{table}[t]
  \begin{center}
    \begin{tabular}
      {l|c|r|r|r|r}
      \hhline{======}
      $L_{2I,2S}$ & mass & $\Gamma_{tot}$ &
      $R_{\pi N}$ & $R_{2\pi N}$ & $R_{K \Sigma}$ \\
      \hhline{======}
      $S_{31}(1620)$ & 1611 & 196 &  34.3 & $ 65.7(-)$ & $ 0.14^a$ \\
      & 1614 & 209 &  34.4 & $ 65.6(-)$ & $ 0.16^a$ \\
      & 1612 & 175 &  36.0 & $ 64.0(-)$ & $ 0.94^a$ \\
      & 1630 & 177 &  43.4 & $ 56.6(+)$ & $ 0.48^a$ \\
      \hline 
      $S_{31}(1900)^P$ & 1984 & 237 &  30.4 & $ 69.5(-)$ & $ 0.1(-)$ \\
      \hhline{======}
      $P_{31}(1750)$ & 1712 & 660 &   0.8 & $ 99.1(+)$ & $ 0.1(+)$ \\
      & 1712 & 626 &   1.0 & $ 98.9(+)$ & $ 0.1(+)$ \\
      & 1752 & 632 &   2.3 & $ 97.2(+)$ & $ 0.6(+)$ \\
      & 1975 & 676 &  19.5 & $ 79.4(+)$ & $ 1.1(-)$ \\
      \hhline{======}
      $P_{33}(1232)$ & 1228 & 106 & 100.0 & $  0.021(-)^b$ & --- \\
      & 1228 & 107 & 100.0 & $0.040(-)^b$ & --- \\
      & 1231 & 101 & 100.0 & $0.002(+)^b$ & --- \\
      & 1230 &  94 & 100.0 & $0.000(+)^b$ & --- \\
      \hline 
      $P_{33}(1600)$ & 1667 & 407 &  13.3 & $ 86.7(+)$ & $ 0.03^a$ \\
      & 1667 & 388 &  13.1 & $ 86.9(+)$ & $ 0.05^a$ \\
      & 1652 & 273 &  13.7 & $ 86.3(+)$ & $ 0.22^a$ \\
      & 1656 & 350 &  13.2 & $ 86.8(+)$ & $ 0.28^a$ \\
      \hline 
      $P_{33}(1920)$ & 2057 & 494 &  15.9 & $ 81.6(-)$ & $ 2.4(-)$ \\
      & 2058 & 557 &  15.0 & $ 83.2(-)$ & $ 1.8(-)$ \\
      & 2057 & 527 &  15.5 & $ 79.5(-)$ & $ 5.0(-)$ \\
      & 2056 & 435 &   9.1 & $ 86.8(-)$ & $ 4.1(-)$ \\
      \hhline{======}
      $D_{33}(1700)$ & 1678 & 591 &  13.9 & $ 86.1(+)$ & $ 0.75^a$ \\
      & 1679 & 621 &  14.1 & $ 85.9(+)$ & $ 0.97^a$ \\
      & 1680 & 591 &  13.6 & $ 86.4(+)$ & $ 2.09^a$ \\
      & 1674 & 678 &  14.6 & $ 85.4(+)$ & $ 3.68^a$ \\
      \hhline{======}
    \end{tabular}
  \end{center}
  \caption{Properties of  $I=\fth$ resonances considered in the
    present calculation. Mass and total width $\Gamma_{tot}$ are given 
    in MeV, the decay ratios $R$ in percent of the total width. In
    brackets, the sign of the coupling is given (all $\pi N$ couplings
    are chosen to be positiv). $^P$: Only found in calculation
    P-p-$\pi +$. $^a$: The coupling is given since the resonance is
    below threshold. $^b$: Decay ratio in
    0.1\permil. First line: C-p-$\gamma +$; second line: C-p-$\gamma -$;
    third line: C-p-$\pi +$; fourth line: P-p-$\pi +$. 
    \label{tabrespropsouri32}}
\end{table}
\begin{table}[t]
  \begin{center}
    \begin{tabular}
      {l|c|r|r|r|r|r|r|r}
      \hhline{=========}
      $L_{2I,2S}$ & mass & $\Gamma_{tot}$ &
      $R_{\pi N}$ & $R_{2\pi N}$ & $R_{\eta N}$ &
      $R_{K \Lambda}$ & $R_{K \Sigma}$ & $R_{\omega N}$ \\
      \hhline{=========}
      $S_{11}(1535)$ & 1524 & 121 & 36.6 & $ 9.8(+)$ & $53.6(+)$ & $-1.28^a$ & $ 0.83^a$ & --- \\
      & 1528 & 137 & 35.6 & $11.2(+)$ & $53.3(+)$ & $-1.62^a$ & $ 1.00^a$ & --- \\
      & 1542 & 148 & 37.7 & $11.5(+)$ & $50.8(+)$ & $ 0.02^a$ & $ 0.27^a$ & --- \\
      & 1545 & 117 & 36.6 & $ 0.9(-)$ & $62.6(+)$ & $-4.46^a$ & $ 0.26^a$ & --- \\
      \hline
      $S_{11}(1650)$ & 1664 & 131 & 67.6 & $28.3(+)$ & $ 1.6(-)$ & $ 2.4(-)$ & $-0.59^a$ & --- \\
      & 1667 & 155 & 61.8 & $34.7(+)$ & $ 0.4(-)$ & $ 3.1(-)$ & $-0.72^a$ & --- \\
      & 1671 & 158 & 65.1 & $22.7(+)$ & $ 5.1(-)$ & $ 7.1(-)$ & $-0.54^a$ & --- \\
      & 1699 & 276 & 68.2 & $14.7(-)$ & $ 3.8(+)$ & $13.3(-)$ & $-0.50^a$ & --- \\
      \hhline{=========}
      $P_{11}(1440)$ & 1512 & 628 & 57.2 & $42.8(+)$ & $ 1.69^a$ & $-2.70^a$ & $ 0.53^a$ & --- \\
      & 1522 & 709 & 57.1 & $42.9(+)$ & $ 1.79^a$ & $-6.65^a$ & $ 6.78^a$ & --- \\
      & 1490 & 463 & 61.5 & $38.5(+)$ & $ 3.27^a$ & $ 3.43^a$ & $-1.01^a$ & --- \\
      & 1515 & 639 & 60.6 & $39.4(+)$ & $ 4.17^a$ & $ 1.97^a$ & $ 3.64^a$ & --- \\
      \hline
      $P_{11}(1710)$ & 1749 & 445 &  7.4 & $38.5(-)$ & $24.9(+)$ & $ 3.4(+)$ & $12.6(-)$ & 13.4 \\
      & 1755 & 327 & 21.7 & $12.1(-)$ & $47.0(+)$ & $ 7.4(+)$ & $ 0.0(-)$ & 11.7 \\
      & 1770 & 430 &  2.0 & $42.7(+)$ & $31.6(-)$ & $ 0.9(+)$ & $ 6.3(-)$ & 16.4 \\
      & 1701 & 348 &  8.5 & $25.7(-)$ & $38.3(+)$ & $26.3(-)$ & $ 1.3(-)$ & --- \\
      \hhline{=========}
      $P_{13}(1720)$ & 1696 & 165 & 19.1 & $69.0(+)$ & $ 0.1(+)$ & $11.8(-)$ & $ 0.0(-)$ & --- \\
      & 1715 & 310 & 14.8 & $79.1(+)$ & $ 0.4(-)$ & $ 5.6(-)$ & $ 0.1(-)$ & --- \\
      & 1724 & 295 & 15.4 & $65.2(+)$ & $ 1.2(+)$ & $ 9.9(-)$ & $ 7.5(-)$ &  0.7 \\
      & 1700 & 148 & 14.2 & $83.1(+)$ & $ 0.0(+)$ & $ 1.7(+)$ & $ 1.0(+)$ & --- \\
      \hline
      $P_{13}(1900)$ & 2003 & 581 & 14.6 & $42.7(-)$ & $ 9.4(-)$ & $ 0.1(-)$ & $ 2.0(-)$ & 31.2 \\
      & 1898 & 664 & 17.9 & $14.7(+)$ & $19.2(-)$ & $ 0.0(+)$ & $ 0.0(-)$ & 48.1 \\
      & 1962 & 683 & 19.1 & $58.2(-)$ & $11.9(+)$ & $ 1.9(-)$ & $ 0.8(+)$ &  8.1 \\
      & 1963 & 694 & 15.7 & $58.2(-)$ & $ 3.0(+)$ & $ 0.1(+)$ & $ 0.0(+)$ & 22.9 \\
      \hhline{=========}
      $D_{13}(1520)$ & 1509 &  99 & 55.8 & $44.2(-)$ & $ 2.0^b(+)$ & $-0.09^a$ & $ 1.13^a$ & --- \\
      & 1510 & 102 & 55.5 & $44.5(-)$ & $ 2.7^b(+)$ & $-0.35^a$ & $ 0.84^a$ & --- \\
      & 1512 &  95 & 58.7 & $41.3(-)$ & $3.1^b(+)$ & $ 0.44^a$ & $ 1.20^a$ & --- \\
      & 1509 &  91 & 60.1 & $39.9(-)$ & $ 2.2^b(+)$ & $ 0.86^a$ & $-3.23^a$ & --- \\
      \hline
      $D_{13}(1700)^P$ & 1745 &  55 &  1.6 & $43.4(+)$ & $ 1.7(+)$ & $ 6.7(-)$ & $ 1.2(-)$ & 45.3 \\
      \hline
      $D_{13}(1950)$ & 1946 & 865 & 12.9 & $67.2(+)$ & $ 5.4(+)$ & $ 0.0(-)$ & $ 0.3(+)$ & 14.1 \\
      & 1946 & 852 & 10.7 & $51.3(+)$ & $ 8.6(+)$ & $ 0.4(-)$ & $ 1.1(-)$ & 27.9 \\
      & 1946 & 885 & 16.2 & $49.1(+)$ & $ 2.2(-)$ & $ 1.2(+)$ & $ 1.9(+)$ & 29.4 \\
      & 1943 & 573 & 13.3 & $50.8(+)$ & $ 0.0(-)$ & $ 2.2(-)$ & $ 0.7(+)$ & 32.9 \\
      \hhline{=========}
    \end{tabular}
  \end{center}
  \caption{Properties of $I=\foh$ resonances considered in the
    calculation. Notation as in Table \ref{tabrespropsouri32}.
    \label{tabrespropsouri12}}
\end{table}
\begin{table}[t]
  \begin{center}
    \begin{tabular}
      {l|r|r|r|r|r|r}
      \hhline{=======}
      $L_{2I,2S}$ & mass & $\Gamma_{tot}$ &
      $R_{\omega N}$ & $R_{\omega N}^0$ & $R_{\omega N}^\foh$ & 
      $R_{\omega N}^\fth$ \\
      \hhline{=======}
      $S_{11}(1535)$ & 1524 & 121 & ---  & $  3.64^{a1}$ & $  6.10^{a2}$ & --- \\
      & 1528 & 137 & ---  & $  1.77^{a1}$ & $  5.66^{a2}$ & --- \\
      & 1542 & 148 & ---  & $ -4.51^{a1}$ & $ -2.61^{a2}$ & --- \\
      & 1545 & 117 & ---  & $  2.50^{a1}$ & $  4.99^{a2}$ & --- \\
      \hline        
      $S_{11}(1650)$ & 1664 & 131 & ---  & $  4.75^{a1}$ & $ -1.78^{a2}$ & --- \\
      & 1667 & 155 & ---  & $  3.24^{a1}$ & $  3.42^{a2}$ & --- \\
      & 1671 & 158 & ---  & $ -0.15^{a1}$ & $  0.00^{a2}$ & --- \\
      & 1699 & 276 & ---  & $  1.84^{a1}$ & $  5.35^{a2}$ & --- \\
      \hhline{=======}
      $P_{11}(1440)$ & 1512 & 628 & ---  & $-18.73^{a1}$ & $ 10.14^{a2}$ & --- \\
      & 1522 & 709 & ---  & $ 15.56^{a1}$ & $ 10.82^{a2}$ & --- \\
      & 1490 & 463 & ---  & $ -1.55^{a1}$ & $  2.09^{a2}$ & --- \\
      & 1515 & 639 & ---  & $ -6.30^{a1}$ & $  3.95^{a2}$ & --- \\
      \hline        
      $P_{11}(1710)$ & 1749 & 445 &  13.4 & $  0.0(-)$ & $ 13.3(-)$ & --- \\
      & 1755 & 327 &  11.7 & $  0.0(-)$ & $ 11.7(-)$ & --- \\
      & 1770 & 430 &  16.4 & $ 10.1(-)$ & $ 6.3(+)$ & --- \\
      & 1701 & 348 &   --- & $  5.2^{a1}$ & $ -5.3^{a2}$ & --- \\
      \hhline{=======}
      $P_{13}(1720)$ & 1696 & 165 &  --- & $-14.0^{a1}$ & $-21.3^{a2}$ & $  5.3^{a3}$ \\
      & 1715 & 310 &   --- & $ -9.4^{a1}$ & $-15.9^{a2}$ & $ -7.5^{a3}$ \\
      & 1724 & 295 &   0.7 & $  1.5(+)^b$ & $  7.8(+)^b$ & $  62.1(+)^b$ \\
      & 1700 & 148 &   0.0 & $  8.8^{a1}$ & $ -2.8^{a2}$ & $ -2.8^{a3}$ \\
      \hline        
      $P_{13}(1900)$ & 2003 & 581 & 31.2 & $  0.0(-)$ & $  7.8(+)$ & $23.4(+)$ \\
      & 1898 & 664 &  48.1 & $ 16.7(-)$ & $ 19.3(+)$ & $ 12.1(+)$ \\
      & 1962 & 683 &   8.1 & $  0.9(+)$ & $  0.0(-)$ & $  7.2(+)$ \\
      & 1963 & 694 &  22.9 & $  5.3(+)$ & $  0.0(+)$ & $ 17.6(+)$ \\
      \hhline{=======}
      $D_{13}(1520)$ & 1509 &  99 & ---  & $-21.33^{a1}$ & $ -7.12^{a2}$ & $ -7.71^{a3}$ \\
      & 1510 & 102 & ---  & $-11.68^{a1}$ & $ 14.67^{a2}$ & $ 16.32^{a3}$ \\
      & 1512 &  95 & ---  & $-13.07^{a1}$ & $ 21.37^{a2}$ & $ -3.91^{a3}$ \\
      & 1509 &  91 & ---  & $ -3.98^{a1}$ & $ -5.36^{a2}$ & $  7.04^{a3}$ \\
      \hline        
      $D_{13}(1700)^P$ & 1745 &  55 &  45.3 & $ 14.2(-)$ & $  7.5(-)$ & $ 23.6(-)$ \\
      \hline        
      $D_{13}(1950)$ & 1946 & 865 & 14.1 & $13.0(+)$ & $ 1.1(-)$ & $  0.0(+)$ \\
      & 1946 & 852 & 27.9 & $  7.0(+)$ & $ 14.7(+)$ & $  6.2(+)$ \\
      & 1946 & 885 & 29.4 & $  9.8(+)$ & $  2.1(+)$ & $ 17.5(+)$ \\
      & 1943 & 573 &  32.9 & $ 12.1(+)$ & $  0.1(+)$ & $ 20.7(+)$ \\
      \hhline{=======}
    \end{tabular}
  \end{center}
  \caption{$\omega N$ helicity decay ratios of $I=\foh$
    resonances. The total widths are given in MeV, all ratios in
    percent. $^{a1}$ ($^{a2}$, 
    $^{a3}$) : The coupling $g_1$ ($g_2$, $g_3$) is given. $^b$: The
    ratio is given in 0.1\permil . $^P$: Only found in calculation
    P-p-$\pi +$. First line: C-p-$\gamma +$; second line: C-p-$\gamma -$;
    third line: C-p-$\pi +$; fourth line: P-p-$\pi +$.
    \label{tabresomprops}}
\end{table}
\begin{table}[t]
  \begin{center}
    \begin{tabular}
      {l|r|r|r|r|r|r|r|r}
      \hhline{=========}
      $L_{2I,2S}$ & $a_{\pi N}$ & $a_{\zeta N}$ & $a_{\eta N}$ & 
      $a_{K\Lambda}$ & $a_{K\Sigma}$ & 
      ${a_{\omega N}}_1$ & ${a_{\omega N}}_2$ & ${a_{\omega N}}_3$ \\ 
      \hhline{=========}
      $P_{13}(1720)$  &  $-0.658$ &   0.832 &  $-4.000$ &   0.573 &  $-0.473$ &   0.679 &  $-3.072$ &   3.495 \\
      & $-0.005$ &   0.768 & $-3.999$ &   0.018  & $-3.998$ &  1.758 & $-4.000$ & $ 2.648$ \\
      & $ 0.183$ &   0.587 &   1.943  & $-0.625$ & $-2.728$ &  1.108 & $-3.499$ & $-1.858$ \\
      & $ 0.258$ &   0.726 & $-1.953$ & $-0.053$ & --- & --- & --- & --- \\
      \hline        
      $P_{13}(1900)$ &  $-1.249$ &  $-0.457$ &  $-0.003$ &   0.852 &  $-3.999$ &   2.920 &   0.897 &  $-3.874$ \\
      &  2.123  & $-0.362$ & $-1.628$ & $-3.828$ & $-4.000$ & $-0.945$ & $-3.647$ & $-0.180$ \\
      & $0.205$ & $ 0.437$ & $-0.739$ &   3.410  & $-3.687$ &   2.195  & $ 0.092$ & $ 1.454$ \\
      & NC & --- & --- & --- & --- & --- & --- & --- \\
      \hhline{=========}
      $D_{13}(1520)$ &   0.872 &  $-0.249$ &   0.366 &   0.794 &   0.501 &  $-2.442$ &  $-4.000$ &  $-4.000$ \\
      &  0.871 &  $-0.407$ &   0.744  &   1.164 &   0.318 &   0.774  & $-3.998$ &   2.562 \\
      &  0.861  & $-0.351$ & $1.796$ &   0.856 &   2.692 & $ 0.344$ & $-0.445$ & $-1.050$ \\
      &  0.819  & $-0.158$ & 1.146 &  & --- & --- & --- & --- \\
      \hline        
      $D_{13}(1950)$ &   0.789 &   0.588 &  0.353 &   1.661 &   2.091 &  $-0.685$ &  $-0.247$ &  $-2.000$ \\
      &  0.663 &   0.365 &   1.025  &   0.503 &   0.215 & $-0.153$ & $-3.986$ &  0.284 \\
      &  0.966 & $ 0.668$ & $0.211$ &   1.019 &   0.663 & $-0.016$ & $-0.976$ & $-1.152$ \\
      &  0.924 &   1.387 &   1.016  &   1.116 & --- & --- & --- & --- \\
      \hhline{=========}
      $P_{33}(1232)$ &   0.222 &  $-1.156$ & --- & --- & --- & --- & --- & --- \\
      &   0.211 &  $-1.006$ & --- & --- & --- & --- & --- & --- \\
      &   0.233 &    4.000  & --- & --- & --- & --- & --- & --- \\
      &   0.148 & & --- & --- & --- & --- & --- & --- \\
      \hline        
      $P_{33}(1600)$ &   1.798 &   0.363 & --- & --- &  $-3.047$ & --- & --- & --- \\
      &   1.937 &   0.363 & --- & --- &  $-4.000$ & --- & --- & --- \\
      &   1.266 &   0.291 & --- & --- &  $-0.783$ & --- & --- & --- \\
      &   0.400 &  $-0.253$ & --- & --- & --- & --- & --- & --- \\
      \hline        
      $P_{33}(1920)$ &  $-2.827$ &   1.244 & --- & --- &  $-1.762$ & --- & --- & --- \\
      &  $-2.492$ &   1.111 & --- & --- &  $-1.683$ & --- & --- & --- \\
      &  $-3.137$ &   1.264 & --- & --- &  $-1.145$ & --- & --- & --- \\
      &  NC  & --- & --- & --- & --- & --- & --- & --- \\
      \hhline{=========}
      $D_{33}(1700)$ &  $-0.282$ &   0.414 & --- & --- &  $-0.156$ & --- & --- & --- \\
      &  $-0.288$ &  0.413  & --- & --- &   0.001 & --- & --- & --- \\
      &  $-0.220$ &  0.425 & --- & --- &   0.473 & --- & --- & --- \\
      &  $-0.181$ &  0.867 & --- & --- & --- & --- & --- & --- \\
      \hhline{=========}
    \end{tabular}
  \end{center}
  \caption{Off-shell parameters $a$ of the spin-$\fth$
    resonances. First line: C-p-$\gamma +$; second line: C-p-$\gamma -$;
    third line: C-p-$\pi +$; fourth line: SM95-pt-3 of Ref.
    \cite{feusti99}. NC: not considered (energy range ended at 
    1.9 GeV). 
    \label{tabresaparas}}
\end{table}
The only exceptions are the $P_{11}(1710)$, $P_{13}(1900)$, and the
exact decomposition of the $\omega N$ strength into the $\omega N$
helicities. Note that the properties of the $P_{11}(1710)$ also
differ largely when comparing the references given in the PDG review
\cite{pdg}. Moreover, Arndt {\it et al.} \cite{arndt02} had similar problems
with fixing the $P_{11}(1710)$ properties. However, in contrast to Ref.
\cite{arndt02}, in the present calculation the properties of the
$S_{11}(1535)$ can be well fixed due to the simultaneous inclusion of
$\eta N$ production data. 

In the $K$-matrix formulation the resonance
properties are identified with the implemented parameters
\cite{vrana}, thus the given decay widths and branching ratios
are calculated at the resonance mass ($\sqrt s = m_R$). Since the
widths are energy dependent (cf. Appendices \ref{appres12lagr} and
\ref{appres32lagr}) and the $RN\phi$ vertices are modified by
form factors, the total decay widths do \textit{not} necessarily
respresent the full width at half maximum (FWHM), which can, e.g., be
observed in the $\pi N$ elastic partial waves. 

Just as the extracted resonance masses and couplings, the spin-$\fth$
off-shell parameters $a$, given in Table \ref{tabresaparas}, are also
very similar in the two 
global calculations with the exception of the $\omega N$ values. Large 
differences only occur when the coupling of the resonance to the
final state is also largely changed, thus keeping the product $g \cdot 
a$ in the same range. Note that our values are also very close to the 
preferred global fit SM95-pt-3 of Ref. \cite{feusti99} and that the observed
discrepancies can be explained by the additional resonances considered
in the present calculation.

In Tables \ref{tabresourpdgvranai32} and \ref{tabresourpdgvranai12} 
\begin{table}[t]
  \begin{center}
    \begin{tabular}
      {l|l|l|l|l|r}
      \hhline{======}
      $L_{2I,2S}$ & mass & $\Gamma_{tot}$ &
      $R_{\pi N}$ & $R_{2\pi N}$ & $R_{K \Sigma}$ \\
      \hhline{======}
      $S_{31}(1620)$ & 1612(2) & 202(7) & 34(1) & 66(1) & \\
                     & 1620 & 150 & 25(5) & 75(5) & \\
                     & 1579 & 153 & 21 & 79 & \\
                     & 1617(15) & 143(42) & 45(5) & \\
      \hline 
      $S_{31}(1900)^P$ & 1984 & 237 & 30 & 70 & 0.1 \\
                     & 1900 & 200 & 20(10) & & \\
                     & NC & & & & \\
                     & 1802(87) & 48(45) & 33(10) & \\
      \hhline{======}
      $P_{31}(1750)$ & 1712(1) & 643(17) & 1(1) & 99(1) & 0.1(0.1) \\
                     & 1750 & 300 & 8 & & \\
                     & NF & & & & \\
                     & 1721(61) & 70(50) & 6(9) & \\
      \hline 
      $P_{31}(1910)^P$ & 1975 & 676 & 19 & 79 & 1.1 \\
                     & 1910 & 250 & 23(7) & & \\
                     & NC & & & & \\
                     & 1995(12) & 713(465) & 29(21) & \\
      \hhline{======}
      $P_{33}(1232)$ & 1228(1) & 106(1) & $\;\;$100(0) & 0.03(0.01)$^a$ & \\
                     & 1232 & 120 & $>99$ & 0 & \\
                     & 1228 & 110 & $\;\;$100 & & \\
                     & 1234(5) & 112(18) & $\;\;$100(1) & \\
      \hline 
      $P_{33}(1600)$ & 1667(1) & 397(10) & 13(1) & 87(1) & \\
                     & 1600 & 350 & 18(7) & 82(8) & \\
                     & 1721 & 485 & 15 & 85 & \\
                     & 1687(44) & 493(75) & 28(5) & & \\
      \hline 
      $P_{33}(1920)$ & 2057(1) & 525(32) & 15(1) & 82(2) & 2.1(0.3) \\
                     & 1920 & 200 & 13(7) & & \\
                     & NC & & & & \\
                     & 1889(100) & 123(53) & $\;\,$5(4) & \\
      \hhline{======}
      $D_{33}(1700)$ & 1678(1) & 606(15) & 14(1) & 86(1) & \\
                     & 1700 & 300 & 15(5) & 85(5) & \\
                     & 1677 & 387 & 14 & 86 & \\
                     & 1732(23) & 119(70) & $\;\,$5(1) & \\
      \hhline{======}
    \end{tabular}
  \end{center}
  \caption{Estimated properties of $I=\fth$ resonances from the present 
    calculation (first line), see text, in comparison with the
    values from Ref. \cite{pdg} (second line), Ref. \cite{feusti99}
    (third line), and Ref. \cite{vrana} (fourth line). In brackets,
    the estimated errors are given. The mass and total width are given
    in MeV, the decay ratios in percent.
    NC: not considered (energy range ended at
    1.9 GeV). $^a$: The decay ratio is given in 0.1\permil. $^P$:
    Calculation P-p-$\pi +$, see text and Tables
    \ref{tabrespropsouri32} and \ref{tabrespropsouri12} above.
    \label{tabresourpdgvranai32}}
\end{table}
\begin{table}[t]
  \begin{center}
    \begin{tabular}
      {l|l|l|l|l|l|l|c|c}
      \hhline{=========}
      $L_{2I,2S}$ & mass & $\Gamma_{tot}$ &
      $R_{\pi N}$ & $R_{2\pi N}$ & $R_{\eta N}$ &
      $R_{K \Lambda}$ & $R_{K \Sigma}$ & $R_{\omega N}$ \\
      \hhline{=========}
      $S_{11}(1535)$ & 1526(2) & 129(8) & 36(1) & 10(2) & 53(1) & & & \\
                     & 1535 & 150 & 45(10) &  $\;\,$6(5) & 43(12) & & & \\
                     & 1549 & 215 & 31 & $\;\,$6 & 63 & & & \\
                     & 1542(3) & 112(19) & 35(8) & & & & & \\
      \hline 
      $S_{11}(1650)$ & 1665(2) & 138(7) & 65(4) & 31(4) & 1.0(0.6) & 2.7(0.4) & & \\
                     & 1650 & 150 & 72(17) & 15(5) &  6(3) & 7(4) & & \\
                     & 1684 & 194 & 73 & 22 & 1 & 5 & & \\
                     & 1689(12) & 202(40) & 74(2) & & & & & \\
      \hhline{=========}
      $P_{11}(1440)$ & 1518(5) & 668(41) & 57(1) & 43(1) & & & & \\
                     & 1440 & 350 & 65(5) & 35(5) & & & & \\
                     & 1479 & 513 & 62 & 38 & & & & \\
                     & 1479(80) & 490(120) & 72(5) & & & & & \\
      \hline 
      $P_{11}(1710)$ & 1752(3) & 386(59) & $\;\,$14(8) & 26(14) & 36(11) & $\;\,$5.4(2) & 7(7) & 13(2) \\
                     & 1710 & 100 & 15(5) & 65(25) & & 15(10) & & \\
                     & 1709 & 284 & $\;\,$0 & 51 & 32 & 17 & & \\
                     & 1699(65) & 143(100) & 27(13) & & & & & \\
      \hhline{=========}
      $P_{13}(1720)$ & 1705(10) & 237(73) & 17(2) & $\;\;\;\:$74(5) & 0.2(0.2) & 9(3) & 0.0(0.1) & \\
                     & 1720 & 150 & 15(5) & $>70$ & & $\;\,$8(7) & & \\
                     & 1801 & 637 & 21 & $\;\;\;\:$75 & 4 & $\;\,$1 & & \\
                     & 1716(112) & 121(39) & $\;\,$5(5) & & & & & \\
      \hline 
      $P_{13}(1900)$ & 1951(53) & 622(42) & 16(2) & 29(15) & 14(5) & 0.1(0.1) & 1.0(1.0) & 39(9) \\
                     & 1900 & 500 & 26 & 45 & & & \\
                     & NC & & & & & & & \\
                     & NF & & & & & & & \\
      \hhline{=========}
      $D_{13}(1520)$ & 1509(1) & $\;\,$100(2) & 56(1) & 44(1) & 2.3(0.4)$^a$ & & & \\
                     & 1520 & 120 & 55(5) & 45(5) & & & & \\
                     & 1512 & $\;\,$93 & 56 & 44 & 4.3$^a$ & & & \\
                     & 1518(3) & 124(4) & 63(2) & & & & & \\
      \hline 
      $D_{13}(1700)^P$ & 1745 & $\;\,$55 & $\;\,$2 & 43 & 1.7 & $\;\;\;\:$7 & 1.2 & 45 \\
                     & 1700 & 100 & 10(5) & 90(5) & & $<3$ & & \\
                     & NF & & & & & & & \\
                     & 1736(33) & 175(133) & $\;\,$4(2) & & & & & \\
      \hline 
      $D_{13}(1950)$ & 1946(1) & $\;\,$859(7) & 12(2) & 59(8) & $\;\,$7(2) & 0.2(0.2) & 0.7(0.4) & 21(7) \\
                     & 2080 & & & & & & & \\
                     & 1940 & $\;\,$412 & 10 & 75 & 14 & 0 & & \\
                     & 2003(18) & 1070(858) & $\;\,$4(2) & & & & & \\
      \hhline{=========}
    \end{tabular}
  \end{center}
  \caption{Comparison of $I=\foh$ resonance properties. Notation as in
    Table \ref{tabresourpdgvranai32}.
    \label{tabresourpdgvranai12}}
\end{table}
we give a direct comparison of the extracted resonance properties of
the present model with the values given by the PDG
\cite{pdg}, extracted by Feuster and Mosel \cite{feusti99}, and
extracted by the $\pi N \ra \pi N /2\pi N$ analysis of Vrana {\it et
al.} \cite{vrana}. Note that in some cases [e.g., $P_{11}(1710)$ mass
and width, $D_{13}(1950)$ and $P_{33}(1920)$ mass, etc.] noticeable
differences to the estimated values of the particle data group
\cite{pdg} are found. The estimated values and errors from the present
model give the average and rms deviation of the values obtained in the 
global calculations C-p-$\gamma +$ and C-p-$\gamma -$, since only
in these two calculations the complete data base including pion- and
photon-induced data has been used. This also means that the given
errors are only rough guidelines and can, sometimes, even be
misleading if both fits are unsatisfactory in a given energy region,
see in particular the discussion below on the properties of the
$P_{11}(1440)$ and $P_{33}(1920)$ resonances. Furthermore, we want to
point out that the results of the global calculation C-p-$\gamma +$,
given in Tables \ref{tabrespropsouri32} -- \ref{tabresaparas},
are to be preferred, since this calculation gives a better description 
in the pion-induced sector (see Table \ref{tabchisquares}), while in
the photon-induced reactions the quality of the two global
calculations are identical, see PMII \cite{pm2}.

In the following, the extracted resonance properties are discussed in
detail for each partial wave. We refer in particular to
Figs. \ref{figppi12} $-$ \ref{figppini12} in the discussion.

\subsubsection{Isospin-$\foh$ resonances}

\noindent $\mathbf{S_{11}:}$\\
For the two four-star resonances in this partial wave [$S_{11}(1535)$
and $S_{11}(1650)$], the parameters can be well fixed in the present
model; the differences between the global and purely hadronic
fit parameters are not very large. The exact properties of
$S_{11}(1535)$ can, however, only be extracted in the simultaneous
analysis of pion- and photon-induced data, which has already been
pointed out by Feuster and Mosel \cite{feusti99}.
The second $S_{11}$ resonance has an almost negligible $\eta N$ width,
but nevertheless interferes destructively in the $\pi N \ra \eta N$
reaction with the $S_{11}(1535)$, see Sec. \ref{secrespe}.
In the purely hadronic fits the extracted properties of the
$S_{11}(1535)$ and $S_{11}(1650)$ are very similar to the values of
Vrana {\it et al.} \cite{vrana} and Batini\'c {\it et al.} \cite{batinic}, who
found the masses $1.542$ ($1.543$) and $1.689$ ($1.668$) GeV and the
widths $112$ ($155$) and 202 (209) MeV. The inclusion of the
photoproduction data, however, requires the lowering of the
$S_{11}(1535)$ mass and total width, in particular for a description
of the $E_{0+}^p$ multipole, see the discussion on pion
photoproduction in PMII \cite{pm2}. Note that
the decay ratios of the $S_{11}(1535)$ are almost identical in the
global and hadronic calculations. Furthermore, it is worth mentioning, 
that the $K\Lambda$ decay ratio of the $S_{11}(1650)$ is considerably
lowered as compared to Feuster and Mosel \cite{feusti99}. This is a
consequence of the fact that in the best global calculation
C-p-$\gamma +$, the $K\Lambda$ production is now explained by a
dominating $P_{13}$ mechanism, while the $S_{11}(1650)$ is only
important very close to threshold, see Sec. \ref{secrespl} above.\\ 
Since in the resonance analyses of Vrana {\it et al.} \cite{vrana},
Batini\'c {\it et al.} \cite{batinic}, and Manley and Saleski
\cite{manley92} a third $S_{11}$ has been found below 2 GeV (i.e., at
$1.82$, $1.705$, and $1.93$ GeV, respectively), we have also checked whether
the inclusion of a third $S_{11}$ below 2 GeV would improve the
results. However, the fit has always decreased all partial-decay
widths of such a resonance to zero. Hence we do not find any hint
for a third $S_{11}$ resonance below 2 GeV in our analysis.
\par

\noindent $\mathbf{P_{11}:}$\\
The mass and width of the Roper $P_{11}(1440)$ resonance turn out to
be rather large in the global fits in comparison with other analyses
(note, however, the range of the width given by Vrana {\it et
al.} \cite{vrana}: $490\pm 120$ MeV, and that
Cutcosky and Wang \cite{cutkosky90} found in analyzing the $\pi N \ra
\pi N$ and $\pi N \ra 2\pi N$ data for the $P_{11}$ partial-wave 
values for the width of 661 and 545 MeV, depending on the $\pi N
\ra \pi N$ single energy partial-wave analysis used). The reason for
these large values is that the $P_{11}(1440)$ parameters are
extremely sensitive to background contributions, i.e., to the
interference pattern between nucleon and $\rho$. Since in the global
fit, the nucleon cutoff has been reduced for a better description of
the $E_{0+}^{p/n}$ photoproduction multipoles (see the discussion on
pion photoproduction in PMII \cite{pm2}),
the description of the $P_{11}$ wave (and also
$S_{11}$) at low energies has become worse. The fit has tried to
compensate for this effect by increasing the $P_{11}(1440)$ mass and
width, which can hence not be reliably extracted in the present
calculation. This problem might also be related to the fact that
there are hints that the $P_{11}(1440)$ resonance is a quasibound
$\sigma N$ state \cite{krehl}, which cannot be generated in the
present $K$-matrix approach. The decay ratios into $\pi N$ and $2\pi
N$, however, turn out to be reliably determined in all calculations.\\ 
Once the photoproduction data are included, the mass of the largely
inelastic $P_{11}(1710)$ resonance is fixed at around 1.75 GeV due to
its important contributions to $\eta N$ and $\omega N$; a mass, which
is 40 MeV above the PDG \cite{pdg} estimate. In all calculations, it
turns out to have a decay 
ratio of more than 10\% to $\omega N$ and more than 25\% to $\eta
N$. The latter result has also been found by Batini\'c {\it et
al.} \cite{batinic}. The $K\Sigma$ decay ratio seems not to be well
determined, since the large value of 12.6\% of C-p-$\gamma +$ is not
confirmed in the calculation C-p-$\gamma -$. However, also in
C-p-$\gamma -$ a large $P_{11}$ contribution to $K\Sigma$ is found,
which can be seen by the increase of the $K\Sigma$ coupling constant
of the $P_{11}(1440)$. Since the switch of the sign of $g_{\omega
  \rho \pi}$ leads to a change of sign of $\kappa_{\omega NN}$ (see
Table \ref{tabborncplgs}) due to interference effects in $\omega N$
production, also the behavior of the $P_{11}$ $K\Sigma$ wave, which 
reacts sensitive on $\omega N$ rescattering, has to be altered. However, 
since the simultaneous description of photon-
and pion-induced data is much better in the calculation C-p-$\gamma +$ 
(see Table \ref{tabchisquares}), the large $P_{11}(1710)$ $K\Sigma$
decay ratio seems to be favored by the experimental data. In contrast
to Feuster and Mosel \cite{feusti99} and the PDG \cite{pdg}, we find a
reduced $K\Lambda$ decay ratio of the $P_{11}(1710)$, which is due to
the shift of this strength to the $P_{13}$ sector. Note that the
increasing $\pi N$ inelasticity of the $P_{11}$ wave above 1.6 GeV
(see Fig. \ref{figppini12}) is caused by the $\eta N$ channel.\\
Manley and Saleski \cite{manley92} have found a third $P_{11}$ around
$1.88$ GeV, while Vrana {\it et al.} \cite{vrana} have identified such a
resonance only around $2.08$ GeV, but with a huge width of more than
$1$ GeV, thus also having a large influence on this partial wave below
$2$ GeV. Therefore we have checked the contribution of an additional
$P_{11}$ around $1.9$ GeV, but just as in the $S_{11}$ wave, its
contribution is always decreased to zero in the fit, and we do not
find any indication for a missing $P_{11}$ contribution below 2 GeV.
\par

\noindent $\mathbf{P_{13}:}$ \\
In all calculations, the mass of the first $P_{13}$ is well fixed
between $1.695$ and $1.725$ GeV. We find important contributions of
this resonance to $K\Lambda$ and also $\omega N$; in the latter case
although the resonance position is below threshold. In comparison to
Feuster and Mosel \cite{feusti99} the $P_{13}(1720)$ plays a less
important role in $\eta N$ (which is mainly due to the inclusion of a
second $P_{13}$, see below), but turns out to be much more important
in $K\Lambda$ production.\\
Guided by the observation of Feuster and Mosel, that there are
contributions missing in this partial wave for higher energies
($\sqrt s >1.8$ GeV), we have included apart from the well established 
$P_{13}(1720)$ the PDG two-star $P_{13}(1900)$ resonance in the
calculation. Although the mass of the second resonance cannot be well 
fixed in the present calculation ($1.9 \leq m_R \leq 2$ GeV), it turns
out that this second resonance gives very important contributions in
all pion-induced reactions -- in particular the $\eta N$, $K\Lambda$,
and $\omega N$ production --, and to some minor degree also in the
photoproduction 
reactions. The inclusion of this second $P_{13}$ also strongly
influences the properties of the $P_{13}(1720)$. As compared to Ref.
\cite{feusti99}, the $P_{13}(1720)$ $\eta N$ decay ratio and the mass
are reduced. Note that the $P_{13}(1720)$ mass now turns out to
be in the PDG region, in contrast to the value found in Ref. 
\cite{feusti99}. In the higher energy region ($\sqrt s > 1.8$ 
GeV), a reasonable fit to the various reactions is virtually
impossible without including a second $P_{13}$ resonance. Especially
in the $\omega N$ production, the resulting $\chi^2$ turns out to be
at least two times worse when such a resonance is excluded. It is
interesting to note that Manley and Saleski \cite{manley92} have also
found a second $P_{13}$ resonance at $1.88$ GeV with a large width of
about 500 MeV, a third of which has been attributed to the (effective)
$\omega N$ channel.\\
As discussed in Sec. \ref{secresp2}, we also find indications for 
missing flux in this partial wave, i.e., contributions of a final state 
which is not included in the present model (e.g., a $3\pi N$ state).
\par

\noindent $\mathbf{D_{13}:}$ \\
In this partial wave, we find discrepancies in the description of the
lower tail of the $D_{13}(1520)$ resonance. The asymmetric behavior
around the $D_{13}(1520)$ partial wave cannot be described within our
model, neither in elastic $\pi N$ scattering, see
Fig. \ref{figppi12}, nor in $\pi N \ra 2 \pi N$,  see
Fig. \ref{figp2both} (nor in the $E_{2-}$ and $M_{2-}$ proton and
neutron multipoles, see the discussion on pion photoproduction in
PMII \cite{pm2}). Even after allowing different cutoff values in the
$\pi N$ and the $2\pi N$ channel or using a different cutoff shape,
i.e., a cutoff $F_t(q^2)$ (\ref{formfact}), 
for this resonance, the slope of the partial wave below the
$D_{13}(1520)$ resonance position cannot be reproduced in either
channel. From the inelasticity and the $2\pi N$ production (see
Figs. \ref{figp2both} and \ref{figppini12} above) one deduces, that
this might be due to the description of the $2\pi N$ channel by an
effective $\zeta$ meson with a fixed mass. Both the inelastic and the
$2\pi N$ production cross sections rise steeper than in the present
calculation. A more physical $2\pi N$ description by including $\pi 
\Delta$ and $\rho N$ might change this behavior because of the
spectral functions of the $\Delta$ and the $\rho$. Furthermore, in the
$J^P = \fth^-$ wave, the $\rho N$ and $\pi \Delta$ states can be
produced in an $S$ wave, leading to a stronger rise of the $2\pi N$
production cross section, while our $\zeta$ meson can only be produced 
in a $P$ wave for $J^P = \fth^-$. \\
This is confirmed by the analyses of 
Manley and Saleski \cite{manley92} and Vrana {\it et al.} \cite{vrana},
since both groups extracted a dominant $2\pi N$ $S$ wave decay of the
$D_{13}(1520)$ into $\rho N$ and $\pi \Delta$. It is also interesting
to note that the rise of the $2\pi N$ partial-wave cross section in
the $P_{33}$ partial wave (see Fig. \ref{figp2both}), 
where $\rho N$ and $\pi \Delta$ cannot be produced in an $S$ wave, is
well described in the present model. Since we have not yet included 
these effects in the calculation, an increase of the errors of the 
$D_{13}$ $2\pi N$ partial-wave cross section by 1 mb up to 1.46
GeV is introduced to prevent the calculation from putting too much
weight into this shortcoming of the present model. Upcoming
investigations will reveal whether the inclusion of more realistic
two-pion nucleon final states, which allow for the correct 
partial-wave behavior and account for the spectral functions of the
two-body states will resolve this problem.\\ 
Furthermore, we confirm the finding of
Refs. \cite{feusti98,feusti99,arndt95} that there is no strong 
evidence -- if at all -- for a resonance in this partial wave 
between 1.7 and 1.9 GeV, see below. Moreover, we corroborate the
importance of a $D_{13}$ resonance between 1.9 and 2 GeV as in Refs. 
\cite{feusti98,feusti99}, especially in $\eta N$ and $\omega N$
production at higher energies; although the $\eta N$ decay ratio is
found to be small as compared to Ref. \cite{feusti99}. Due to rescattering,
this resonance also gives large background contributions at higher
energies in the 
$\pi N$ elastic amplitude. It is also interesting to note that when
only the pion-induced data are considered, the importance of this
resonance is even stronger in the $\omega N$ channel and becomes also
visible in $K\Sigma$ production. We have checked this finding by
also performing fits without this resonance, but always ended up with
much higher $\chi^2$, no matter which spin-$\fth$ couplings and
$g_{\rho \omega \pi}$ coupling sign have been initialized. The final
structure of this resonance is always very broad, having a width of
more than $600$ MeV and being located close to the upper boundary of
the considered energy range, which makes the exact determination of
its total width difficult. Note that also
other resonance analyses identified a very broad $D_{13}$ resonance in
this energy region: For example, Batini\'c {\it et al.} \cite{batinic}
(analyzing $\pi N \ra \pi N$ for $I=\foh$ and $\pi N \ra \eta N$) and
Vrana {\it et al.} \cite{vrana} (analyzing $\pi N \ra \pi N$, $\pi N \ra
2\pi N$, and using the results from Ref. \cite{batinic}) both have found a
$D_{13}$ resonance at 2 GeV with a large width of about 1 GeV. \\
When we allow for another $D_{13}$ resonance in the energy region
between $1.7$ and $1.9$ GeV for the calculation using the conventional 
spin-$\fth$ couplings, the fit systematically decreases the
resonance's width until it is only be visible via its off-shell
contributions in the spin-$\foh$ channels. The outcome is a very
narrow ($\Gamma_{tot} \leq 30$ MeV) resonance, and the best $\chi^2$
in this situation is still worse than in the calculation when such a
resonance is neglected. However, the situation is slightly different
in the case when using the Pascalutsa couplings. Adding a
$D_{13}(1700)$ in this case improves the overall $\chi^2$ by about
$5-10\%$. The resulting total width is 50-55 MeV, half of which
are due to $2\pi N$ and the other half due to $\omega N$. The $\pi N$
decay ratio is only about 2\%, hence the resulting resonance is
similarly inelastic as in the analysis of Vrana {\it et al.} \cite{vrana}
and Batini\'c {\it et al.} \cite{batinic}. Since we only find small $\chi^2$ 
improvements due to this resonance in the Pascalutsa calculations, the
indication for a $D_{13}(1700)$ in the experimental data seems to be
only weak and not of resonant nature, and can thus also be described
by nonresonant contributions generated by spin-$\fth$ off-shell (or 
additional other background) contributions. It is interesting to note
that the slight hump around $1.76$ GeV in the imaginary part of the
$\pi N \ra \pi N$ partial wave is close to the $\omega N$ and
$K\Sigma$ thresholds and could therefore be due to kinematic effects
of these two channels.

\subsubsection{Isospin-$\fth$ resonances}

In the isospin-$\fth$ sector, a very good agreement among the
resonance parameters extracted from the different calculations can be
observed, cf. Table \ref{tabrespropsouri32} above. Even the inclusion of
the photoproduction data basically only changes the $K\Sigma$
couplings and decay ratios.

\noindent $\mathbf{S_{31}:}$ \\
In all our calculations, the first $S_{31}$ resonance is found 
around $1.62$ GeV with a width of about 175 MeV. Depending on the
spin-$\fth$ prescription, the value for its mass is either $1.61$
or $1.63$ GeV, for the conventional and the Pascalutsa, respectively,
couplings. The former value is corroborated upon taking into account
the pion-photoproduction multipoles. The $E_{0+}^\fth$ 
multipole helps to pin down the exact resonance properties, in
particular the mass, see the discussion on pion photoproduction in 
PMII \cite{pm2}. In the global fits, the
mass is fixed at $1.611$ GeV, in agreement with the value of the 
pion-photoproduction analysis of Arndt {\it et al.} \cite{arndt02}, but
smaller than the PDG \cite{pdg} value. \\
The particle data group \cite{pdg} lists a second $S_{31}$ resonance
around $1.9$ GeV with a two-star 
status, which has been found by Manley and Saleski \cite{manley92} and 
Vrana {\it et al.} \cite{vrana}. However, in the latter analysis, this
resonance turns out to be very narrow with large uncertainties in the
width: $\Gamma_{tot} = 48\pm 45$ MeV. We have also checked the
importance of such a resonance in the present model, and only found
very weak indications for its existence. Upon inclusion of a second
$S_{31}$ above $1.85$ GeV, the $\chi^2$ is greatly enhanced in the
$\pi N$ elastic and $\pi N \ra 2 \pi N$ channels for the case of the 
conventional spin-$\fth$ couplings. Using the Pascalutsa spin-$\fth$
couplings, additional strength is needed in the $S_{31}$ partial wave
above $1.9$ GeV, and thus a second $S_{31}$ resonance improves the
$\chi^2$ slightly. The mass is found in P-p-$\pi +$ and P-p-$\pi -$
between $1.9$ and $1.99$ GeV, 
while the width is $180 - 240$ MeV, about 30\% of which are due to
$\pi N$ and the other 70\% due to $2\pi N$. This shows, similarly to
the $D_{13}(1700)$ case, that the indications for a second $S_{31}$
resonance are only weak and rather of nonresonant nature. Hence the
needed $S_{31}$ strength above $1.85$ GeV can also be explained easily
by background contributions. Note that Arndt {\it et al.} \cite{arndt95}
have not found a $S_{31}(1900)$ either.
\par

\noindent $\mathbf{P_{31}:}$ \\
In this partial wave, the particle data group \cite{pdg} lists two
resonances below 2 GeV, a one-star at $1.75$ GeV and a four-star at
$1.91$ GeV. Therefore we have checked the importance of these two
resonances, which have not been considered by Feuster and Mosel
\cite{feusti98,feusti99}. As in the $S_{31}$ partial wave, we do not
find a resonance in the energy region above $1.85$ GeV when using the 
conventional spin-$\fth$ couplings. Again, the inclusion of such a
resonance deteriorates the $\chi^2$ tremendously in the $\pi N$
elastic and $\pi N \ra 2 \pi N$ channel. However, there is a strong
need for a very inelastic $P_{31}(1750)$ resonance below $1.8$ GeV to
be able to correctly reproduce the change of slope in the real part of
the $\pi N$ elastic partial wave. This is in stark contrast to the
four-star rating of the $P_{31}(1910)$ and the one-star rating of the
$P_{31}(1750)$ PDG \cite{pdg}. Only in the calculation with the
Pascalutsa couplings, the $P_{31}$ resonance moves to approximately
$1.98$ GeV with a broad inelastic width of around 700 MeV. But as is
obvious from Fig. \ref{figppi32}, this resonance can rather be seen as
a compensation of missing background in the high-energy region, since
the high-energy tail of the $P_{31}$ partial wave starts deviating
from the data in this calculation. In the conventional coupling
calculation, this additional strength is generated by spin-$\fth$
off-shell contributions. Thus also the indication for a $P_{31}(1910)$
is very weak in the experimental data and seems to be only of
nonresonant nature. This finding is confirmed upon inclusion of the
photoproduction data, which allows to additionally nail down the
$P_{31}(1750)$ properties. The change of slope of the imaginary part
of the $M_{1+}^\fth$ multipole (see the discussion on pion
photoproduction in PMII \cite{pm2}) leads to a 
reduction of the $P_{31}$ mass by about 40 MeV, while its total
width and inelasticity stay about the same.
\par

\noindent $\mathbf{P_{33}:}$ \\
In all calculations, the extracted properties of the $P_{33}(1232)$
are almost identical. A striking difference, however, is seen in the
total width extracted in the Pascalutsa calculation, which is rather
low with 94 MeV. However, this value is not surprising. As a result of 
the additional factor $s / m_\Delta^2$ in the amplitude (see Sec.
\ref{secrescontr}), the effective width of the resonance is increased 
above the resonance position. To prevent large discrepancies with the
$\pi N$ partial-wave data, the width at the resonance position has to
be reduced. This effect is only visible for this resonance, since the
higher the resonance mass, the smaller is the variation of $s / 
m_R^2$ around the corresponding resonance position.\\
Besides the well fixed $P_{33}(1232)$ resonance, we can also confirm
the need for a $P_{33}(1600)$ as in Refs. \cite{feusti98,feusti99},
\cite{vrana}, and \cite{manley92}. While the width and decay ratios
are similar to the values of the PDG \cite{pdg} and of Feuster and
Mosel \cite{feusti99}, the mass is fixed due to the $2\pi N$
production at 1.665 GeV, which is considerably higher than the PDG
value, but lower than the value of Feuster and Mosel.\\
Furthermore, in the present
calculation, there is a need for additional ($\pi N$) strength in this
partial wave at higher energies, which is not generated by the
implemented background. This gives rise to the necessity of the
inclusion of a third $P_{33}$. Although its mass is fixed above 2 GeV
(see Table \ref{tabrespropsouri32}), its resonant structure already shows
up below 2 GeV, see Fig. \ref{figppi32}. However, as a result of this
high mass, the extracted properties of this third $P_{33}$ resonance
can only be of qualitative nature, i.e., that the resonance is located
above 2 GeV, that it has a large
inelastic decay fraction, and also gives important contribution in
$K\Sigma$ production. The inclusion of the third $P_{33}$ also affects
the properties of the $P_{33}(1600)$. In particular, the $P_{33}(1600)$
mass is lowered in all calculations to about $1.66$ GeV, as compared
to the results of Feuster and Mosel, who have found in their global
fit a mass of $1.72$ GeV.\\ 
Similarly as in the $P_{13}$ wave, we find indications for a missing
inelastic contribution of about 1 mb in the $P_{33}$ partial wave
above $1.7$ GeV (cf. Fig. \ref{figp2both}) in the present model, 
i.e., the contribution of a $3\pi N$ state as $\rho \Delta$. While the 
$2\pi N$ partial-wave cross section decreases to about 2 mb, the
inelastic partial-wave cross section stays almost constant at 3
mb. The missing inelasticity can only be compensated in our model
above $1.91$ GeV, since there are no $2\pi N$ data points any more and
thus inelastic strength can be shifted to the $2\pi N$ channel.
\par

\noindent $\mathbf{D_{33}:}$ \\
In the $D_{33}$ partial wave, we only need one resonance below 
2 GeV for a satisfying description of the experimental data. In all
calculations, the resulting properties are very similar. The width is
found to be about 600 MeV, 86\% of which coming from the $2\pi N$
decay. Due to the $\pi N \ra 2\pi N$ partial-wave cross section data,
already in the hadronic fits the mass of the $D_{33}(1700)$ is well
fixed between $1.675$ and $1.68$ GeV. This mass is confirmed in the
global fit, where the resulting value of $1.678$ is also in accordance
with the value of $1.668$ GeV of Arndt {\it et
al.} \cite{arndt02}. Moreover, the  inelasticity is in good agreement 
with Ref. \cite{arndt02} and also with Manley and Saleski \cite{manley92},
while Vrana {\it et al.} \cite{vrana} found a much narrower ($\Gamma = 
120$ MeV) and even more inelastic (95\%) resonance at $1.73$
GeV. Although the resonance position is just below the $K\Sigma$
threshold, it gives important contributions to pion- and photon-induced 
$K\Sigma$ production, see Sec. \ref{secresps} and the discussion on
$K\Sigma$ photoproduction in PMII \cite{pm2}.\\
As in the $D_{13}$ case, the resulting $2\pi N$ production cross
section does not rise steeply enough from 1.3 GeV up to the
$D_{33}(1700)$ resonance position. For the same reasons as discussed
for the $D_{13}(1520)$, this is probably due to the deficiency of
the effective treatment of the $2\pi N$ final state in the present
model.

\section{Summary of Pion-Induced Results}

A very good description of all pion-induced data on $\pi N$, $2\pi N$,
$\eta N$, $K\Lambda$, $K\Sigma$, and $\omega N$ with one parameter set
is possible within the present model, where unitarity is guaranteed by
solving the scattering equation via the $K$-matrix approximation. 
This shows that all important contributions up to 2 GeV are included
and also, that the experimental data of all channels are consistent
with each other. Since the driving potential is built up by the use of
effective Lagrangians for Born-, $t$-channel, spin-$\foh$, and
spin-$\fth$ resonance contributions, also the background contributions
are generated consistently for all partial waves and the number of
parameters is greatly reduced. The extension of 
the energy range and model space has required the inclusion of
additional resonances [$P_{13}(1900)$, $P_{31}(1750)$, $P_{33}(1920)$] 
as compared to the previous analysis of Feuster and Mosel
\cite{feusti99}, where the former two are particularly important in
the production mechanisms of the higher-lying final states $K\Lambda$, 
$K\Sigma$, and $\omega N$. These
extensions lead to differences in the 
descriptions of some final states, as, e.g., the $K\Lambda$
production, which is now dominated by a $IJ^P=\foh\fth^+$ ($P_{13}$)
in contrast to the $IJ^P=\foh\foh^+$ ($P_{11}$) dominance of earlier
analyses \cite{feusti99,manley92}. Since a good description of all
channels is possible although no spin-$\ffh$ resonances are considered 
in our model, this indicates, that higher-spin ($\geq \ffh$)
resonances are only of minor importance in the production of $\eta N$,
$K\Lambda$, $K\Sigma$, and $\omega N$. This point is investigated
further at present \cite{vitali}.

Due to the inclusion of all important final states below 2 GeV, all
threshold effects are included correctly. As compared to the
calculation of Feuster and Mosel \cite{feusti98,feusti99}, this leads
especially to an improvement of the description of the $K\Lambda$
channel, which is influenced by both the $K\Sigma$ and the
$\omega N$ thresholds. Thus, in contrast to the speculation of
Refs. \cite{feusti98,feusti99}, the inclusion of $u$-channel
contributions from hyperon resonances is far less important for a
good description of the associated strangeness channels $\pi N \ra
K\Lambda/K\Sigma$ than the correct treatment of all unitarity effects. 

The effects of chiral symmetry have been
checked by allowing for a chirally symmetric or a chiral symmetry
breaking $\sigma \pi \pi$ coupling vertex. The chiral symmetric one
has proven superior not only for the low, but also for the
intermediate energy region in $\pi N$ elastic scattering.

The description of the pion-induced data is also still possible, when
we further reduce the freedom of our background contributions by using 
Pascalutsa spin-$\fth$ vertices instead of the conventional
ones. These couplings remove the off-shell spin-$\foh$ contributions
of the spin-$\fth$ resonance processes, thus reducing the background
contributions in the spin-$\foh$ sector. This reduction automatically
leads to an increase of the importance of the $t$-channel diagrams,
resulting in a much harder cutoff value $\Lambda_t$. Thereby, the
contributions of the $t$-channel diagrams become more important in the
lower partial waves and agreement with the experimental data is
achieved. However, the increase of the total $\chi^2$ from the
conventional to the Pascalutsa  prescription ($2.66 \ra 3.53$) shows
that indeed additional background terms are necessary for a better
description of the experimental data.

As a result of the additional inclusion of the photoproduction data on 
all channels, the description of the pion-induced reactions becomes
worse. This is not unexpected, since due to the more recent
photoproduction data of high quality, the reaction process is much
more constrained and thus allows for less freedom. However, the 
pion-induced data are still well described in a global calculation
including all pion- and photon-induced data. The largest changes are
observed in the $I,J=\foh$ ($S_{11}$ and $P_{11}$) waves, where the
properties of the $S_{11}(1535)$, $S_{11}(1650)$, and $P_{11}(1710)$
can be better controlled once the photoproduction data --- in
particular on $\eta N$, $K\Lambda$, and $\omega N$ --- are
included. Differences are also found in the background $\rho NN$ 
coupling, which turns out to be close to the KSRF value in the
hadronic calculations. The differences in the global fits can be
traced back to the necessity of changing the 
nucleon form factor cutoff $\Lambda_N$ for the description of the 
pion-photoproduction multipoles, see also PMII \cite{pm2}. The
Born couplings extracted from the global fits are close to SU(3)
values.

The influence of the sign of $g_{\omega \rho \pi}$ can be best
summarized when comparing the results of the two global calculations
C-p-$\gamma +$ and C-p-$\gamma -$. Switching the sign of $g_{\omega
  \rho \pi}$ leads to basically the same extracted couplings and
resonance parameters. The main difference is a switch of signs of some 
$\omega N$ couplings, i.e., $\kappa_{NN\omega}$, $g_{\omega_1}$ of the
$P_{11}(1440)$, and $g_{\omega_2}$ and $g_{\omega_3}$ of the
$D_{13}(1520)$, while almost all other $\omega N$ contributions are
similar. This indicates that the same interference pattern between
these specific contributions and the $t$-channel contribution is 
preferred in the pion-induced reaction, while the remaining
contributions are rather unaffected. Comparing the quality of the
fits, there is a tendency of preferring the positive $g_{\omega \rho
  \pi}$ sign in line with SU(3) flavor symmetry. This becomes most
obvious in the $\chi^2$ of the $\omega N$ production channels, while
all other channels remain basically unchanged. Especially the
pion-induced $\omega N$ production can be much better described with
the positive sign, when the photoproduction data are included.

There are also some indications for room for improvement of the
model. Assuming that the $2\pi N$ data \cite{manley84} are correct,
there are evidences for important additional $3\pi N$ final 
state contributions, which are not considered up to now, in the
$J^P=\fth^+$ partial waves. We also find evidences for the necessity
of a more correct treatment of the $2\pi N$ state in the low-energy
tails of the  $D_{13}(1520)$ and $D_{33}(1700)$ resonance. As a
consequence of the generalization of the partial-wave 
decomposition, which has become necessary in the present model for the 
inclusion of the $\omega N$ final state, a more realistic description
of the $2\pi N$ final state in terms of $\rho N$ and $\pi \Delta$ is
now possible. The inclusion of these final states allows to
mimic the three particle phase space while still dealing with two body 
unitarity. The accounting for the spectral function of the $\rho$
meson and the $\Delta$ baryon would then allow for the complete
description of $2\pi N$ production within the present model. This
extension will probably improve the description of the $D_{I3}$ waves
below the first resonance.

In PMII \cite{pm2}, the results of the two global fits C-p-$\gamma +$
and C-p-$\gamma -$ on all photoproduction reactions are presented and
discussed in detail.

\begin{acknowledgments}
We like to thank W. Schwille for making the preliminary SAPHIR data on
$\omega$ photoproduction \cite{barthom} available to us. One of the
authors (G.P.) is grateful to C. Bennhold for the hospitality at the
George Washington University, Washington, D.C., in the early stages of
this work. This work was supported by DFG and GSI Darmstadt.
\end{acknowledgments}

\begin{appendix}

\section{Notations}
\label{appnot}

We work in the c.m. frame and use the metric of Bjorken and Drell
\cite{bjorkendrell}, i.e., $g_{\mu \nu} = \mathrm{diag}
(1,-1,-1,-1)$. Four-momenta are denoted by italic letters ($p$, $k$,
$q$, etc.), three-momenta by bold letters ($\vec p$, $\vec k$, $\vec
q$, etc.)\footnote{Note that three-vectors are denoted in general by
  bold letters.}, their absolute values by upright letters ($\avec p$,
$\avec k$, $\avec q$, etc.), and their unit vectors by $\hat {\vec
  p}$, $\hat {\vec k}$, $\hat {\vec q}$, etc. In general, incoming,
outgoing, and intermediate meson (baryon) momenta are denoted by
$k$, $k'$, and $k_q$ ($p$, $p'$, and $p_q$), respectively. 

Two-particle momentum states with helicity $\lambda \equiv \lambda_k -
\lambda_p$ are normalized in the following way:
\bea
\langle f| i\rangle &\equiv&
\langle \vec p' \vec k', \lambda' | \vec p \vec k ,\lambda \rangle 
\label{momstatenorm} \\
&=& \delta^4 (P' - P) 
\frac{\sqrt s}{\avec k E_B E_M} 
\delta (\Omega'_k - \Omega_k) \delta_{\lambda' \lambda}
\nonumber \\
&=& \delta^4 (P' - P) \frac{\sqrt s}{\avec k E_B E_M} 
\langle \vartheta' \varphi', \lambda' | 
\vartheta \varphi, \lambda \rangle
\nonumber \; .
\eea

The helicity notation for the $\omega N$ and $\gamma N$ helicity states 
is: $\pm 0$: $\lambda = \lambda_V - \lambda_B = 0 \pm \foh$, $\pm
\foh$: $\lambda = \pm 1 \mp \foh$, and $\pm \fth$: $\lambda =\pm 1 \pm
\foh$.

The relation between the scattering matrix $S$ and the transition
matrix $T$ is defined as
\bea
S \equiv 1 + 2 \mi T \; .
\eea
With the two-particle states \refe{momstatenorm}, the matrix $M$ is
given by
\bea
\langle f | S | i \rangle = \delta_{fi} 
- \mi (2\pi)^4 \delta^4 (P_f - P_i) 
\left( \prod_{j=1}^4 N_j \right) 
\langle f | M | i \rangle
\eea
with the usual normalization factors (see, e.g., Ref. \cite{bjorkendrell})
and hence
\bea
\langle f | T | i \rangle = 
- \frac{1}{2} (2\pi)^4 \delta^4 (P_f - P_i) 
\left( \prod_{j=1}^4 N_j \right) 
\langle f | M | i \rangle  \; . \label{reltandm}
\eea
The scattering amplitude $\mct^{fi}_{\lambda' \lambda} (\vt)$ and the
$K$-matrix amplitude $\mck^{fi}_{\lambda' \lambda} (\vt)$ are defined
by 
\bea
\mct^{fi}_{\lambda' \lambda} &\equiv& 
- \frac{\sqrt{\avec p \avec p' m_{B'} m_B}}{(4\pi)^2 \sqrt s} 
\langle f | M | i \rangle \; ,
\label{tandm} \\
\mck^{fi}_{\lambda' \lambda} &\equiv&
- \frac{\sqrt{\avec p \avec p' m_{B'} m_B}}{(4\pi)^2 \sqrt s} 
\langle f | K | i \rangle \; ,
\label{kandv}
\eea
where $K = V$ in the $K$-matrix Born approximation and $\langle f|$ 
and $|i\rangle$ denote two-particle momentum states as defined above.

\section{Partial-Wave Decomposition}
\label{apppwd}

Using the rotational invariance of the interaction and the properties
of the Wigner functions ($d$ functions), the c.m. scattering amplitude
$\mct^{fi}_{\lambda' \lambda} (\vt)$  can be decomposed into
amplitudes with total angular momentum $J$: 
\bea 
\mct^{fi}_{\lambda' \lambda} &=& 
\sum_{J} \frac{2J+1}{4\pi} \mct^J_{\lambda'\lambda} (\sqrt s)
d^J_{\lambda \lambda'} (\vt) \; , 
\label{defamplihelij}
\eea
where we have defined $\mct^J_{\lambda'\lambda} (\sqrt s) \equiv
\langle J, \lambda' | T(\sqrt s) | J, \lambda
\rangle$. The $d^J_{\lambda \lambda'} (\vt)$ play the role of the
Legendre polynomials, but for half integer
spin. Equation \refe{defamplihelij} can be inverted to
\bea
\mct^J_{\lambda'\lambda} (\sqrt s) &=& 
2 \pi \int_{-1}^{+1} \difd (\cos \vt) d^J_{\lambda \lambda'} (\vt) 
\mct^{fi}_{\lambda'\lambda} \; .
\label{defhelij}
\eea
The helicity states $|J,\lambda \rangle \equiv |J,\lambda_k \lambda_p
\rangle$ fulfill the parity property \cite{jacobwick}:
\bea
\hat P |J,\lambda \rangle = \eta_k \eta_p (-1)^{J-s_k-s_p} |J,-\lambda
\rangle \; .
\label{parjlambda}
\eea
Here, $\eta_k$, $\eta_p$, and $s_k$, $s_p$ are the intrinsic parities
and spins, respectively of the two particles. The construction of normalized
states with parity $(-1)^{J\pm \foh}$ is now straightforward:
\bea
|J,\lambda;\pm \rangle &\equiv&
\frac{1}{\sqt} 
\left( |J,+\lambda\rangle \pm \eta |J,-\lambda \rangle \right)
\nonumber \\
\Rightarrow
\hat P |J,\lambda; \pm\rangle &=& (-1)^{J\pm\foh} 
|J,\lambda; \pm\rangle \; ,
\label{paritystates}
\eea
where we have defined
\be
\eta \equiv \eta_k \eta_p (-1)^{s_k+s_p+\foh} \; . 
\label{intrinseta}
\ee
For parity conserving interactions $T = \hat P^{-1} T \hat P$ one has
\be
\langle 
J, \lambda' | T (\sqrt s) | J,-\lambda \rangle 
= \eta (\eta')^{-1}
\langle J,\lambda' | T (\sqrt s) | J,\lambda \rangle
\label{helipariprop}
\ee
and one can use the states \refe{paritystates} to project out helicity
partial-wave amplitudes with a definite parity of $(-1)^{J\pm \foh}$: 
\bea
\mct^{J\pm}_{\lambda' \lambda} &\equiv& 
\langle J,\lambda';\pm | T | J,\lambda;\pm \rangle 
=
\mct^J_{\lambda' \lambda} \pm \eta \mct^J_{\lambda' -\lambda} \; .
\label{parheli}
\eea
These helicity partial-wave amplitudes $\mct^{J\pm}_{\lambda'
  \lambda}$ have definite, identical $J$ and definite, but opposite
parity. It is quite obvious that this method is valid for any
meson-baryon final state combination, even cases as, e.g., $\omega N
\rightarrow \pi \Delta$.

The parity properties of the angle dependent c.m. helicity scattering 
amplitudes $\mct^{fi}_{\lambda' \lambda} (\vt)$ follow:
\bea 
\mct^{fi}_{-\lambda', -\lambda} (\vt) &=&
\eta (\eta')^{-1} (-1)^{\lambda-\lambda'} 
\mct^{fi}_{\lambda' \lambda} (\vt) \; .
\label{anglepariprop}
\eea
Now the rescattering part of the BS equation \refe{tkrelwint} can be
decomposed into partial waves:
\bea
\mct^{fi}_{\lambda' \lambda} &=&
\mck^{fi}_{\lambda' \lambda} + \mi 
\int \difd \Omega_q \sum_{\lambda_q} \mct_{\lambda' \lambda_q} 
\mck_{\lambda_q \lambda}
\nonumber \\
 &=&
\mck^{fi}_{\lambda' \lambda} + \mi 
\sum_{\lambda_q} 
\sum_{J} \frac{2J+1}{4 \pi}
d^J_{\lambda \lambda'} (\vt')
\mct^J_{\lambda' \lambda_q}
\mck^J_{\lambda_q \lambda} ,
\nonumber
\eea
where the $\mct^J_{\lambda' \lambda_q}$ and $\mck^J_{\lambda_q
 \lambda}$ are defined in the same way as in Eq.
\refe{defhelij}. Inserting this into the BS equation and integrating
over $2 \pi \int \difd (\cos \vt')$, we arrive at an algebraic BS
equation for each partial wave:
\be
\mct^J_{\lambda' \lambda} =
\mck^J_{\lambda' \lambda} + \mi 
\sum_{\lambda_q} 
\mct^J_{\lambda' \lambda_q}
\mck^J_{\lambda_q \lambda} \; .
\label{bsejdec}
\ee
Using the parity conserving states we finally have
\bea
\mct^{J\pm}_{+\lambda',\lambda} 
&=&
\mck^{J\pm}_{+\lambda',\lambda} + \mi 
\sum_{\lambda_q>0}
\mct^{J\pm}_{+\lambda',\lambda_q}
\mck^{J\pm}_{+\lambda_q,\lambda} \; .
\label{bsejpdec}
\eea

Apart from the recursion formulae for the $d$ functions $d^J_{\foh,
  \pm \foh}$, $d^J_{\foh, \pm \fth}$, which can be found in many
textbooks, there is also a need for a recursion formula for
$d^J_{\fth, \pm \fth}$:  
\bea
d^{J-1}_{+\fth +\fth} (\vartheta) 
&=& \frac{-1}{(1 + \cos \vartheta)}
\left[ 2 d^J_{+\foh +\foh} (\vartheta) + 
\sqrt{\frac{J+\foh}{J-\fth}} \left(
  2 \sin \vartheta d^{J-1}_{+\foh +\fth} (\vartheta) -
  (1-\cos \vartheta) d^{J-1}_{-\foh +\fth} (\vartheta) \right) \right] 
\; , 
\nonumber \\
d^{J-1}_{+\fth -\fth} (\vartheta) 
&=& \frac{1}{(1 - \cos \vartheta)}
\left[ 2 d^J_{-\foh +\foh} (\vartheta) +
  \sqrt{\frac{J+\foh}{J-\fth}} \left(
  2 \sin \vartheta d^{J-1}_{-\foh +\fth} (\vartheta) -
  (1+\cos \vartheta) d^{J-1}_{+\foh +\fth} (\vartheta) \right) \right]
\; .
\eea

\section{Lagrangians, Widths, and Couplings}
\label{applagr}

All interaction Lagrangians given below in this appendix also contain
an isospin part, which is discussed in Appendix \ref{appisoop}.

\subsection{Background interactions}
\label{appbornlagr}

The asymptotic particles and intermediate $t$-channel mesons entering
the potential interact in the hadronic reactions via the background
Lagrangian,
\bea
\mcl_{Born} + \mcl_t &=& - \bar u_{B'} (p') \left[ 
  \frac{g_{\tilde \varphi}}{m_B+m_{B'}} 
  \gamma_5 \gamma_\mu (\partial^\mu \tilde \varphi ) 
  + g_\eta \mi \gamma_5 \eta + g_S S 
+ g_V \left( \gamma_\mu V^\mu + 
  \frac{\kappa_V}{2m_N} \sigma_{\mu \nu} V^{\mu \nu} \right)
\right] u_B (p)
\nonumber \\ &&
- \frac{g_S}{2 m_\pi} (\partial_\mu \varphi')
(\partial^\mu \varphi ) S 
- g_V \varphi' (\partial_\mu \varphi ) V^\mu
- \frac{g}{4 m_\varphi} \ve_{\mu \nu \rho \sigma} V^{\mu \nu}
{V'}^{\rho \sigma} \varphi \; ,
\label{lagback} 
\eea
with the asymptotic baryons $B,B'=(N,\Lambda,\Sigma)$, the
(pseudo) scalar mesons $\tilde \varphi=\pi,K$,
$(\varphi,\varphi')=(\pi,\eta,K)$, $S=(\sigma,a_0,K_0^*)$, the vector
mesons $V =(\rho,\omega,K^*)$ and
\bea
V^{\mu \nu} = \partial^\mu V^\nu - \partial^\nu V^\mu \; .
\eea
Note that for comparison, also a nonderivative $S\varphi \varphi$
coupling $\mcl = - g'_S m_S \varphi' \varphi S$ is used in one
calculation. Here, $g'_S$ is related to $g_S$ via $g'_S = - g_S (m_S^2
- m^2_\varphi - m^2_{\varphi'})/(4 m_S m_\pi)$. 

Using the values for the decay widths from Ref. \cite{pdg}, the following
couplings are extracted:
\bea
\ba{lcrclcr}
g_{\rho \pi \pi}       &=&  6.020 \; , & & 
g_{\omega \rho \pi}    &=&  2.060 \; , \\
g_{K^* K \pi}          &=& -6.500 \; , & & \\
g_{K^*_0 K \pi}        &=& -0.900 \; , & & 
g_{a_0 \eta \pi}       &=& -2.100 \; . \\
\ea
\label{mesdeccons}
\eea
The $\omega \rho \pi$ coupling constant is determined from the $\omega
\ra \rho \pi \ra \pi^+ \pi^- \pi^0$ decay width of $\approx 7.4$ MeV.

\subsection{Spin-$\foh$ baryon resonance interactions}
\label{appres12lagr}

\subsubsection{(Pseudo) scalar meson decay}

For negative-parity spin-$\foh$ resonances, PS coupling is used:
\bea
\mcl^{PS}_{\foh B\varphi} = - g_{RB\varphi} \bar u_R
\left( \ba{c} 1 \\ -\mi \gamma_5 \ea \right)
u_B \varphi \; .
\eea
For the positive-parity spin-$\foh$ resonances, PV coupling is used:
\bea
\mcl^{PV}_{\foh B\varphi} = 
- \frac{g_{RB\varphi}}{m_R \pm m_B} \bar u_R
\left( \ba{c} \gamma_5 \\ \mi \ea \right)
\gamma_\mu u_B \partial^\mu \varphi \; .
\eea
In both cases, the upper (lower) sign and operator hold for
pseudoscalar (scalar) mesons $\varphi$.

For negative-parity resonances (PS coupling), this leads to the decay
width
\bea
\Gamma^{PS}_\pm = 
f_I \frac{g^2_{RB\varphi}}{4\pi} \avec k_\varphi \frac{E_B \mp
  m_B}{\sqrt s} 
\label{onehalfpsdecay}
\eea
and for positive-parity resonances (PV coupling) to
\bea
\Gamma^{PV}_\pm &=& 
f_I \frac{g^2_{RN\varphi}}{4\pi} \avec k_\varphi 
\frac{E_B \mp m_B}{\sqrt s} 
\left( \frac{\sqrt s \pm m_B}{m_R \pm m_B} \right)^2 
\nonumber \\
&\stackrel{\sqrt s = m_R}=& 
\Gamma^{PS}_\pm
\label{onehalfpvdecay}
\eea
with the absolute value of the meson three-momentum $\avec
k_\varphi$. The upper (lower) sign always corresponds to a parity-flip
(parity-nonflip) transition, e.g., $P_{11}(1440) \ra \pi N$
[$S_{11}(1535) \ra \pi N$]. The isospin factor $f_I$ is equal to $1$
for isospin-$\fth$ resonances, equal to $3$ for the decay of
isospin-$\foh$ resonances into a $I=1\oplus \foh$ final state, and
equal to $1$ for the decay of isospin-$\foh$ resonances into
$I=0\oplus \foh$.

\subsubsection{Vector meson decay}

For the $\omega N$ decay we apply the Lagrangian
\bea
\mcl_{\foh N\omega} = -\bar u_R
\left( \begin{array}{c} 1 \\ - \mi \gamma_5 \end{array} \right)
\left( g_1 \gamma_\mu - \frac{g_2}{2 m_N} \sigma_{\mu \nu}
  \partial^\nu_\omega \right) u_N \omega^\mu \; ,
\eea
The upper (lower) operator corresponds to a positive- (negative-)
parity resonance.

The resulting helicity decay amplitudes are: 
\bea
A^{\omega N}_{\foh} 
&=& \mp \frac{\sqrt{E_N \mp m_N}}{\sqrt{m_N}} \left( g_1 + g_2
  \frac{m_N \pm m_R}{2 m_N} \right) \; , 
\label{heliampli12} \\
A^{\omega N}_0 
&=& \mp \frac{\sqrt{E_N \mp m_N}}{m_\omega \sqrt{2 m_N}} \left( 
  g_1 (m_N \pm m_R) + g_2 \frac{m_\omega^2}{2 m_N} \right) \; .
\nonumber
\eea
The lower indices correspond to the $\omega N$ helicities and are
determined by the $\omega$ and nucleon spin-$z$ components as in
Appendix \ref{appnot}: $\foh$: $1 - \foh = \foh$ and $0$: $0 + \foh =
\foh$. The resonance $\omega N$ decay widths are then given by
\be
\Gamma^{\omega N} = \frac{2}{2J+1}\sum\limits_{\lambda = 0}^{\lambda = 
  +J} \Gamma^{\omega N}_\lambda \; , \hspace{4mm} 
\Gamma^{\omega N}_\lambda = \frac{\avec k_\omega m_N}{2 \pi m_R} \left|
  A^{\omega N}_\lambda \right|^2 \label{heliwidth} \; .
\ee

\subsection{Spin-$\fth$ baryon resonance interactions}
\label{appres32lagr}

For all the conventional spin-$\fth$ couplings given below, the
corresponding Pascalutsa couplings can be extracted by the
replacement
\bea
\Gamma_\mu u_R^\mu \ra 
\Gamma_\mu \gamma_5 \gamma_\nu \tilde U_R^{\nu \mu} \; ,
\eea
where the dual of the resonance field tensor is given by: 
$\tilde U_R^{\mu \nu} = \foh \ve^{\mu \nu \alpha \beta} {U_R}_{\alpha
  \beta} = \foh \ve^{\mu \nu \alpha \beta} (\partial_\alpha {u_R}_\beta -
\partial_\beta {u_R}_\alpha)$. At the same time, the off-shell
projectors $\Theta_{\mu \nu}(a)$ [cf. Eq. \refe{offshellapp}] are
dropped.

\subsubsection{(Pseudo) scalar meson decay}

The interaction with (pseudo) scalar mesons for positive-parity
spin-$\fth$ resonances is
\bea
\mcl_{\fth B\varphi} = \frac{g_{RB\varphi}}{m_\pi} \bar u_R^\mu
\Theta_{\mu \nu} (a_{RB\varphi}) 
\left( \ba{c} 1 \\ -\mi \gamma_5 \ea \right) 
u_B \partial^\nu \varphi 
\eea
and for negative-parity resonances
\bea
\mcl_{\fth B\varphi} = - \frac{g_{RB\varphi}}{m_\pi} \bar u_R^\mu
\Theta_{\mu \nu} (a_{RB\varphi}) 
\left( \ba{c} \mi \gamma_5 \\ 1 \ea \right)
u_B \partial^\nu \varphi \; .
\eea
As in the spin-$\foh$ case, the upper (lower) operator holds for 
pseudoscalar (scalar) mesons $\varphi$. $\Theta_{\mu \nu}$ is the
off-shell projector: 
\bea
\Theta_{\mu \nu} (a) = 
g_{\mu \nu} - a \gamma_\mu \gamma_\nu \; ,
\label{offshellapp}
\eea
where $a$ is related to the commonly used off-shell parameter $z$
by $a = (z + \sfoh )$.

These couplings lead to the decay width:
\bea
\Gamma^\fth_\pm = 
f_I \frac{g^2_{RB\varphi}}{12\pi m_\pi^2} 
\avec k_\varphi^3 \frac{E_B \pm m_B}{\sqrt s} \; .
\label{threehalfdecay}
\eea
The upper (lower) sign corresponds to the decay of a resonance into
a meson with opposite (identical) parity, e.g., $P_{33}(1232) \ra \pi
N$ [$D_{13}(1520) \ra \pi N$]. The isospin factor $f_I$ is the same as
in Eqs. \refe{onehalfpsdecay} and \refe{onehalfpvdecay}.

\subsubsection{Vector meson decay}

For the $\omega N$ decay we use
\bea
\mcl_{\fth N\omega} &=& -\bar u_R^\mu 
\left( \begin{array}{c} \mi \gamma_5 \\ 1 \end{array} \right)
\left( \frac{g_1}{2m_N} \gamma^\alpha + \mi \frac{g_2}{4 m_N^2} 
  \partial^\alpha_N + \mi \frac{g_3}{4 m_N^2} 
  \partial^\alpha_\omega \right) 
\left( \partial^\omega_\alpha g_{\mu \nu} - \partial^\omega_\mu
  g_{\alpha \nu} \right) u_N \omega^\nu \; . \hspace*{6mm}
\eea
The upper (lower) operator corresponds to a positive- (negative-)
parity resonance. Note that for clarity, the off-shell projectors
$\Theta_{\mu \nu} (a)$ [cf. Eq. \refe{offshellapp}], which are
contracted with each coupling operator, are not displayed. This leads
to the $\omega N$ helicity decay amplitudes:
\bea
A^{\omega N}_{\fth} 
&=& - \frac{\sqrt{E_N \mp m_N}}{\sqrt{2 m_N}}\frac{1}{2m_N}
\left( g_3 \frac{m_\omega^2}{2 m_N} - g_1 (m_N \pm m_R) +
  g_2 \frac{m_R^2 - m_N^2 - m_\omega^2}{4 m_N}
  \right)  \; , \nonumber \\
A^{\omega N}_{\foh} 
&=& \pm \frac{\sqrt{E_N \mp m_N}}{\sqrt{6 m_N}}\frac{1}{2m_N}
\left( g_3 \frac{m_\omega^2}{2 m_N} \pm g_1 \frac{m_N (m_N
    \pm m_R) - m_\omega^2}{m_R} + g_2 \frac{m_R^2 - m_N^2 - m_\omega^2}{4 m_N}
  \right)  \; , \label{heliampli32} \\
A^{\omega N}_0 
&=& \pm m_\omega \frac{\sqrt{E_N \mp m_N}}{\sqrt{3 m_N}}\frac{1}{2m_N}
\left( g_1 \mp g_2 \frac{m_R^2 + m_N^2 - m_\omega^2}{4 m_R
    m_N} \mp g_3 
  \frac{m_R^2 - m_N^2 + m_\omega^2}{4 m_R m_N} \right) \; .
\eea
The helicity notation is the same as in the spin-$\foh$ case; in
addition, there is the helicity state $\fth$: $1 + \foh = \fth$. The
resonance $\omega N$ decay widths is given by Eq. \refe{heliwidth}.

\section{Calculation of Amplitudes}
\label{appampli}

The calculation of the amplitudes $\mcv^{fi} \equiv \langle f | V | i
\rangle$ which enter Eq. \refe{kandv} are extracted from the Feynman
diagrams via
\bea
\mcv_{\lambda' \lambda}^{fi} &=&
\bar u(p',\lambda_{B'}) \Gamma (s, u) u(p,\lambda_B) 
\nonumber \\
&=&
\frac{4 \pi \sqrt s }{\sqrt{m_B m_{B'}}} \chi^\dagger_{\lambda_{B'}}
\mcf(s, u) \chi_{\lambda_B} \; .
\label{basicdecomp}
\eea

\subsection{Spin-$0$ spin-$\foh$ scattering}

The Dirac operator $\Gamma$ is given by 
\bea
\Gamma (s, u) = \Theta \cdot ( A \umat_4 + B \slash{\bar k} ) \; ,
\label{pipidecomp}
\eea
where $\bar k$ is the average of the meson momenta: $\bar k = (k +
k')/2$ and $\Theta = \umat_4$ for incoming and outgoing mesons of
identical parity and $\Theta = \mi \gamma_5$ for mesons of opposite
parity. Realizing that 
\bea
\bar u (p', s') (\mi \gamma_5) 
= \pm \mi \bar u (p', s', E_{B'} + m_{B'} \ra E_{B'} - m_{B'})
\label{ubarsigmak} 
\eea
for $s' = \pm \foh$, the Pauli operator $\mcf$ results in
\bea
\mcf = \theta \cdot ( \tilde A \umat_2 + 
\tilde B \vec \sigma \cdot \vec{\hat{k'}} 
\vec \sigma \cdot \vec{\hat k} )
\hspace*{3mm} \label{fpipidecomp}
\eea
with $\theta = \umat_2$ for mesons of identical parity and $\theta =
\mi \vec \sigma \cdot \vec{\hat{k'}}$ for mesons of opposite parity.
Here, $\tilde A$ and $\tilde B$ are related to $A$ and $B$ in the
following way: 
\bea
\tilde A &=& + \frac{\sqrt{R_+ R'_\pm}}{8 \pi \sqrt s} 
\left( A + \sfoh B ( S_- + S'_\mp ) \right) \; ,
\nonumber \\
\tilde B &=& - \frac{\sqrt{R_- R'_mp}}{8 \pi \sqrt s} 
\left( A - \sfoh B ( S_+ + S'_\pm ) \right)
\; , \label{ababpipi}
\eea
where the upper sign is for mesons of identical and the lower one for
mesons with opposite parity and
\bea
R_\pm &=& E_B \pm m_B \; , \hspace{5mm} R'_\pm = E_{B'} \pm m_{B'}
 \; , \nonumber \\
S_\pm &=& \sqrt s \pm m_B \; , \hspace{5mm} 
S'_\pm = \sqrt s \pm m_{B'}
\; . \label{randsdef}
\eea
Using $\vec \sigma \cdot \vec{\hat{k'}} \chi^f_{\pm\foh} = \pm
\chi^f_{\pm\foh}$ and $\vec \sigma \cdot \vec{\hat k} \chi^i_{\pm\foh}
= \pm \chi^i_{\pm \foh}$ the helicity dependent amplitudes result in:
\bea
\mcv_{+\foh +\foh} &=& \pm \mcv_{-\foh -\foh} =
f \frac{4 \pi \sqrt s }{\sqrt{m_B m_{B'}}} \cos \fvh
\left( \tilde A + \tilde B \right) \; ,
\nonumber \\
\mcv_{+\foh -\foh} &=& \pm \mcv_{-\foh +\foh} =
f \frac{4 \pi \sqrt s }{\sqrt{m_B m_{B'}}} \sin \fvh
\left( \tilde A - \tilde B \right) 
\label{fandmpipi}
\eea
with the upper sign and $f = 1$ for mesons of identical and the lower
sign and $f = \mi$ for mesons of opposite parity.

\subsection{Spin-$1$ spin-$\foh$ $\ra$ spin-$0$ spin-$\foh$}

Replacing the Dirac operator $\Gamma \ra \Gamma_\mu
\varepsilon^\mu_{\lambda_V}$ the general form of $\Gamma_\mu$ is 
\bea
\Gamma_\mu (s, u) = \Theta &\cdot& \left( 
A_p p_\mu + A_{p'} p'_\mu + (B_p p_\mu + B_{p'} p'_\mu) \slash k +
C \gamma_\mu + D \slash k \gamma_\mu \right) \; ,
\label{pivectorgam}
\eea
with $\Theta = \mi \gamma_5$ for pseudoscalar and $\Theta = \umat_4$
for scalar outgoing mesons. $\mcf$ is constructed in analogy to 
the virtual photon case \cite{ben95}:
\bea
\mcf &=& 
  \mi \vec \sigma \cdot \vec \varepsilon \mcf_1 + 
  \vec \sigma \cdot \vec{\hat{k'}} \vec \sigma \cdot (\vec{\hat{k}} 
  \times \vec \varepsilon) \mcf_2 +
  \mi \vec \sigma \cdot \vec{\hat{k}} 
  \vec \varepsilon \cdot \vec{\hat{k'}} \mcf_3 + 
  \mi \vec \sigma \cdot \vec{\hat{k'} }
  \vec \varepsilon \cdot \vec{\hat{k'}} \mcf_4 
  - \mi \varepsilon^0 
  (\vec \sigma \cdot \vec{\hat{k'}} \mcf_5 +
   \vec \sigma \cdot \vec{\hat{k}}  \mcf_6) \; ,
\label{fpivecdecomp}
\eea
with $\varepsilon_{\lambda_V}^\mu = (\varepsilon^0,
\vec{\varepsilon})$. Obviously, $\mcf_5$ and $\mcf_6$ only contribute 
for longitudinal polarizations. This has to be replaced for scalar
meson production by $\mcf \ra -\mi \vec \sigma \cdot \vec{\hat{k'}}
\mcf$. Equations \refe{pivectorgam} and \refe{fpivecdecomp} are related via 
\bea
\mcf_1 &=& \frac{1}{8 \pi \sqrt s} \sqrt{R'_\pm R_+}
  \left( C - S_- D \right) \; , \nonumber \\
\mcf_2 &=& \frac{1}{8 \pi \sqrt s} \sqrt{R'_\mp R_-}
  \left( C + S_+ D \right) \; , \nonumber \\
\mcf_3 &=& \frac{\avec k'}{8 \pi \sqrt s} \sqrt{R'_\pm R_-}
  \left( -A_{p'} + S_+ B_{p'} \right) \; , 
\label{pivecfi} \\
\mcf_4 &=& \frac{\avec k'}{8 \pi \sqrt s} \sqrt{R'_\mp R_+}
  \left( A_{p'} + S_- B_{p'} \right) \; , \nonumber \\
\mcf_5 &=& - \frac{1}{\avec k'} \widetilde \mcf_4 -
  \frac{1}{8 \pi m_M \sqrt s} \sqrt{R'_\mp R_-}
  \left( S_+ C + m_M^2 D \right) \; , \nonumber \\
\mcf_6 &=& - \frac{1}{\avec k'} \widetilde \mcf_3 -
  \frac{1}{8 \pi m_M \sqrt s} \sqrt{R'_\pm R_+}
  \left( S_- C - m_M^2 D \right) 
\nonumber
\eea
with
\bea
\widetilde \mcf_i &=& \etpp \mcf_i + 
\etp \mcf_i(A_{p'} \ra A_p, B_{p'} \ra B_p) \; , 
\nonumber \\
 \etp &\equiv& \ve_0^\mu p_\mu =
\frac{{\avec k} \sqrt s}{m_M} \; , 
\nonumber \\
\etpp &\equiv& \ve_0^\mu p'_\mu =
\frac{1}{m_M} \left(E_{B'} {\avec k} + {\avec k'} E_M \cos \vt
\right) \; .
\nonumber
\eea
In the c.m. system the $\mcf_i$ are related to the
helicity dependent amplitudes via\footnote{Note that there is a
  misprint in Eq. (B12) in Ref. \cite{feusti99}: the $H_4$ term should
  start with $-\sqrt 2 \sin \fvh$.}
\bea
\mcv_{+\frac{1}{2} +\frac{3}{2}} &= 
  \pm \mcv_{-\frac{1}{2} -\frac{3}{2}} =& 
  f \frac{4 \pi \sqrt s }{\sqrt{m_B m_{B'}}}
  \frac{1}{\sqrt 2} \sin \vartheta \cos \frac{\vartheta}{2} 
  (-\mcf_3 - \mcf_4) \; , \nonumber \\
\mcv_{+\frac{1}{2} - \frac{3}{2}} &= 
  \mp \mcv_{- \frac{1}{2} +\frac{3}{2}} =& 
  f \frac{4 \pi \sqrt s }{\sqrt{m_B m_{B'}}}
  \frac{1}{\sqrt 2} \sin \vartheta \sin \frac{\vartheta}{2} 
  (-\mcf_3 + \mcf_4) \; , \nonumber \\
\mcv_{+\frac{1}{2} +\frac{1}{2}} &= 
  \mp \mcv_{-\frac{1}{2} -\frac{1}{2}} =& 
  f \frac{4 \pi \sqrt s }{\sqrt{m_B m_{B'}}}
  \sqrt 2 \cos \frac{\vartheta}{2} \left[ -\mcf_1 + \mcf_2 
    + \sin^2 \frac{\vartheta}{2} (\mcf_3 - \mcf_4) \right] \; , 
  \nonumber \\
\mcv_{+\frac{1}{2} - \frac{1}{2}} &= 
  \pm \mcv_{- \frac{1}{2} +\frac{1}{2}} =& 
  f \frac{4 \pi \sqrt s }{\sqrt{m_B m_{B'}}}
  \sqrt 2 \sin \frac{\vartheta}{2} \left[ \mcf_1 + \mcf_2 
    + \cos^2 \frac{\vartheta}{2} (\mcf_3 + \mcf_4) \right] \; , 
  \nonumber \\
\mcv_{+\frac{1}{2} +0} &= \mp \mcv_{-\frac{1}{2} -0} =& 
  f \frac{4 \pi \sqrt s }{\sqrt{m_B m_{B'}}}
  \varepsilon^0 \cos \frac{\vartheta}{2} (-\mcf_5 - \mcf_6) \; , 
  \nonumber \\
\mcv_{+\frac{1}{2} -0} &= \mp \mcv_{-\frac{1}{2} +0} =& 
  f \frac{4 \pi \sqrt s }{\sqrt{m_B m_{B'}}}
  \varepsilon^0 \cos \frac{\vartheta}{2} (-\mcf_5 + \mcf_6) \; ,
  \label{fandmpivec}
\eea
where the upper (lower) sign and $f = \mi$ ($f = 1$) hold for
pseudoscalar (scalar) meson production. Here, we have used the helicity 
notation introduced in Appendix \ref{appnot}.

\subsection{Spin-$1$ spin-$\foh$ $\ra$ spin-$1$ spin-$\foh$}

Replacing the Dirac operator $\Gamma (s, u)$ by $\Gamma_{\mu \nu} (s,
u) \varepsilon^\mu_{\lambda_V} \varepsilon_{\lambda_{V'}}^{\nu^\dagger}$, it is
straightforward to rewrite $\Gamma_{\mu \nu}$ by 
\bea
\Gamma_{\mu \nu}(s, u) &=& 
      A_{\mu \nu} + B_{\mu \nu} \slash k + 
      C_\nu \gamma_\mu + D_\nu \slash k \gamma_\mu + 
      E_\mu \gamma_\nu + F_\mu \slash k \gamma_\nu + 
      G \gamma_\mu \gamma_\nu + H \slash k \gamma_\mu \gamma_\nu
\hspace*{5mm} \label{gammunuvnvn}
\eea
with
\bea
A_{\mu \nu} &=& A_{pp} p_\mu p_\nu + A_{pp'} p_\mu p'_\nu +
  A_{p'p} p'_\mu p_\nu + 
A_{p'p'} p'_\mu p'_\nu + A_g g_{\mu \nu}, 
\; \mbox{ similarly for } B_{\mu \nu} \; , 
  \nonumber \\
C_\nu &=& C_p p_\nu + C_{p'} p'_\nu, \; \mbox{ similarly for } D_\nu 
\; , \nonumber \\
E_\mu &=& E_p p_\mu + E_{p'} p'_\mu, \; \mbox{ similarly for } F_\mu
  \; . \label{minvnvn}
\eea
Note that this is not a minimal set of Lorentz tensors, since by
applying parity  considerations the minimal set must consist of $3
\times 2 \times 3 \times 2 / 2 = 18$ elements, whereas the above set
contains 20 elements. This is due to the mixing of Lorentz and Dirac
space. An alternative approach would be to span the Lorentz space
first via a basis $n_\mu \equiv \{p_\mu, p'_\mu, k_\mu,
\varepsilon_{\mu \alpha \beta \delta} p^\alpha p'^\beta k^\delta\}$,
and then combining this basis with the nonreducible contractions of
the $\gamma$ matrices with the basis' elements: $\Gamma_{\mu \nu} =
n_\mu n_\nu \otimes \{ \slash k, 1 / \gamma_5 \slash k, \gamma_5 \}$,
where the $\gamma_5$ is needed when exactly one Levi-Civita tensor is
involved. By comparing these two sets one can deduce how to rewrite
the set \refe{minvnvn} in terms of a minimal set of 18 Lorentz 
tensors. However, since it is more straightforward to decompose the
Feynman amplitudes in terms of the set given via Eq. \refe{minvnvn} the 
corresponding formulas are presented for this set. In the notation 
\be
\mcv_{\lambda' \lambda} \equiv \frac{1}{\sqrt {4 m_B m_{B'}} R_+ R'_+} 
\mca_{\lambda' \lambda}
\ee
one finds
\bea
\mca_{+\fth +\fth} 
  &=& - \cos^3 \fvh \left\{ 
  Q_- \left[ 2 \avec k \avec k' \sin^2 \fvh (A_{p'p} - 2
    F_{p'}) + A_g + 2 G \right] + 
Q_+^s 
    \left[ 2 \avec k \avec k' \sin^2 \fvh B_{p'p} + B_g + 2 H \right]
    \right\} \; , \nonumber \\
\mca_{+\fth -\fth} 
  &=& \sin^3 \fvh \left\{ 
  Q_+ \left[ 2 \avec k \avec k' \cos^2 \fvh (A_{p'p} - 2
    F_{p'}) - A_g - 2 G \right] + 
Q_-^s 
    \left[ 2 \avec k \avec k' \cos^2 \fvh B_{p'p} - B_g - 2 H \right]
    \right\} \; , \nonumber \\
\mca_{+\foh +\fth} 
  &=& \cos^2 \fvh \sin \fvh \left\{ 
  Q_+ \left[ 2 \avec k \avec k' \sin^2 \fvh (A_{p'p} - 2
    F_{p'}) + A_g + 2 G \right] + 
  Q_-^s 
    \left[ 2 \avec k \avec k' \sin^2 \fvh B_{p'p} + B_g + 2 H \right]
    +
\right. \nonumber \\
 &&  \hspace{22mm} \left. 
  2 \avec k' \left[ P_- E_{p'} + P_+^s F_{p'} \right] \bigbracket \; 
\right\} \; , \nonumber \\
\mca_{+\foh -\fth} 
  &=& - \sin^2 \fvh \cos \fvh \left\{ 
  Q_{-} \left[ 2 \avec k \avec k' \cos^2 \fvh (A_{p'p} - 2
    F_{p'}) - A_g - 2 G \right] + 
  Q_{+}^s  
    \left[ 2 \avec k \avec k' \cos^2 \fvh B_{p'p} - B_g - 2 H \right]
    +
\right. \nonumber \\
 &&  \hspace{26mm} \left.
  2 \avec k' \left[ P_+ E_{p'} + P_-^s F_{p'} \right] 
\bigbracket \; \right\} \; , \nonumber \\
\mca_{+\fth +\foh} 
  &=& - \cos^2 \fvh \sin \fvh \left\{ 
  Q_+ \left[ 2 \avec k \avec k' \sin^2 \fvh (A_{p'p} - 2
    F_{p'}) + A_g + 2 G \right] + 
  Q_-^s 
  \left[ 2 \avec k \avec k' \sin^2 \fvh B_{p'p} + B_g + 2 H \right] -
\right. \nonumber \\
 &&  \hspace{26mm} \left.
   2 \avec k \left[ P_- (C_p + 2 H) + P_+^s D_p \right] 
       \bigbracket \; \right\} \; , \nonumber \\
\mca_{+\fth -\foh} &=& \sin^2 \fvh \cos \fvh \left\{ 
  Q_- \left[ 2 \avec k \avec k' \cos^2 \fvh (A_{p'p} - 2
    F_{p'}) - A_g - 2 G \right] + 
 Q_+^s 
 \left[ 2 \avec k \avec k' \cos^2 \fvh B_{p'p} - B_g - 2 H \right] +
\right. \nonumber \\
 &&  \hspace{22mm}
 \left. 
    2 \avec k \left[ P_+ (C_p + 2 H) + P_-^s D_p \right] 
    \bigbracket \; \right\} \; , \nonumber \\
\mca_{+\foh +\foh} 
  &=& -\cos \fvh \left\{ 
  Q_- \left[ \left( 2 \avec k \avec k' \sin^2 \fvh (A_{p'p} - 2
    F_{p'}) + A_g \right) \cos^2 \fvh - 2 \sin^2 \fvh G \right] + 
  \right. \nonumber \\
 &&  \hspace{15mm} 
  Q_+^s 
    \left[ \left( 2 \avec k \avec k' \sin^2 \fvh B_{p'p} + B_g \right) \cos^2 \fvh 
      - 2 \sin^2 \fvh H \right] +
    \nonumber \\
 &&  \hspace{15mm} \left.
  2 \sin^2 \fvh \left[ P_+ \left\{ \avec k (C_p + 2 H) + \avec k' E_{p'}
  \right\} + P_-^s ( \avec k D_p + \avec k' F_{p'}) \right] 
  \hspace{8mm} \right\} \; , \nonumber \\
\mca_{+\foh -\foh} 
  &=& -\sin \fvh \left\{ 
  Q_+ \left[ \left( 2 \avec k \avec k' \cos^2 \fvh (A_{p'p} - 2
    F_{p'}) - A_g \right) \sin^2 \fvh + 2 \cos^2 \fvh G \right] + 
  \right. \nonumber \\
 &&  \hspace{14.6mm} 
  Q_-^s 
    \left[ \left( 2 \avec k \avec k' \cos^2 \fvh B_{p'p} - B_g \right) \sin^2 \fvh 
      + 2 \cos^2 \fvh H \right] -
    \nonumber \\
 &&  \hspace{14.6mm} \left. 
  2 \cos^2 \fvh \left[ P_- \left\{ \avec k (C_p + 2 H) - \avec k' E_{p'}
  \right\} + P_+^s ( \avec k D_p - \avec k' F_{p'}) \right] 
  \hspace{8mm} \right\} \; , \nonumber \\
\mca_{+\fth +0} 
  &=& \sqrt 2 \cos^2 \fvh \sin \fvh \left\{ Q_- \left[ 
      \bigbracket \;
      \avec k \left\{ \etp (A_{pp} - 2 F_p) + \etpp (A_{p'p} - 2
        F_{p'}) \right\} - 
      \frac{E_M}{m_M} ( A_g + 2 G) \right] + \right. \nonumber \\
  &&  \hspace{27.5mm} 
  \left. Q_+^s \left[ 
    \avec k \left( \etp B_{pp} + \etpp B_{p'p} \right) - 
    \frac{E_M}{m_M} B_g \right] + 
     \frac{\avec k P_+^s }{m_M} C_p + m_M (\avec k P_- D_p - 2 Q_+ H)
     \right\} \; , \nonumber \\
\mca_{+\fth -0} 
  &=& \sqrt 2 \sin^2 \fvh \cos \fvh \left\{ Q_+ \left[ \bigbracket \;
      \avec k \left\{ \etp (A_{pp} - 2 F_p) + \etpp (A_{p'p} - 2
        F_{p'}) \right\} - 
      \frac{E_M}{m_M} ( A_g + 2 G) \right] + \right. \nonumber \\
  &&  \hspace{27.5mm} 
  \left. Q_-^s \left[ 
    \avec k \left( \etp B_{pp} + \etpp B_{p'p} \right) - 
    \frac{E_M}{m_M} B_g \right] + 
     \frac{\avec k P_-^s}{m_M} C_p + m_M (\avec k P_+ D_p - 2 Q_- H)
     \right\} \; , \nonumber \\
\mca_{+\foh -0} 
  &=& + \mca_{+\fth +0} + \sqrt 2 \sin \fvh \left\{ 
  P_+ \left( \etp E_p + \etpp E_{p'} \right) +
  P^s_- \left( \etp \avec k F_p + \etpp F_{p'} \right) +
  \frac{Q^s_-}{m_M} G + m_M Q_+ H \right\} \; , \nonumber \\
\mca_{+\foh +0} 
  &=& - \mca_{-\fth +0} - 
  \sqrt 2 \cos \fvh \left\{ 
  P_- \left( \etp E_p + \etpp E_{p'} \right) +
  P^s_+ \left( \etp F_p + \etpp F_{p'} \right) 
  + \frac{Q^s_+}{m_M} G + m_M Q_- H  \right\} \; , \nonumber \\
\mca_{+0 +\fth} 
  &=& - \sqrt 2 \cos^2 \fvh \sin \fvh \left\{ 
  Q_- \left[  
    \avec k' ( \eptpp A_{p'p'} +  \eptp A_{p'p} + 2 \eptk F_{p'}) -
    \frac{E_{M'}}{m_{M'}} ( A_g + 2 G) \right] + \right. \nonumber \\
  &&  \hspace{29mm} 
\left.  Q_+^s \left[ 
    \avec k' ( \eptpp B_{p'p'} + \eptp B_{p'p} ) - 
    \frac{E_{M'}}{m_{M'}} (B_g + 2 H) \right] + 
     \frac{\avec k'}{m_{M'}} ( {P'}_+^s E_{p'} + P^{ss}_- F_{p'} ) 
     \right\} \; , \nonumber \\
\mca_{+0 -\fth} 
  &=& \sqrt 2 \sin^2 \fvh \cos \fvh \left\{ 
  Q_+ \left[ 
    \avec k' (\eptpp A_{p'p'} + \eptp A_{p'p} + 2 \eptk F_{p'}) -
    \frac{E_{M'}}{m_{M'}} ( A_g + 2 G) \right] + \right. \nonumber \\
  &&  \hspace{27mm} 
\left.  Q_-^s \left[ \avec k' ( \eptpp B_{p'p'} + \eptp B_{p'p} ) - 
    \frac{E_{M'}}{m_{M'}} (B_g + 2 H) \right] + 
     \frac{\avec k'}{m_{M'}} ( {P'}_-^s E_{p'} - P^{ss}_+ F_{p'} ) 
     \right\} \; , \nonumber \\
\mca_{+0 -\foh} 
  &=& - \mca_{+0 +\fth} + \sqrt 2 \sin \fvh \left\{ \bigbracket \; \; 
  P_+ \left( \eptp C_p + \eptpp C_{p'} - 2 \eptk H \right) +
  \right. \nonumber \\
  && \hspace{37mm} \left. \bigbracket
  P^s_- \left( \eptp D_p + \eptpp D_{p'} \right) +
  \frac{1}{m_{M'}} \left( {Q'}^s_- G + Q^{ss}_+ H \right) \right\} \; , \nonumber \\
\mca_{+0 +\foh} 
  &=& - \mca_{+0 -\fth} + \sqrt 2 \cos \fvh \left\{ \bigbracket \; \; 
  P_- \left( \eptp C_p + \eptpp C_{p'} - 2 \eptk H \right) +
  \right. \nonumber \\
  && \hspace{37mm} \left. \bigbracket
  P^s_+ \left( \eptp D_p + \eptpp D_{p'} \right) -
  \frac{1}{m_{M'}} \left( {Q'}^s_+ G + Q^{ss}_- H \right) \right\} \; , \nonumber \\
\mca_{+0 +0} 
  &=& \cos \fvh \left\{ \bigbracket \; \; 
  Q_- \left[ 
    \etp (\eptp A_{pp} + \eptpp A_{pp'} + 2 \eptk F_p) + 
    \right. \right. \nonumber \\
  &&  \hspace{18.2mm} \left.
    \etpp (\eptpp A_{p'p'} + \eptp A_{p'p} + 2 \eptk F_{p'}) +
    \etep ( A_g + 2 G) \right] + \nonumber \\
  &&  \hspace{11.5mm} 
  Q_+^s \left[ 
    \etp (\eptp B_{pp} + \eptpp B_{pp'} ) + 
    \etpp (\eptpp B_{p'p'} + \eptp B_{p'p} ) +
    \etep ( B_g + 2 H) \right] + \nonumber \\
  && \hspace{11.5mm}  
    \frac{P^s_+}{m_M} ( \eptpp C_{p'} + \eptp C_p - 2 \eptk H ) +
    m_M P_- ( \eptpp D_{p'} + \eptp D_p ) + \nonumber \\
  && \hspace{11.5mm}
    \left. \frac{{P'}^s_+}{m_{M'}} ( \etp E_p + \etpp E_{p'} ) +
    \frac{P^{ss}_-}{m_{M'}}  ( \etp F_p + \etpp F_{p'} ) + 
     \frac{1}{m_{M'} m_M} ( Q_-^{ss} G + m_M^2 {Q'}^s_+ H ) 
     \right\} \; , \nonumber \\
\mca_{+0 -0} 
  &=& \sin \fvh \left\{ \bigbracket \; \; 
  Q_+ \left[ 
    \etp (\eptp A_{pp} + \eptpp A_{pp'} + 2 \eptk F_p) + 
    \right. \right. \nonumber \\
  &&  \hspace{18.2mm} \left.
    \etpp (\eptpp A_{p'p'} + \eptp A_{p'p} + 2 \eptk F_{p'}) +
    \etep ( A_g + 2 G) \right] + \nonumber \\
  &&  \hspace{11.5mm} 
  Q_-^s \left[ 
    \etp (\eptp B_{pp} + \eptpp B_{pp'} ) + 
    \etpp (\eptpp B_{p'p'} + \eptp B_{p'p} ) +
    \etep ( B_g + 2 H) \right] + \nonumber \\
  && \hspace{11.5mm}  
    \frac{P^s_-}{m_M} ( \eptpp C_{p'} + \eptp C_p - 2 \eptk H ) +
    m_M P_+ ( \eptpp D_{p'} + \eptp D_p ) + \nonumber \\
  && \hspace{11.5mm}
    \left. \frac{{P'}^s_-}{m_{M'}} ( \etp E_p + \etpp E_{p'} ) -
    \frac{P^{ss}_+}{m_{M'}}  ( \etp F_p + \etpp F_{p'} ) - 
     \frac{1}{m_{M'} m_M} ( Q_+^{ss} G + m_M^2 {Q'}^s_- H ) 
     \right\}
\eea
with
\bea
\ba{lclclcl}
Q_\pm &=& R'_+ R_+  \pm \avec k \avec k' \; , & &
P_\pm &=& \avec k R'_+ \pm \avec k' R_+ \; , \nonumber \\
Q^s_\pm &=& R'_+ R_+ S_- \pm \avec k \avec k' S_+ \; , & &
P^s_\pm &=& \avec k R'_+ S_+ \pm \avec k' R_+ S_- \; , \nonumber \\
{Q'}^s_\pm &=& R'_+ R_+ S'_- \pm \avec k \avec k' S'_+ \; , & &
{P'}^s_\pm &=& \avec k R'_+ S'_+ \pm \avec k' R_+ S'_- \; , \nonumber \\
Q^{ss}_\pm &=& R'_+ R_+ S_- S'_- \pm \avec k \avec k' S_+ S'_+  \; , &
\hspace{1cm} &
P^{ss}_\pm &=& \avec k R'_+ S_+ S'_- \pm \avec k' R_+ S_- S'_+ 
\ea
\nonumber
\eea
and for $\lambda_V$, $\lambda_{V'} = 0$:
\bea
\eptp &=& \frac{1}{m_{M'}} \left( E_B \avec k' + E_{M'} \avec k \cos
  \vartheta \right) \; , 
\hspace{1cm} 
\eptpp = \frac{\sqrt s \avec k'}{m_{M'}} \; , \nonumber \\
\eptk &=& \frac{1}{m_{M'}} \left( E_M \avec k' - E_{M'} \avec k \cos
  \vartheta \right) \; , 
\hspace{1cm} 
\etep = \frac{1}{m_M m_{M'}} \left( \avec k \avec k' - E_M E_{M'} \cos
  \vartheta \right) \; .
\eea
The other helicity amplitudes follow via
\be
\mca_{\lambda' \lambda} = 
(-1)^{\lambda' - \lambda} \mca_{-\lambda' -\lambda} \; .
\ee
We have checked these formulas numerically against the calculation
method developed by \cite{schaefer}, where the combinations $\bar u
\Gamma_{\mu \nu} u \ve^\mu {\ve'}^{\nu^\dagger}$ have been calculated
by a decomposition of $\Gamma_{\mu \nu}$ into the 16 $4\times4$
Clifford algebra elements.

\section{Partial Waves and Helicity Amplitudes}
\label{apppartials}

In this appendix the relation between the helicity partial waves and
the standard partial waves for $\pi N \ra \pi N$ is given.

Using Eqs. \refe{defhelij}, \refe{fandmpipi} and the well-known
relations between the Wigner $d$ functions and the Legendre
polynomials $P_{\ell_\pi} (x)$, $x=\cos \vt$ one recovers the standard
partial waves:
\bea
\mct^{J\pm}_{\foh \foh} 
&=& \mct^J_{+\foh +\foh} \pm \mct^J_{+\foh -\foh} 
\nonumber \\
&=& - \frac{\sqrt{\avec p \avec p' m_{B'} m_B}}{8\pi \sqrt s} 
\int \difd x 
\left( 
d^J_{+\foh +\foh} (\vt) \mcv^J_{+\foh +\foh} \pm 
d^J_{-\foh +\foh} (\vt) \mcv^J_{+\foh -\foh} \right) 
\nonumber \\
&=& - \frac{\sqrt {\avec p \avec p'}}{2}
\int \difd x \left[
d^J_{+\foh +\foh} (\vt) \cos \vth \left( \tilde A + \tilde B \right)
\pm
d^J_{-\foh +\foh} (\vt) \sin \vth \left( \tilde A - \tilde B \right)
\right]
\nonumber \\
&=& - \frac{\sqrt {\avec p \avec p'}}{2}
\int \difd x (\tilde A P_{\ell_\pi} (x) + \tilde B
P_{\ell_\pi\pm 1} (x) ) 
\nonumber \\
&=& T_{\ell_\pi \pm}^{\pi \pi} \; ,
\eea
where the pion angular momentum $\ell_\pi$ is related to the total
angular momentum by $J = \ell_\pi \pm \foh$.

\section{Isospin Decomposition of Hadronic Reactions}
\label{appiso}

\subsection{Scattering of $(I = 1 \oplus \foh)$ into $(I = 1 \oplus
  \foh)$}

The isospin projection operators for scattering of $(I = 1 \oplus
\foh)$ into $(I = 1 \oplus \foh)$ can be written in a cartesian basis
as
\bea
\left[ \hat P_\foh \right]_{kj} &\equiv&
\langle \varphi_k | \hat P_\foh | \varphi_j \rangle 
= \fot \tau_k \tau_j \; , \nonumber \\
\left[ \hat P_\fth \right]_{kj} &\equiv&
\langle \varphi_k | \hat P_\fth | \varphi_j \rangle 
= \delta_{kj} - \fot \tau_k \tau_j 
\; , \label{isoproj}
\eea
where $| \varphi_j \rangle $ and $\langle \varphi_k |$ refer to the
incoming and outgoing asymptotic isospin-$1$ particles. The possible
charge amplitudes can hence be decomposed into isospin amplitudes 
\bea
\langle \vp_k ; I=\sfoh | \; T_{fi} \; | \vp_j; I=\sfoh \rangle &=& 
\langle \vp_k ; I=\sfoh | \; \hat P_\foh T^\foh_{fi} + 
\hat P_\fth T^\fth_{fi} \; | \vp_j ; I=\sfoh \rangle
\nonumber \\
&=& \langle I=\sfoh | \; \sfot \tau_k \tau_j T^\foh_{fi} + 
(\delta_{kj} - \sfot \tau_k \tau_j ) T^\fth_{fi} \; | I=\sfoh \rangle \; ,
\label{amplii32}
\eea
where $|I=\foh \rangle$ and $\langle I=\foh |$ have to be replaced by
the isospinors $\chi_\pm = |\foh,\pm \foh \rangle$. Using the pion
phase convention $| \pi^\pm \rangle = \frac{\mp 1}{\sqt} | 1, \pm \mi, 
0 \rangle$, this leads explicitly to
\bea
\langle 1, +1; \; \sfoh, +\sfoh | \; T_{fi} \; | 1, +1; \; \sfoh, +\sfoh
=
\langle 1, -1; \; \sfoh, -\sfoh | \; T_{fi} \; | 1, -1; \; \sfoh, -\sfoh
\rangle &=& T^\fth_{fi} \; , \nonumber \\
\langle 1, -1; \; \sfoh, +\sfoh | \; T_{fi} \; | 1, -1; \; \sfoh, +\sfoh
=
\langle 1, +1; \; \sfoh, -\sfoh | \; T_{fi} \; | 1, +1; \; \sfoh, -\sfoh
\rangle &=& \fot ( T^\fth_{fi} + 2 T^\foh_{fi} ) \; , \nonumber \\
\langle 1, \; \; \, 0; \; \sfoh, -\sfoh | \; T_{fi} \; 
| 1, -1; \; \sfoh, +\sfoh \rangle 
=
\langle 1, \; \; \, 0; \; \sfoh, +\sfoh | \; T_{fi} \; 
| 1, +1; \; \sfoh, -\sfoh \rangle &=& \frac{\sqt}{3} 
( T^\fth_{fi} - T^\foh_{fi} ) \; , 
\label{amplii32spec} \\ 
\langle 1, \; \; \, 0; \; \sfoh, +\sfoh | \; T_{fi} \; 
| 1, \; \; \, 0; \; \sfoh, +\sfoh \rangle 
=
\langle 1, \; \; \, 0; \; \sfoh, -\sfoh | \; T_{fi} \; 
| 1, \; \; \, 0; \; \sfoh, -\sfoh \rangle &=& \fot 
( 2 T^\fth_{fi} + T^\foh_{fi} ) 
\; , \nonumber
\eea
which is is in line with the Condon-Shortley convention \cite{condon}
and the usual Clebsch-Gordan coefficients \cite{pdg}.

\subsection{Scattering of $(I = 1 \oplus \foh)$ into $(I = 0 \oplus
  \foh = \foh)$} 

Choosing the $I=\foh$ projection operator in accordance with the
Condon-Shortley convention and hence correctly normalized,
\bea
\left[ \hat P_\foh \right]_j = \frac{-1}{\sqth} \tau_j \; ,
\nonumber
\eea
the isospin decomposed amplitudes are
\bea
\langle I = 0 ; \; I=\sfoh | \; T_{fi} \; | I = 1 ; \; I=\sfoh \rangle = 
\langle I=\sfoh | \; -\smallfrac{1}{\sqth} \tau_j T^\foh_{fi} \; |
I=\sfoh \rangle
\label{amplii12}
\eea
and explicitly
\bea
\langle 0, 0; \; \sfoh, -\sfoh | \; T_{fi} \; | 1, -1; \; \sfoh, +\sfoh
\rangle &=& -\frac{\sqt}{\sqth} T^\foh_{fi} \; , \nonumber \\
\langle 0, 0; \; \sfoh, +\sfoh | \; T_{fi} \; | 1, +1; \; \sfoh, -\sfoh
\rangle &=& \frac{\sqt}{\sqth} T^\foh_{fi} \; , \nonumber \\
\langle 0, 0; \; \sfoh, +\sfoh | \; T_{fi} \; | 1, \; \; \, 0; \;
\sfoh, +\sfoh \rangle &=& \frac{-1}{\sqth} T^\foh_{fi} \; , \nonumber \\
\langle 0, 0; \; \sfoh, -\sfoh | \; T_{fi} \; | 1, \; \; \, 0; \;
\sfoh, -\sfoh \rangle &=& \frac{1}{\sqth} T^\foh_{fi} \; . 
\label{amplii12spec}
\eea

\subsection{Isospin operators in the interaction Lagrangians}
\label{appisoop}

The isospin operators in the hadronic interaction Lagrangians for
$1+2\ra 3$ are given in Table \ref{tabisoop},
\begin{table}
  \begin{center}
    \begin{tabular}
      {r|r|r|l}
      \hhline{====}
      $I_1$ & $I_2$ & $I_3$ & operator \\ 
      \hline 
      $0$ & $\foh$ & $\foh$ & $\chi_3^\dagger \chi_2$ \\ 
      $0$ & $1$ & $1$ & $\vec \varphi_3^\dagger \vec \varphi_2$ \\ 
      $1$ & $\foh$ & $\foh$ & 
      $\chi_3^\dagger \vec \tau \cdot \vec \varphi_1 \chi_2$ \\ 
      $1$ & $1$ & $1$ & 
      $\mi \vec \varphi_3^\dagger \cdot (\vec \varphi_1 \times \vec
      \varphi_2)$ \\ 
      $1$ & $\foh$ & $\fth$ & 
      $\vec T_3^\dagger \cdot \vec \varphi_1 \chi_2$ \\ 
      \hhline{====}
    \end{tabular}
  \end{center}
  \caption{Isospin operators in the interaction Lagrangians for $1+2
    \ra 3$. For the notation, see text. The missing normalization
    factor of $\foh$ for $1 \oplus \foh \ra \foh$ is absorbed in the
    coupling constant. Note that in the last case, the coefficient
    resulting from the transition operator is just the Clebsch-Gordan
    coefficient $(\sfth , {I_3}_z | 1, {I_1}_z ;\sfoh
    ,{I_2}_z)$.\label{tabisoop}}
\end{table}
where the vector $\vec T$ for $I=\fth$ particles is given by
\bea
\vec T (M)^\dagger = \sum_{r,m} (\sfth , M | 1, r ; \sfoh ,m) \vec
\varphi_r^\dagger \chi_m^\dagger
\nonumber
\eea
with $(\sfth , M | 1, r ; \sfoh ,m)$ the usual Clebsch-Gordan
coefficients (see, e.g., Ref. \cite{pdg}).

\section{Observables}
\label{appobs}

\subsection{Cross sections}

The uniform differential cross section expression for all reactions is:
\bea
\frac{d \sigma}{d \Omega} &=& 
\frac{4 m_B m_B'}{4 (4\pi)^2 s} 
\frac{{\avec k}'}{\avec k} \frac{1}{s_i} 
\sum \limits_{\lambda , \lambda'} 
\left| \mcm_{\lambda' \lambda} (\vt) \right|^2
\nonumber \\
&=& \frac{(4\pi)^2}{\avec k^2} \frac{1}{s_i}
\sum \limits_{\lambda , \lambda'} 
\left| \mct_{\lambda' \lambda} (\vt) \right|^2 \; ,
\eea
where Eq. \refe{tandm} was used and the sum extends over all values of
$\lambda$ and $\lambda'$. $s_i$ is the usual spin averaging
factor for the initial state. The amplitude $\mct_{\lambda' \lambda}
(\vt)$ is given by (e.g., for $\lambda,\lambda' > 0$):
\bea
\mct_{\lambda' \lambda} (\vt) 
&=& \frac{1}{2 \pi}
\sum \limits_J (J + \sfoh) d^J_{\lambda \lambda'} (\vt) 
\mct_{\lambda' \lambda}^J
\nonumber \\
&=& \frac{1}{4 \pi}
\sum \limits_J (J + \sfoh) d^J_{\lambda \lambda'} (\vt) 
\left( 
\mct_{\lambda' \lambda}^{J+} + \mct_{\lambda' \lambda}^{J-}
\right)
\label{amplitdif} \; .
\eea
The total cross section reads for all reactions
\bea
\sigma = \frac{4 \pi}{\avec k^2} 
\frac{1}{s_i} \sum_{J,P} \sum_{\lambda ,\lambda'} (J + \sfoh)
\left| \mct_{\lambda' \lambda}^{JP} \right|^2 \; ,
\eea
where the second sum extends only over positive $\lambda$ and
$\lambda'$.

\subsection{Recoil polarization}

The recoil asymmetry results in
\bea
\mci (\vt) \mathcal P &=& 
2 \mbox{Im} \mct_{+\foh +\foh} \mct_{-\foh +\foh}^*
\label{pipipol} \; ,
\eea
where we have used the amplitude of Eq. \refe{amplitdif} and the
cross section intensity
\bea
\mci (\vt) \equiv 
\foh \sum \limits_{\lambda , \lambda'} 
\left| \mct_{\lambda' \lambda} (\vt) \right|^2 \; .
\label{crossint}
\eea
Here the sum extends over all possible values for $\lambda$ and 
$\lambda'$.

\end{appendix}

\end{document}